\documentclass[fleqn,usenatbib]{mnras}
\usepackage{newtxtext,newtxmath}
\usepackage[T1]{fontenc}
\usepackage{ae,aecompl}

\usepackage{graphicx}
\usepackage{amsmath}
\usepackage{xspace}
\usepackage[usestackEOL]{stackengine}

\usepackage{xcolor}

\hypersetup{
pdftitle = {Measuring the evolution of intergalactic gas from z=0 to 5 using the kinematic Sunyaev-Zel'dovich effect},
pdfsubject = {Cosmology},
pdfauthor = {Chaves-Montero, Jonas},
pdfkeywords = {Large-scale structure of the Universe}
}

\newcommand{\Planck}{\textit{Planck}\xspace }

\newcommand{\Healpix}{{\sc healpix}\xspace }
\newcommand{\Cola}{{\sc cola}\xspace }

\newcommand{\Lcola}{{\sc l-picola}\xspace }
\newcommand{\Commander}{{\sc commander}\xspace }
\newcommand{\Nilc}{{\sc nilc}\xspace }
\newcommand{\Sevem}{{\sc sevem}\xspace }
\newcommand{\Sevema}{{\sc sevem-}{\scriptsize 100}\xspace }
\newcommand{\Sevemb}{{\sc sevem-}{\scriptsize 143}\xspace }
\newcommand{\Sevemc}{{\sc sevem-}{\scriptsize 217}\xspace }
\newcommand{\Smica}{{\sc smica}\xspace }
\newcommand{\Smicanosz}{{\sc smica-nosz}\xspace }
\newcommand{\tauAP}{\ensuremath{\tau_{\rm AP}}\xspace }
\newcommand{\simtauAP}{\ensuremath{\mathbf{T}_{\rm sim}}\xspace }
\newcommand{\bsimtauAP}{\ensuremath{\bar{\mathbf{T}}_{\rm sim}}\xspace }

\newcommand{\wfig}{0.320\textwidth}
\newcommand{\hfig}{0.245\textwidth}
\newcommand{\wfigb}{0.245\textwidth}
\newcommand{\hfigb}{0.205\textwidth}

\newcommand{\tauAPth}{\ensuremath{\tau_{\rm AP}^{\rm th}}\xspace }

\newcommand{\Mpc}{\ensuremath{\mathrm{Mpc}}}
\newcommand{\hMpc}{\ensuremath{h\,\mathrm{Mpc}^{-1}}}
\newcommand{\Mpch}{\ensuremath{h^{-1}\mathrm{Mpc}}}

\newcommand{\Msunh}{\ensuremath{h^{-1}\,\mathrm{M}_\odot}}

\newcommand{\degc}{\ensuremath{\mathrm{deg}^2}}

\newcommand{\angr}{ \ensuremath{\hat{\bf \Omega} }}
\newcommand{\rhom}{ \ensuremath{n_\Omega} }
\newcommand{\zm}{ \ensuremath{z_\Omega} }

\newcommand{\thetaAP}{ \ensuremath{\theta_{\rm AP}} }
\newcommand{\deltaang}{ \ensuremath{\delta^K_{\angr_j \angr_i}} }
\newcommand{\arfksz}{ \ensuremath{C_\ell^{{\rm ARF}-{\rm kSZ}}} }
\newcommand{\hatarfksz}{ \ensuremath{\hat{C}_\ell^{{\rm ARF}-{\rm kSZ}}} }

\newcommand{\pgas}{ \ensuremath{\Delta_\mathrm{gas} } }

\title[Intergalactic gas and the kSZ effect]{Measuring the evolution of intergalactic gas from $z=0$ to $5$ using the kinematic Sunyaev-Zel'dovich effect}

\author[J. Chaves-Montero et al.]{
\parbox[h]{0.95\textwidth}{Jon\'as Chaves-Montero$^{1,2,3}$\thanks{E-mail: \href{mailto:jonas.chaves@dipc.org}{jonas.chaves@dipc.org}}, Carlos Hern\'andez-Monteagudo$^{4}$, Ra\'ul E. Angulo$^{3,5}$ and J.D. Emberson$^{6}$}
\vspace*{6pt}
\\$^{1}$ HEP Division, Argonne National Laboratory, 9700 South Cass Avenue, Lemont, IL 60439, USA.
\\$^{2}$ Centro de Estudios de F\'isica del Cosmos de Arag\'on, Plaza San Juan 1, Planta-3, 44001, Teruel, Spain.
\\$^{3}$ Donostia International Physics Centre, Paseo Manuel de Lardizabal 4, 20018 Donostia-San Sebastian, Spain.
\\$^{4}$ Centro de Estudios de F\'isica del Cosmos de Arag\'on, Unidad Asociada CSIC, Plaza San Juan 1, Planta-3, 44001, Teruel, Spain.
\\$^{5}$ IKERBASQUE, Basque Foundation for Science, E-48013 Bilbao, Spain.
\\$^{6}$ CPS Division, Argonne National Laboratory, 9700 South Cass Avenue, Lemont, IL 60439, USA.
}

\date{Accepted XXX. Received YYY; in original form ZZZ}
\pubyear{2020}

\begin{document}
\label{firstpage}
\pagerange{\pageref{firstpage}--\pageref{lastpage}}
\maketitle

\begin{abstract}
A complete census of baryons in the late universe is a long-standing challenge due to the intermediate temperate and rarefied character of the majority of cosmic gas. To gain insight into this problem, we extract measurements of the kinematic Sunyaev-Zel'dovich (kSZ) effect from the cross-correlation of angular redshift fluctuations maps, which contain precise information about the cosmic density and velocity fields, and CMB maps high-pass filtered using aperture photometry; we refer to this technique as ARF-kSZ tomography. Remarkably, we detect significant cross-correlation for a wide range of redshifts and filter apertures using 6dF galaxies, BOSS galaxies, and SDSS quasars as tracers, yielding a 11 sigma detection of the kSZ effect. We then leverage these measurements to set constraints on the location, density, and abundance of gas inducing the kSZ effect, finding that this gas resides outside dark matter haloes, presents densities ranging from 10 to 250 times the cosmic average, and comprises half of cosmic baryons. Taken together, these findings indicate that ARF-kSZ tomography provides a nearly complete census of intergalactic gas from $z=0$ to 5.
\end{abstract}



\section{Introduction}
\label{sec:introduction}

Over the last decades, precise observations of primordial CMB anisotropies \citep[e.g.,][]{Planck2018VI} and Big Bang nucleosynthesis studies \citep[e.g.,][]{Cooke2018} have set strict constraints on the abundance and distribution of baryons in the early universe. However, a complete census of baryons at late times remains elusive; this is principally due to the intermediate temperature and rarefied character of nearly all cosmic gas, which hinders the detection of baryons outside high-density regions and leaves invisible the majority of cosmological volume. Such is the case that until recently, low redshift studies only detected 70\% of the expected amount of baryons \citep{Fukugita04, Nicastro08, Shull12}.

Nonetheless, recent studies have successfully detected baryons outside high-density areas conducting kinematic Sunyaev-Zel'dovich (kSZ) effect observations \citep{chm15, HillprjkSZ16}, thermal Sunyaev-Zel'dovich (tSZ) studies \citep{deGraaff17, Tanimura19}, low-redshift Lyman-$\alpha$ surveys \citep{Gallego2018}, and deep X-ray campaigns \citep{Nicastro18, Kovacs2019}. Despite their success, these works only set constraints on the distribution of baryons at either a few specific redshifts or across a reduced number of line-of-sights, failing to provide a complete picture of cosmic gas in the late universe.

Of the different strategies listed above, throughout this work we set constraints on the properties of intergalactic gas using measurements of the kSZ effect, which refers to the Doppler boosting of CMB photons as these scatter off free electrons moving relative to the CMB rest frame \citep{Sunyaev1972, Sunyaev1980}. The motivation of using these measurements is that the kSZ effect is sensitive to free electrons independently of the temperature and density of the medium in which these reside, and thus it is uniquely suited to study the large-scale distribution of baryons at low redshift.

Even though the kSZ effect presents significant advantages to observe cosmic gas, it is challenging to extract this effect from observations because the amplitude of kSZ fluctuations is approximately two orders of magnitude smaller than that of primordial CMB fluctuations and the spectral shape of both signals is practically the same. Furthermore, most extraction methods require estimating the peculiar velocity field of intervening matter \citep[e.g.,][]{planck_unbound, schaan16}, which is also challenging and adds notable uncertainties. Other approaches circumvent such estimation but require either using cosmological simulations for calibration or modelling and subtracting other effects \citep{Ferreira99, Hand2012, HillprjkSZ16, FerraroprjkSZ16}, which also introduces substantial uncertainties.

In this scenario, the cross-correlation angular redshift fluctuations \citep[ARF,][HM19 in what follows]{CHM_ARF_2019}, which encode precise information about the cosmic density and velocity fields, and CMB observations provides a clean window towards a tomographic detection of the kSZ effect given that systematic uncertainties affecting either of these observables do not present significant correlation. This approach, which we refer to as ARF-kSZ tomography, requires redshift information from either spectroscopic or spectro-photometric surveys as well as theoretical predictions for the large-scale cross-correlation between ARF and the kSZ effect. Interestingly, in contrast with the majority of kSZ estimators, ARF-kSZ tomography involves a new observable that cannot be reduced to the bispectrum of density fluctuations and temperature anisotropies \citep{simthetal_kSZbisp19}.

We start by deriving theoretical predictions for the cross-correlation of ARF and the kSZ effect, and then we generate ARF maps at different redshifts using galaxies from the 6dF Galaxy Survey \citep[6dF;][]{jones04}, galaxies from the Baryon Oscillation Spectroscopic Survey \citep[BOSS;][]{eisenstein11, dawson13}, and quasars from the extended Baryon Oscillation Spectroscopic survey \citep[eBOSS;][]{myers15}. We continue by extracting the kSZ signal induced by gas surrounding these galaxies and quasars from the cross-correlation ARF maps and \Planck maps high-pass filtered using aperture photometry. Remarkably, we find statistically significant correlation for a wide range of redshifts and filter apertures. Lastly, we leverage these measurements to set constraints on the location, density, and abundance of kSZ gas from redshift $z=0$ to 5.

The remainder of this paper is organised as follows. We derive the dependence of the power spectrum of both ARF and the kSZ effect on cosmological parameters in \S\ref{sec:theory}, and then we use this information establish the foundations of ARF-kSZ tomography in \S\ref{sec:foundations}. In \S\ref{sec:thsim}, we use cosmological simulations to quantify the precision of our theoretical derivations and study the large-scale distribution of cosmic gas. We continue by cross-correlating ARF and filtered CMB maps to extract measurements of the kSZ effect in \S\ref{sec:resobv}, and then we analyse these to set constraints on the location, properties, and abundance cosmic gas in \S\ref{sec:gprop}. We study the robustness of our results in \S\ref{sec:robustness} and we summarise our main findings and conclude in \S\ref{sec:conclusion}.

Throughout this work, we use \Planck 2015 cosmological parameters \citep{planck14b}: $\Omega_{\rm m}= 0.314$, $\Omega_\Lambda = 0.686$, $\Omega_{\rm b} = 0.049$, $\sigma_8 = 0.83$, $h_0 = 0.67$, and $n_{\rm s} = 0.96$.


\section{Theory preambles}
\label{sec:theory}

In this section, we seek to derive the dependence of angular redshift fluctuations and the kinematic Sunyaev-Zel'dovich effect on cosmology.


\subsection{Angular redshift fluctuations}
\label{sub:theory_arf}

As recently shown in HM19, angular fluctuations in sky maps of galaxy redshifts contain precise information about the cosmic density and velocity fields. To derive the dependence of the power spectrum of these fluctuations on cosmological parameters, we start by projecting galaxy redshifts weighted by a radial selection function $W$ onto a sky map

\begin{equation}
\label{eq:redshift_field}
\mathcal{Z}^{2D}(\angr) = \int \mathrm{d}s\, s^2 \left[z + (1+z)\frac{v}{c}\right] W(s)\, n({\bf s}),
\end{equation}

\noindent where $s=r+(1+z)[v/H(z)]$ indicates redshift-space coordinates, ${\bf r}$ denotes comoving coordinates, $\angr$ stands for a unitary angular vector, $v$ refers to the radial component of galaxy peculiar velocities, $H(z)=H_0\sqrt{\Omega_m(1+z)^3+\Omega_\Lambda}$ is the Hubble parameter in a flat universe dominated by matter and dark energy, and $n$ denotes the number density of tracers. In the equation above, the first and second terms enclosed in brackets account for the dependence of galaxy redshifts on the Hubble flow and peculiar velocities, respectively. Note that we do not consider relativistic corrections owing to the small amplitude of these; we refer the reader to HM19 for further details.

We continue by expanding the selection function around configuration space coordinates $W(s)\simeq W(r)+(1+z)[v/H(z)](\mathrm{d}W/\mathrm{d}r)$ while restricting ourselves to first-order terms in perturbations; this is motivated by the modest signal-to-noise ratio of kSZ measurements in \S\ref{sec:resobv}. In this manner, we find

\begin{multline}
\mathcal{Z}^{2D}(\angr) = \zm + \int \mathrm{d}r\, r^2\\ \bar{n}(r) \left\{z W(r) \delta_g({\bf r}) + (1+z) \left[W(r) + \frac{z c}{H(z)} \frac{\mathrm{d}W}{\mathrm{d}r}\right] \frac{v}{c} \right\},
\end{multline}

\noindent where $\zm=\int\mathrm{d}z\,z\,W(z)N(z)$ is the integral of galaxy redshifts weighted by the selection function, $N(z)$ refers to the number of sources at redshift $z$, $\delta_g({\bf r})=[n({\bf r})-\bar{n}(r)]/\bar{n}(r)$ indicates the density contrast of galaxies, and $\bar{n}(r)=r^{-2}(\mathrm{d}z/\mathrm{d}r) N(z)$ denotes the average number density of tracers at comoving distance $r$. It is important to note that the first and second terms between brackets depend on the cosmic density and velocity fields, respectively.

Before pursuing our derivation, we assume that there is no systematic shift between the velocity of galaxies and dark matter haloes and that the large-scale bias of galaxies $b$ is approximately scale independent $\hat{\delta}_g=b\,\hat{\delta}_k$ \citep[e.g.,][]{mo96}, where $\hat{\delta}_g$ and $\hat{\delta}_k$ are the Fourier transforms of the density contrast of galaxies and the matter density field, respectively. Furthermore, we consider the linearised continuity equation to connect the velocity and density fields: $\hat{\bf v}_k= -i\,f(z)H(z)/(1+z) \hat{\delta}_k (\hat{\rm k}/k)$, where $\hat{\rm k}$ indicates the radial vector in Fourier space, $f(z)=-(1+z)(\mathrm{d}\log D/\mathrm{d}z)$ refers to the linear growth of velocities, and $D(z)=H_0^{-1} H(z)\int_z^\infty \mathrm{d}z' H^{-3}(z') (1+z') \left[\int_0^\infty \mathrm{d}z' H^{-3}(z')(1+z')\right]^{-1}$ accounts for the linear growth of density perturbations. 

To gain insight into the cosmological information encoded in the projected redshift map, we decompose relative fluctuations around the mean of $\mathcal{Z}^{2D}$ into a series of spherical harmonics while considering the aforementioned approximations, finding that the ARF power spectrum can be assembled by introducing the kernels 

\begin{equation}
\label{eq:wdeltaz}
K^{\delta_z}_\ell(k) = \zm^{-1} \int \mathrm{d}z \,z\, D(z) b(z) W(z) N(z) j_\ell[k r(z)],
\end{equation}

\begin{multline}
\label{eq:wvz}
K^{v_z}_\ell(k) = \zm^{-1} \int \mathrm{d}z\, 
\frac{H(z)}{c} D(z) f(z) \Sigma_v(k) \\
\left(1 + \frac{\mathrm{d}\log W}{\mathrm{d}\log z} \right)
W(z) N(z) \frac{j'_\ell[k r(z)]}{k},
\end{multline}

\noindent into the following equation 

\begin{equation}
\label{eq:template}
C_\ell^{\alpha \beta} = \frac{2}{\pi} \int \mathrm{d}k \, k^2 P(k, z=0) K^{\alpha}_\ell(k) K^{\beta}_\ell(k),
\end{equation}

\noindent where $P(k, z=0)$ indicates the matter power spectrum at $z=0$, $\alpha$ and $\beta$ run over the density and velocity kernels $\delta_z$ and $v_z$, respectively, $j_\ell$ and $j'_\ell$ indicate the spherical Bessel function of order $\ell$ and its derivative, and $\Sigma_v=\exp(-k^2\sigma_v^2)$ refers to the suppression of the power spectrum due to small-scale peculiar velocities. Throughout this work, we use the publicly available Boltzmann solver {\sc camb} \citep{lewis00} to compute the matter power spectrum.


\subsection{Kinematic Sunyaev-Zel'dovich effect}
\label{sub:theory_ksz}

The kinematic Sunyaev-Zel'dovich effect is one of the most important sources of CMB secondary anisotropies \citep{Sunyaev1980}. In the non-relativistic limit, temperature fluctuations induced by this effect are given by the following expression

\begin{equation}
\label{eq:kSZ}
T_{\rm kSZ}(\angr) \equiv \frac{\delta T_{\rm kSZ}(\hat{\bf \Omega})}{T_{\rm CMB}} = - \sigma_T^{} \int \mathrm{d}l\, n_e \frac{v_e}{c},
\end{equation}

\noindent where $T_{\rm CMB}$ indicates the average temperature of primordial CMB radiation, $\sigma_T^{}$ stands for the Thomson scattering cross section, $n_e$ and $v_e$ refer to the physical number density and radial peculiar velocity of free electrons, respectively, $c$ denotes the speed of light, and the integral is performed along the proper line-of-sight $l$. As it is standard in the spherical coordinate system, we assign a positive (negative) radial velocity to gas moving away from (towards) the observer. We do not consider relativistic corrections because these are only significant for clusters of galaxies \citep{Nozawa1998, Sazonov1998}, while we extract the kSZ effect using as tracers galaxies and quasars hosted by moderate mass haloes.

We begin our derivation by projecting the kSZ signal induced by gas surrounding a set of tracers onto a sky map

\begin{equation}
\label{eq:maptt}
T_{\rm kSZ}(\angr) = \rhom^{-1} \int \mathrm{d}r \, r^2 \bar{n}(r) W(s) \int \mathrm{d}r' g(r-r') \frac{v_e}{c},
\end{equation}

\noindent where $\rhom=\int\mathrm{d}z\,W(z)N(z)$ is the integral of the number density of tracers weighted by the selection function, $g(r-r')=-\sigma_T^{}\,n_e(r-r')(1+z)^{-1}$ refers to the strength of the kSZ effect generated by a distribution of gas $n_e(r-r')=\bar{n}_e\,\pgas(r-r')$, $\bar{n}_e=(f_e/m_p)\rho_{\rm cr}\,\Omega_{\rm b}(1+z)^3$ is the physical cosmic number density of electrons, $\pgas(r-r')$ denotes a spherically symmetric overdensity of gas at comoving distance $r-r'$ from tracers, $\rho_{\rm cr}$ stands for the critical density at present time, and $m_p$ is the proton mass. In the previous expression, the number of electrons per unit of baryonic mass is given by

\begin{equation}
f_e = \frac{1-Y[1-N_{\rm He}(z)/4]}{\mu_e (1-Y/2)},
\end{equation}

\noindent where $N_{\rm He}$ indicates the number of helium ionizations, $Y=0.248$ denotes the primordial helium abundance, and $\mu_e = 1.14$ is the effective number of electrons per nucleon. Throughout this work, we assume that the ionization of hydrogen is completed before $z=5$\footnote{Depending on the model, the ionization of hydrogen occurs between $z=7.8$ and 8.8 \citep{Planck16_reionization}.} and that the reionization of {\small HeII} occurs instantaneously at $z=3$ \citep[e.g.,][]{Becker2015_reionization}; therefore, the number of helium ionizations is $N_{\rm He}=1$ and 2 before and after $z=3$, respectively.

To continue, we proceed in the same manner as in \S\ref{sub:theory_arf}: we decompose Eq.~\ref{eq:maptt} into a series of spherical harmonics while considering analogous simplifying assumptions. After some algebra, we find that the power spectrum of the kSZ effect can be assembled by introducing the kernel

\begin{multline}
\label{eq:wksz}
K^{\rm kSZ}_\ell(k) = \frac{-\tau_{\rm eff}}{\rhom} \int \mathrm{d}z \,(1+z) \frac{H(z)}{c} D(z) f(z) \Sigma_v(k)\\ 
W(z) N(z) \frac{j'_\ell[k r(z)]}{k},
\end{multline}

\noindent in Eq.~\ref{eq:template}. In the previous expression, $\tau_{\rm eff}=\sigma_T^{}\bar{n}_{e,0} \int\mathrm{d}r\,\pgas(r)$ provides the amplitude of the kSZ effect as a function of the overdensity of gas surrounding tracers, while $\bar{n}_{e,0}$ refers to the comoving cosmic number density of electrons. Throughout the remainder of this paper, we refer to $\tau_{\rm eff}$ as kSZ optical depth. Note that to derive the previous expression we assume that velocities remain correlated throughout the line-of-sight integral defining $\tau_{\rm eff}$, which is naturally expected because the correlation length of velocities is approximately one order of magnitude larger than the average extension of gas overdensities.


\section{Foundations of ARF-kSZ tomography}
\label{sec:foundations}

In this section, we set the basis of ARF-kSZ tomography, a new technique that leverages the cross-correlation of ARF and high-pass filtered CMB maps to extract measurements of the kSZ effect. We start by describing our strategy to produce ARF and filtered CMB maps in \S\ref{sub:foundations_mapARF} and \ref{sub:foundations_filtered_maps}, respectively, we derive the dependence of the cross-correlation of these maps on cosmology in \S\ref{sub:foundations_cross}, and we estimate the range of scales across which these maps are strongly correlated in \S\ref{sub:foundations_scales}.


\subsection{Angular redshift fluctuations maps}
\label{sub:foundations_mapARF}

To generate ARF maps\footnote{Note that Eq.~\ref{eq:map_arf} differs slightly from the the estimator used in HM19; nevertheless, both produce similar results on the scales of interest for ARF-kSZ tomography.}, we project the redshifts of a sample of galaxies or quasars onto the pixelated surface of a sphere using the following expression

\begin{equation}
\label{eq:map_arf}
\mathcal{M}_{\rm ARF}(\angr_j) = \log \left[\frac{\zm^{-1} \sum_i z_i\, W(z_i) \deltaang}{\rhom^{-1} \sum_i W(z_i) \deltaang}\right],
\end{equation}

\noindent where $i$ and $j$ run over tracers and pixels, respectively, while $\delta^K_{ij}$ is the Kronecker delta function. In contrast with the ARF map presented in Eq.~\ref{eq:redshift_field}, which is equivalent to the numerator of the previous expression, the pixels of the $\mathcal{M}_{\rm ARF}$ map capture the average redshift of all sources falling in these weighted by the selection function. As a result, observational systematics affecting the angular number density of sources present limited influence on this map (see also HM19 and \S\ref{sub:robustness_numang}), while the impact of this type of systematics is substantial on traditional clustering studies \citep[e.g.,][]{ross17}.


\subsection{Filtered CMB maps}
\label{sub:foundations_filtered_maps}

In this section, we present our strategy to high-pass filter CMB maps aiming to decouple the kSZ effect from primordial CMB anisotropies.

To alleviate the impact of primordial CMB contamination on the kSZ signal generated by gas surrounding a set of galaxies, we apply high-pass filters of constant aperture on the sky coordinates of these galaxies. Even though the efficiency of this technique, which is commonly known as aperture photometry \citep[e.g.,][]{hernandez-monteagudo2004}, is inferior relative to others such as matched filtering, aperture photometry removes primordial CMB fluctuations constant over the filter aperture without the need of providing the angular dependence of the target signal. This last property is critical for kSZ studies because it is very challenging to estimate the large-scale distribution of gas inducing this effect; actually, this is one of the main goals of this work.

To apply aperture photometry we proceed as follows. First, we compute the average temperature of the CMB to within both a circle of radius $\thetaAP$ and an annulus of radii $\thetaAP$ and $\thetaAP\sqrt{2}$ centred at each tracer that we use to generate ARF maps. Then, we subtract the average temperature in the annulus from that in the circle to eliminate temperature fluctuations constant over the filter aperture, thereby removing primordial CMB anisotropies larger than the aperture. This process also erase kSZ fluctuations wider than the filter aperture; we model the impact of aperture photometry on kSZ measurements as follows

\begin{equation}
\label{eq:tauap}
\tauAPth\left(\thetaAP\right) = \tau\left(0,\,\thetaAP\right) - \tau\left(\thetaAP,\,\thetaAP\sqrt{2}\right),
\end{equation}

\noindent where $\tau(x, y) = 2 \sigma_T^{} \bar{n}_{e,0} \thetaAP^{-2} \int_x^y \theta \, \mathrm{d}\theta \int \mathrm{d}l \, \pgas(r)$ indicates the kSZ signal induced by gas enclosed to within an annulus of radii $\theta=x$ and $y$ centred at each tracer, $\theta$ and $l$ refer to the radial and vertical coordinates of an imaginary cylinder at each tracer, respectively, and $r=\sqrt{\theta^2+l^2}$ stands for the radial distance in spherical coordinates.

The previous equation shows that the impact of aperture photometry on kSZ measurements depends on the distribution of gas inducing this effect. Given that galaxies are surrounded by gas, we expect the kSZ signal to increase with the filter aperture as more gas is enclosed by this, reach a maximum for apertures encompassing the majority of gas moving coherently with tracers, and approach zero for even larger apertures as the two terms of Eq.~\ref{eq:tauap} converge to the same asymptotically value. An interesting corollary of this angular dependence is that kSZ measurements must be positive in the absence of other sources of uncertainty.

Given that aperture photometry does not subtract CMB fluctuations smaller than the filter aperture, this technique provides a noisy estimate of the kSZ effect in the sky direction of each tracer. To further reduce the impact of primordial CMB anisotropies as well as other sources of contamination, we combine kSZ estimates from different galaxies to produce a filtered CMB map

\begin{equation}
\label{eq:map_AP}
\mathcal{M}_{\rm kSZ} (\angr_j) = \frac{\sum_i T_{{\rm kSZ},i}^{\rm AP} W(z_i) \deltaang}{\sum_i W(z_i) \deltaang},
\end{equation}

\noindent where $i$ and $j$ run over tracers and pixels, respectively, while $T_{{\rm kSZ},i}^{\rm AP}$ indicates the result of applying aperture photometry to the tracer $i$. Each pixel of the map contains the average kSZ estimate of all tracers falling in such pixel weighted by the same selection function that we use to generate ARF maps; consequently, filtered CMB maps are robust against systematic uncertainties affecting the angular number density of tracers for the same reasons as ARF maps.

For the sake of definiteness, throughout this section we assume that millimetre observations only contain primordial CMB anisotropies and kSZ fluctuations; we discuss the impact of other sources of contamination on kSZ studies in \S\ref{sec:robustness}.


\subsection{Cross-correlation of ARF and filtered CMB maps}
\label{sub:foundations_cross}

We extract kSZ optical depths from the cross-correlation of ARF and filtered CMB maps using the following expression

\begin{equation}
\label{eq:computetau}
\tauAP(\thetaAP) = \frac{\sum_{\ell\ell'} \hatarfksz(\thetaAP) \mathcal{C}_{\ell\ell'}^{-1}(\thetaAP) C_{\ell'}^{\rm cr}} {\sum_{\ell\ell'} C_\ell^{\rm cr} \mathcal{C}_{\ell\ell'}^{-1}(\thetaAP) C_{\ell'}^{\rm cr}},
\end{equation}

\noindent where $\mathcal{C}_{\ell \ell'}$ stands for the covariance matrix of the cross-correlation, $C_\ell^{\rm cr} = [\tauAPth(\thetaAP)]^{-1}\arfksz$ isolates the dependence of the cross-correlation on cosmology, and $\arfksz$ and $\hatarfksz$ indicate a theoretical predictions and an actual measurement of the cross-correlation, respectively.

To derive the dependence of this cross-correlation on cosmology, we first need to consider angular density fluctuation maps (ADF), which correspond to standard clustering maps. To generate these, we introduce the same tracers and selection function as we use to create ARF and filtered CMB maps in

\begin{equation}
\label{eq:map_adf}
1 + \mathcal{M}_{\rm ADF}(\angr_j) = \rhom^{-1} \sum_i W(z_i) \deltaang,
\end{equation}

\noindent where $i$ and $j$ run over tracers and pixels, respectively. Then, we follow the same approach as in \S\ref{sub:theory_arf} to derive the power spectrum of ADF maps, finding that we can assemble this by introducing the kernels

\begin{equation}
\label{eq:wdeltar}
K^{\delta_d}_\ell(k) = \rhom^{-1} \int \mathrm{d}z \, D(z) b(z) W(z) N(z) j_\ell[k r(z)],
\end{equation}

\begin{equation}
\label{eq:wvr}
K^{v_d}_\ell(k) = \rhom^{-1} \int \mathrm{d}z \, \frac{H(z)}{c} D(z) f(z) \Sigma_v(k)
\frac{\mathrm{d}W}{\mathrm{d}z} N(z) \frac{j'_\ell[k r(z)]}{k},
\end{equation}

\noindent in Eq.~\ref{eq:template}, where $\delta_d$ and $v_d$ encode the dependence of ADF maps upon the cosmic density and velocity fields, respectively.

Before pursuing our derivation, it is useful to notice that the numerator and denominator of the ARF estimator presented in Eq.~\ref{eq:map_arf} correspond to the redshift and clustering maps given by Eqs.~\ref{eq:redshift_field} and \ref{eq:map_adf}, respectively. We can thus derive the dependence of the cross-correlation of ARF and filtered CMB maps on cosmology by computing cross-terms between the ADF, ARF, and kSZ kernels introduced in Eqs.~\ref{eq:wdeltar} and \ref{eq:wvr}, \ref{eq:wdeltaz} and \ref{eq:wvz}, and \ref{eq:wksz}, respectively. In this manner, we find

\begin{equation}
    \label{eq:cr}
    C_\ell^{\rm cr} = \tau_{\rm eff}^{-1}\left(C_\ell^{{\rm kSZ}-\delta_z}-C_\ell^{{\rm kSZ}-\delta_d}+C_\ell^{{\rm kSZ}-v_z}-C_\ell^{{\rm kSZ}-v_d}\right),
\end{equation}

\noindent where the first (last) two terms in brackets depend on the cross-correlation of the kSZ effect and both ADF and ARF density (velocity) terms.


\begin{figure}
\includegraphics[width=\columnwidth]{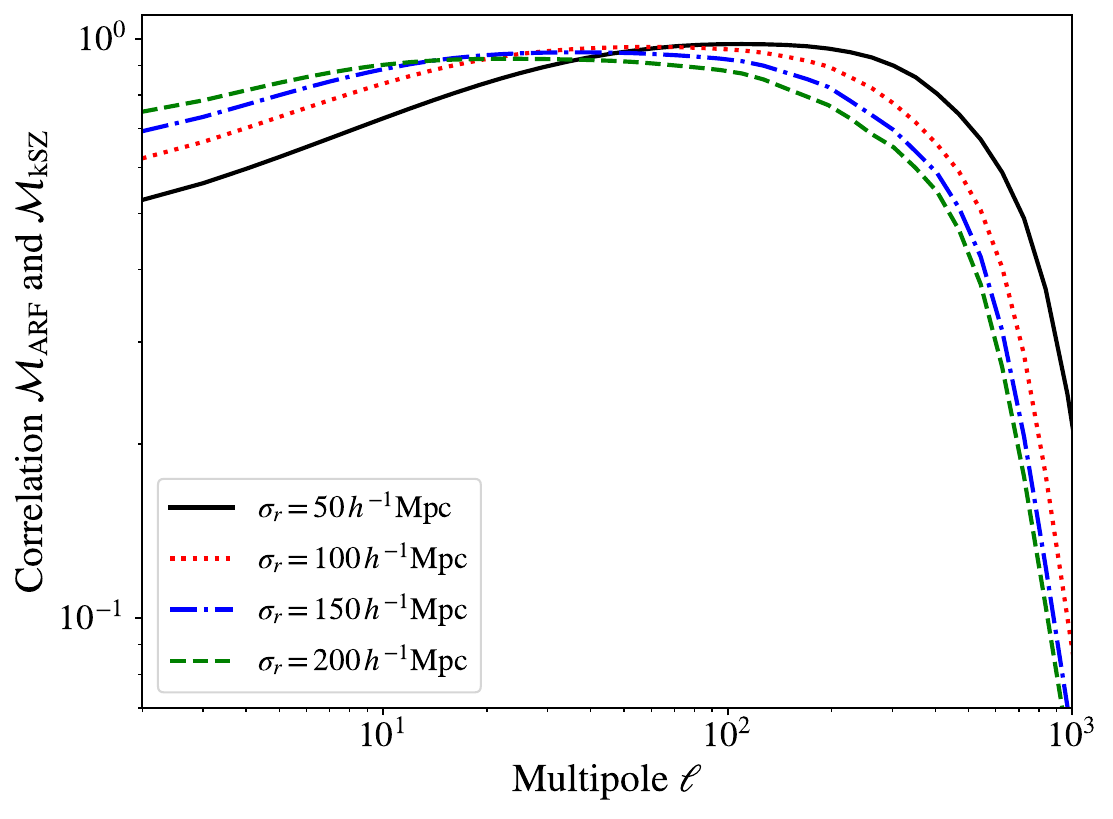}
\caption{
\label{fig:corr_arf_ksz}
Optimal range of scales for ARF-kSZ tomography as a function of the width of the selection function. Black, red, blue, and green lines indicate theoretical predictions for the correlation of ARF and the kSZ effect using Gaussian shells centred at $z_{\rm cen}=0.5$ and with widths $\sigma_r=50$, 100, 150, and $200\,\Mpch$, respectively. As we can see, the range of scale across which the correlation is substantial decreases with the width of the selection function.
}
\end{figure}

\subsection{Optimal range of scales for ARF-kSZ tomography}
\label{sub:foundations_scales}

In this section, we compute the range of scales across which ARF and filtered CMB maps are correlated. This estimation is crucial to determine the samples, selection functions, and CMB observations that we need to use in order to extract statistically significant measurements of the kSZ effect.

In Fig.~\ref{fig:corr_arf_ksz}, we display theoretical predictions for the correlation of ARF and filtered CMB maps as a function of the width of the selection function for tracers with constant number density. Black, red, blue, and green lines indicate the results for Gaussian shells centred at $z_\mathrm{cen}=0.5$ and with comoving widths $\sigma_r=50$, 100, 150, and $200\,\Mpch$, respectively. We find that the correlation is stronger on small angular scales for narrower shells and that the range of scales across which these maps present substantial correlation increases with the central redshift of the selection function (see HM19 for further details).

Note that to generate these predictions, we neglect statistical uncertainties arising from the limited number of galaxies accessible in realistic studies, which decreases for narrower shells. As a result, the optimal range of scales for ARF-kSZ tomography is a trade-off between the number density of tracers and the width of the selection function.



\section{Insights from simulations}
\label{sec:thsim}

To gain further insight into ARF-kSZ tomography, in this section we use cosmological simulations to address the precision of our theoretical derivations and to study the distribution of gas surrounding galaxies.


\subsection{Precision of theoretical derivations}
\label{sub:thsim_precision}

Ideally, we would use cosmological hydrodynamical simulations to evaluate the precision of theoretical expressions for ARF-kSZ tomography; however, even the largest simulations of this kind cannot sample the large scales across which ARF and the kSZ effect present substantial correlation. On the other hand, gravity-only simulations can access these volumes at a fraction of the computational cost of hydrodynamical simulations, but cannot provide direct predictions for the kSZ effect. It is nonetheless important to note that the cross-correlation of ARF and radial peculiar velocities and that of ARF and the kSZ effect present the same dependence upon cosmology at first order in perturbations, which enables using gravity-only simulations to estimate the precision of our derivations. We proceed to describe the simulations that we use, detail our approach to produce sky maps and compute their cross-correlation, and compare theoretical predictions and results from simulations.

\subsubsection{Gravity-only simulations}

To test our methodology, we carry out an ensemble of 100 gravity-only $N$-body lightcone simulations using \Lcola \citep{howlett15}, an efficient parallel implementation of the Comoving Lagrangian Acceleration method \citep[\Cola;][]{tassev13}. This technique presents a significant improvement in execution speed relative to full $N$-body simulations at the expense of loss of precision on small scales: \Lcola recovers the power spectrum of the cosmic density and velocity fields as predicted by a full $N$-body simulation to within 2 and 3\% up to $k=0.3$ and $0.15\,\hMpc$ \citep{howlett15, koda15}, respectively, which correspond to multipoles $\ell=197$ and 99 at $z=0.5$. The generation of each lightcone adds additional errors because it requires four replications of the simulation box; however, the impact of these uncertainties on the scales of interest for ARF-kSZ tomography is minimal \citep{Klypin2018}. We provide further details about these simulations in \citet{chavesmontero18}.

\subsubsection{Generation of sky maps}
\label{sub:thsim_precision_genmaps}

To project simulation data onto sky maps, we use the publicly available package \Healpix\footnote{\url{http://healpix.sourceforge.net}} \citep{Gorski2005, Zonca2019}, which includes routines for the discretisation of functions on the surface of a sphere. In particular, we generate ARF, ADF, and radial peculiar velocity maps by introducing particles from our lightcone simulations into Eqs.~\ref{eq:map_arf}, \ref{eq:map_adf}, and 

\begin{equation}
    \mathcal{M}_{\rm VEL}(\angr_j)=\frac{\sum_i (v_i/c) W(z_i)\deltaang}{\sum_i W(z_i)\deltaang},
\end{equation}

\noindent respectively, where $i$ and $j$ run over particles and pixels. Note that each pixel of the velocity map captures the average radial peculiar velocity of all sources falling in this pixel weighted by the selection function; consequently, these maps are robust against systematic uncertainties in the angular number density of tracers for the same reasons as ARF and filtered CMB maps.

Motivated by the moderate mass resolution of our simulations, which only permits resolving $\sim10^{14}\Msunh$ haloes precisely, and to ensure that the number of objects falling in each pixel is large enough for statistical studies, we generate sky maps using dark matter particles instead of haloes and we set the resolution of these maps to $N_{\rm side}=64$. This resolution is equivalent to dividing maps into 49,152 pixels of area $\simeq0.84\,\degc$, which enables resolving auto- and cross-correlations on multipoles below $\ell\simeq200$ precisely. Note that considering particles instead of haloes has minimal impact on the scales of interest for ARF-kSZ tomography because it only manifests as a variation on the large-scale bias of tracers.

\begin{figure}
\begin{center}
\hspace*{-1cm}\includegraphics[width=\columnwidth]{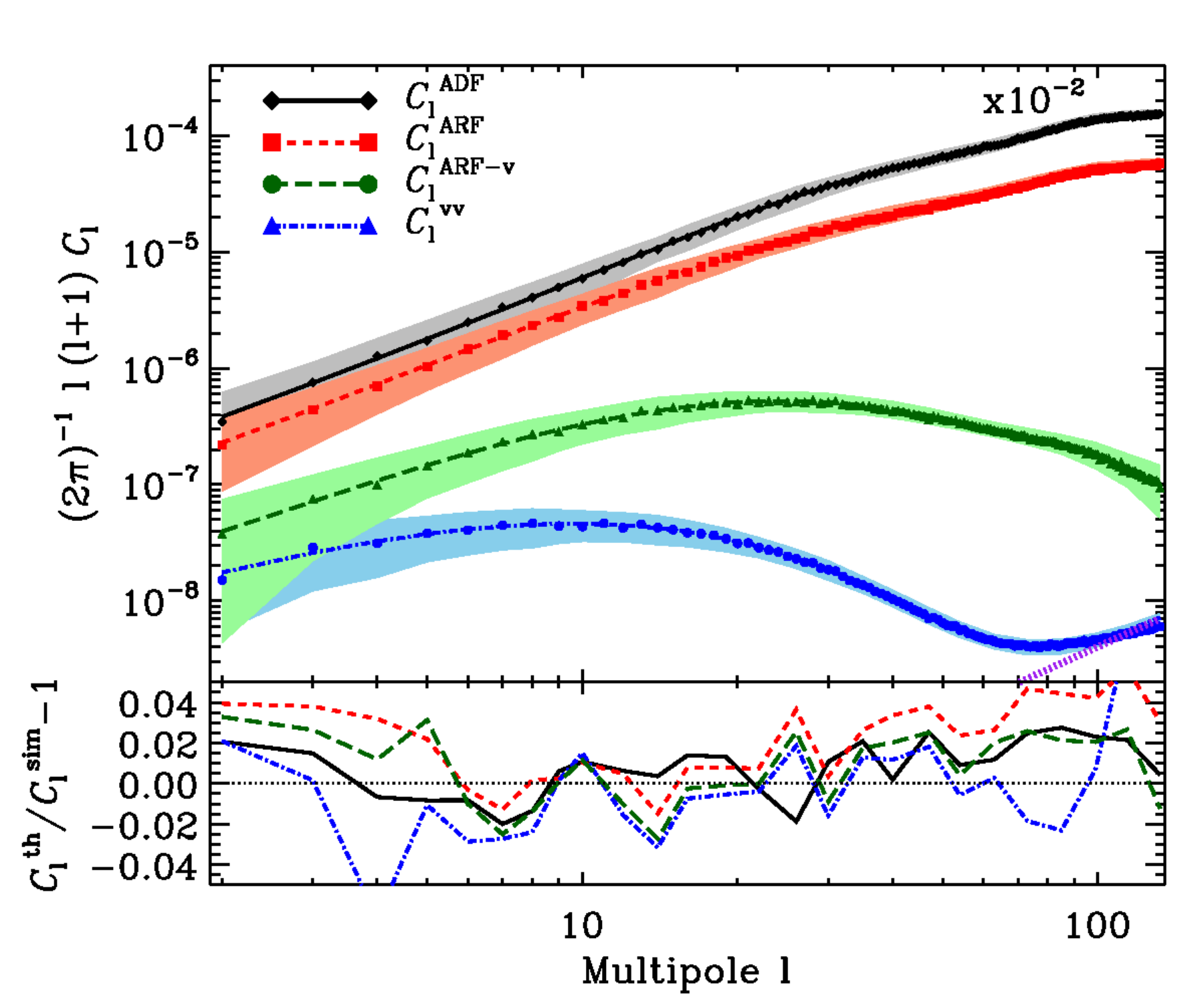}
\end{center}
\caption{
\label{fig:cls}
Power spectrum of ADF, ARF, and radial peculiar velocity maps, and cross-correlation of the last two. Lines and symbols indicate theoretical predictions and results from simulations, respectively, while shaded areas denote $1\,\sigma$ mock-to-mock uncertainties. In the bottom panel, we display the relative difference between theoretical predictions and results from simulations. As we can see, these agree to within 5\% across the range of multipoles shown.
}
\end{figure}

\begin{figure}
\hspace*{-0.5cm}\includegraphics[width=\columnwidth]{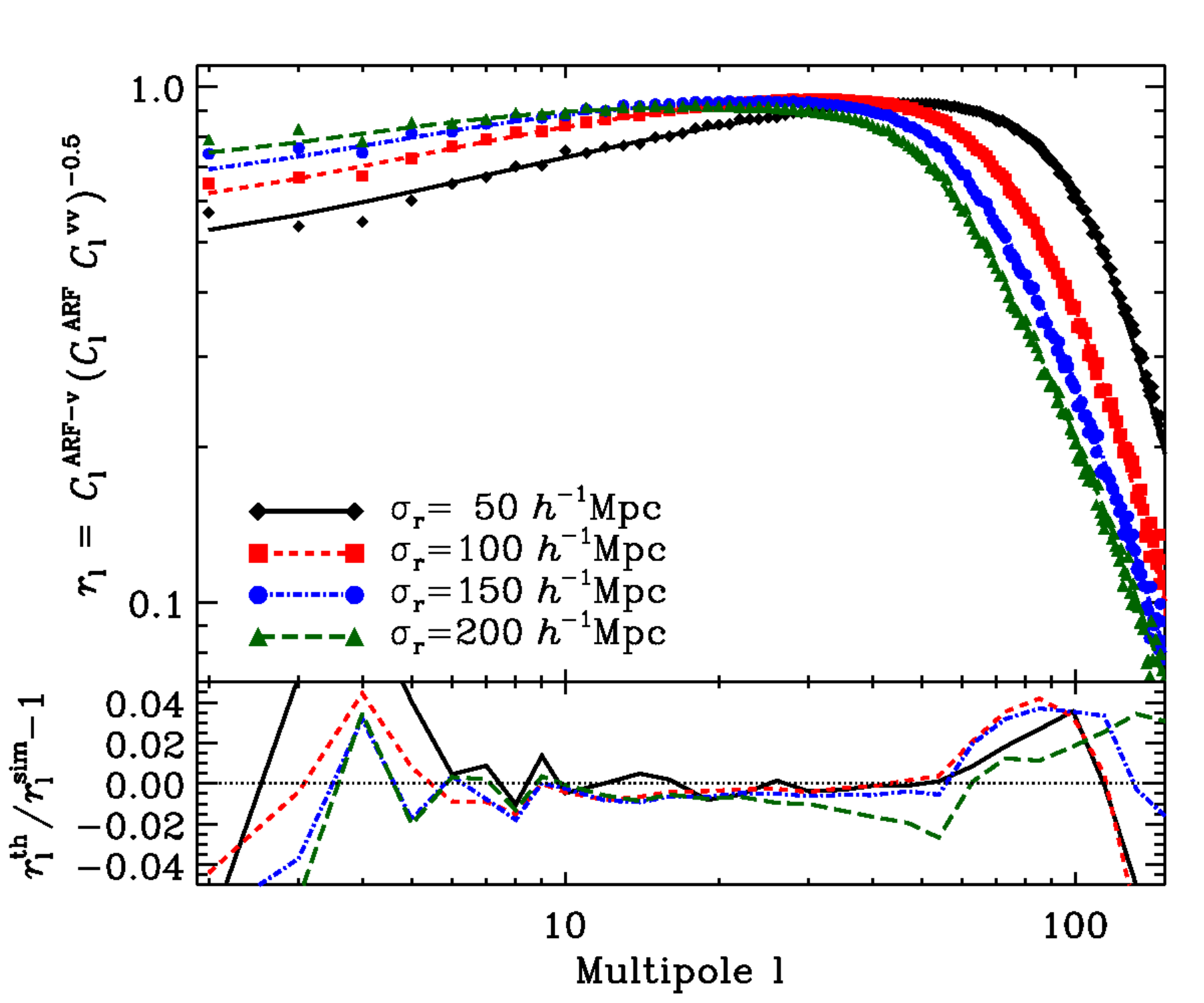}
\caption{
\label{fig:arf_vel_width}
Correlation of ARF and radial peculiar velocity maps at $z_{\rm cen}=0.5$ as a function of the width of the selection function. Lines indicate theoretical predictions, symbols display results from simulations, and the bottom panel shows the relative difference between these. We find that ARF and radial peculiar velocity maps present substantial correlation across a broader range of scales for narrower shells.
}
\end{figure}

\subsubsection{Auto- and cross-correlation of sky maps}
\label{sub:thsim_precision_corr}

We compute the auto- and cross-correlation of sky maps using the publicly available program {\sc polspice} \citep{szapudi01, chon04}, which includes routines for the analysis of sky maps while accounting for the impact of a mask, inhomogeneous weights, or both. This last possibility is quite useful for ARF-kSZ tomography; even though ARF and filtered CMB maps are robust against uncertainties affecting the angular number density of sources, the precision of these maps depends on the number of sources falling in each pixel. Motivated by this, we compute the cross-correlation of ARF and filtered CMB maps after weighting these by a sky map containing the number density of tracers weighted by the selection function, $\mathcal{M}_{\rm W}(\angr_j) = \rhom [1 + \mathcal{M}_{\rm ADF}(\angr_j)]$.

In the top panel of Fig.~\ref{fig:cls}, we display the power spectrum of ADF, ARF, and radial peculiar velocity maps, and the cross-correlation of the last two. Note that we normalise the power spectrum of ADF maps by 100 for clarity and that we produce these maps using a Gaussian redshift shell with centre $z_{\rm cen}=0.5$ and comoving width $\sigma_r=100\,\Mpch$. Lines and symbols indicate theoretical predictions and average results from 100 different lightcones simulations, respectively, while shaded areas denote mock-to-mock $1\sigma$ uncertainties. Overall, we find that the power spectra of ADF and ARF maps look alike; the amplitude of these grows monotonically with $\ell$ across the whole range of scales shown, and their shapes present wiggles induced by baryonic acoustic oscillations. On the other hand, the power spectrum of radial peculiar velocity maps and the correlation of these and ARF maps increase with $\ell$ on large scales, peak at $\ell=10-30$, and the first (second) increases (decreases) after that. We model the increase in the power spectrum of velocity maps, which is not captured by our theoretical derivations, using an experimental power law indicated by the purple dotted line at the bottom-right of the top panel.

In the bottom panel, we show the relative difference between theoretical predictions and results from simulations. As we can see, the precision of our theoretical derivations for the power spectrum of ADF, ARF, and the cross-correlation of ARF and radial peculiar velocities is superior to 5\% across the whole range of multipoles shown. We remind the reader that this cross-correlation and that of ARF and the kSZ effect present analogous dependence on cosmology at first order, and thus we expect our theoretical expression for the latter to present a similar level of precision as that for the former.

In Fig.~\ref{fig:arf_vel_width}, we display the correlation of ARF and radial peculiar velocity maps at $z_{\rm cen}=0.5$ as a function of the width of the selection function. Lines indicate theoretical predictions, symbols display results from simulations, and the bottom panel shows the relative difference between these. We find that these maps are only correlated on large scales and that the range of multipoles across which the correlation is substantial decreases for wider shells; these findings support that the kSZ effect can only be extracted from large angular scales ($\ell < 128$) using ARF-kSZ tomography.

Note that we account for the impact of small-scale velocities in our predictions using $\sigma_v=7\,\Mpch$, which is the value of $\sigma_v$ that provides the best fit to data from simulations. It is however important to note that the impact of this parameter on ARF-kSZ tomography is minimal as it only modifies theoretical predictions on small scales.



\begin{figure}
\includegraphics[width=\columnwidth]{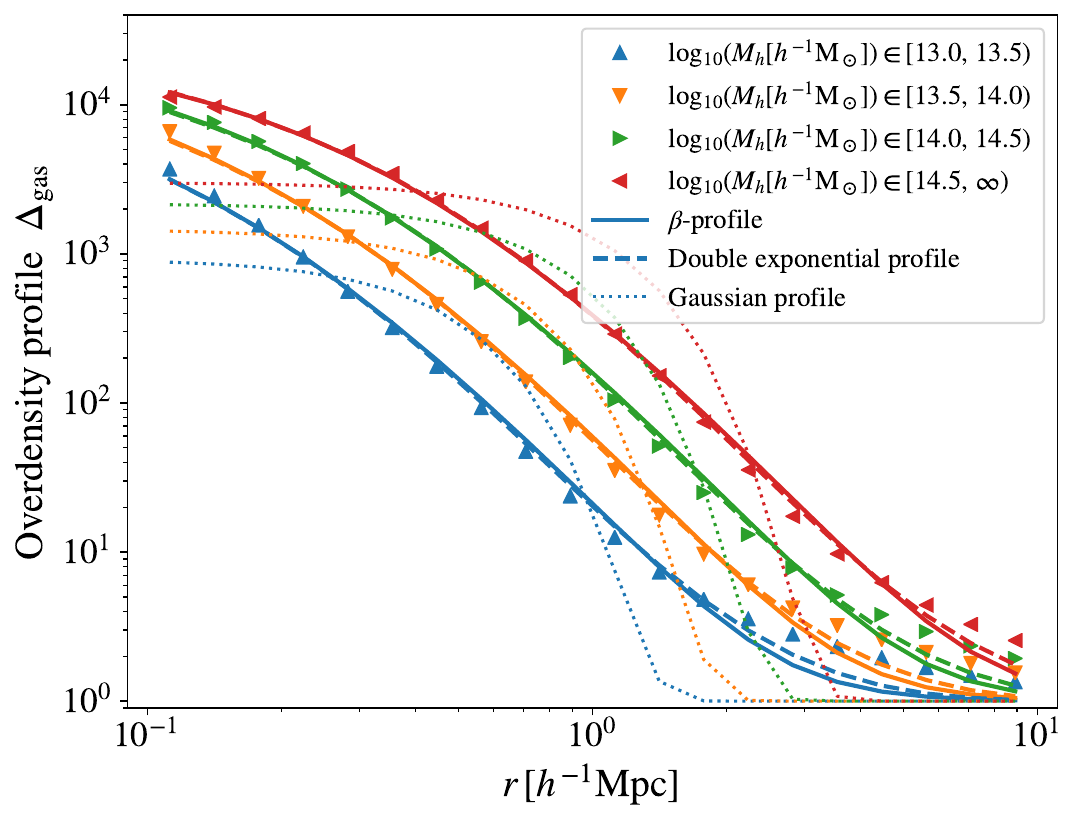}
\caption{
\label{fig:profile_bcube}
Large-scale distribution of gas surrounding haloes of different masses as predicted by the cosmological hydrodynamical simulation Borg Cube. Symbols indicate results from the simulation at $z=0.24$, while solid, dashed, and dotted lines denote the best-fitting solution to these using $\beta$-, double exponential, and Gaussian profiles, respectively. We can readily see that the $\beta$- and double exponential profiles capture simulation results precisely, while Gaussian profiles provide a lousy fit to data.
}
\end{figure}

\subsection{Distribution of gas surrounding tracers}
\label{sub:thsim_gprofile}

To gain insight into the large-scale distribution of cosmic gas, we use the cosmological hydrodynamical simulation Borg Cube \citep{Emberson2018}. This simulation evolved $2\times2304^3$ dark matter plus baryonic particles of masses 2.56 and $0.52\times10^9\,\Msunh$, respectively, in a periodic comoving box of $800\,\Mpch$ on a side while treating baryons in the non-radiative regime. Although the Borg Cube simulation only incorporates non-radiative baryonic processes, we do not expect other baryonic effects to present significant impact on the large-scale distribution of cosmic gas owing to the precise agreement among the power spectra of gravity-only, non-radiative, and full-physics simulations on scales larger than $1\,\Mpch$ \citep{Springel2018, Emberson2018}.

In Fig.~\ref{fig:profile_bcube}, we display the large-scale distribution of gas surrounding spherical overdensity haloes of different masses in Borg Cube. Symbols indicate results from the simulation, while lines denote the best-fitting solution to these using three distinct functional forms: a $\beta$-profile $\pgas(r)=\Delta_b\left[1+r^2/r_s^2\right]^{-7/2}+1$, a double exponential profile $\log \pgas^\mathrm{exp}(r)=\Delta_b'\,\exp\left[r^{3/4}/r_s'^{3/4}\right]$, and a Gaussian profile $\pgas^\mathrm{gauss}(r)=\Delta_b''\,\exp\left[-0.5\, r^2/r_s''^2\right]+1$, where $\Delta_b$ and $r_s$ regulate the amplitude and breadth of the profiles, respectively. As we can see, the $\beta$- and double exponential profiles are flexible enough to capture the distribution of gas predicted by the simulation, while Gaussian profiles provide a lousy fit to data. Using simulation data at other redshifts, we find that these results hold at least up to redshift $z=4$.



\section{Applying ARF-kSZ tomography to observations}
\label{sec:resobv}

In this section, we use ARF-kSZ tomography to extract measurements of the kSZ effect from observations. We start by producing ARF and filtered CMB maps in \S\ref{sub:resobv_maps}, then we estimate the uncertainty in the cross-correlation of these in \S\ref{sub:resobv_cov}, we continue by extracting kSZ measurements from this cross-correlation in \S\ref{sub:resobv_cross}, and we conclude computing the significance of kSZ measurements in \S\ref{sub:resobv_significance}.


\subsection{Creation of ARF and filtered CMB maps}
\label{sub:resobv_maps}

In this section, we produce ARF and filtered CMB maps using different samples, selection functions, apertures, and foreground-reduced \Planck maps. We proceed to describe all these ingredients and our strategy to produce these maps.


\subsubsection{Tracers}
\label{sub:resobv_maps_tracers}

The most natural sources to produce sky maps at low and high redshift are galaxies and quasars due to their abundance and luminosity, respectively. We describe the galaxy and quasar samples that we use below.

\begin{itemize}

\item {\bf 6dF-G sample.} We study the low redshift universe using galaxies from the 6dF Galaxy Survey \citep[6dF;][]{jones04}, which provides spectroscopic redshifts for 108\,030 galaxies spanning $\sim17\,000\;\degc$ of the southern sky \citep{jones09}. These galaxies present a median redshift of $z=0.053$, host halo masses ranging from $10^{11}$ to $10^{12}\Msunh$, and an average large-scale bias of $b\simeq1.48$ \citep{beutler12}.

\item {\bf SDSS-G sample.} We access intermediate redshifts using galaxies from the Baryon Oscillation Spectroscopic Survey \citep[BOSS;][]{eisenstein11, dawson13}, which obtained precise redshifts for 1\,325\,856 galaxies covering $9\,376\;\degc$ of the sky. According to the target selection criteria, BOSS galaxies are usually divided into two main groups: the LOWZ sample, which includes the brightest and reddest galaxies at $z<0.43$, and the CMASS sample, which contains slightly bluer galaxies at $0.43<z<0.7$. The average halo mass and large-scale bias of these two samples are similar and equal to $M_h\simeq10^{13}\Msunh$ and $b\simeq2$, respectively \citep{parejko13, saito16, rodriguez-torres16}.

\item {\bf SDSS-Q sample.} We sample the high redshift universe using 526\,356 quasars with secure redshifts from the 14th data release of the SDSS quasar catalogue \citep{paris18}, which comprises sources observed during more than 16 years as part of any of the stages of SDSS \citep{york00, eisenstein11, blanton17}. These quasars present redshifts to within the interval $z\in[0.5, 6]$, host halo masses ranging from $10^{12}$ to $10^{13}\Msunh$, and a large scale bias given by $b=0.278[(1+z)^2-6.565]+2.393$ \citep{laurent17, Ata2018}, which translates into $b=1.2$ and 10 at low and high redshift, respectively. It is important to note that the target selection of these quasars evolved over time, and thus the properties of these sources may vary across the $9\,376\;\degc$ of survey footprint.

\end{itemize}


\subsubsection{Selection function}
\label{sub:resobv_maps_sel}

The kSZ effect results from scattering of CMB photons off free electrons moving relative to the CMB rest frame, and thus it is just natural to attempt to detect this effect across the entire redshift range spanned by our tracers. Taken together with the tomographic nature of our approach, we generate ARF and filtered CMB maps using 6dF-G, SDSS-G, and SDSS-Q sources and redshift shells centred at $z_{\rm cen}=0.18$; 0.27, 0.42, 0.59, and 0.78; and 0.72, 0.92, 1.15, 1.41, 1.71, 2.07, 2.50, 3.01, 3.64, 4.43, and 5.42, respectively. To better characterise these, we use the effective redshift of the selected tracers $z_{\rm eff} = \zm/\rhom$ instead of the central redshift of the shell; the motivation is that the first quantity captures better the average redshift of tracers. In this manner, we find $z_{\rm eff}=0.09$; 0.29, 0.44, 0.56, and 0.65; and 0.73, 0.92, 1.14, 1.41, 1.70, 2.07, 2.43, 2.85, 3.35, 4.03, and 4.62 for the 6dF-G, SDSS-G, and SDSS-Q shells, respectively.

The width of these shells needs to be chosen as a compromise between two effects: broader shells select more sources, which reduces statistical uncertainties, while narrower shells increase the range of scales across which ARF and filtered CMB maps present substantial correlation. To balance both, we adopt a comoving width of $\sigma_r=180\,\Mpch$. Note that we leave a separation of $\sigma_r\sqrt{2}$ between contiguous shells to reduce correlations.


\subsubsection{CMB observations}
\label{sub:resobv_maps_cmb}

The extraction of the kSZ effect via ARF-kSZ tomography requires accessing large angular scales; motivated by this, we use publicly available temperature maps from the \Planck survey, which represents the largest CMB dataset to date \citep{Planck2018I}. In particular, we use the four maps generated by applying each of the foreground-cleaning algorithms \Commander, \Nilc, \Sevem, and \Smica \citep{planck16IX, Planck2018IV}, one map produced using an improved version of \Smica that also attempts to reduce tSZ contamination, and three single-frequency maps cleaned using \Sevem. We refer to these eight foreground-reduced maps as \Commander, \Nilc, \Sevem, \Smica, \Smicanosz, \Sevema, \Sevemb, and \Sevemc. Even though the \Planck collaboration uses different algorithms to generate these maps, these show excellent consistency at the pixel level \citep{Planck2018IV}.


\subsubsection{Map generation}
\label{sub:resobv_maps_gen}

The resolution of sky maps needs to be chosen as a compromise between two effects: a lower resolution increases the average number of tracers falling in each pixel, which reduces statistical uncertainties, while a higher resolution increases the range of scales accessible from their auto- and cross-correlation. Given that ARF and filtered CMB maps present substantial correlation just on large angular scales, we produce sky maps of resolution $N_{\rm side}=64$, which corresponds to pixels of area $0.84\,\degc$ and enables resolving multipoles below $\ell\simeq200$ precisely. Although this is the resolution that we use to generate ARF and filtered CMB maps, we compute aperture photometry estimates using \Planck maps of resolution $N_{\rm side}=2048$.

To generate sky maps using data from observations, we follow the same procedure as with simulation data in \S\ref{sec:thsim}. Using each redshift shells described in \S\ref{sub:resobv_maps_sel}, we generate a ARF map and $8\times33$ filtered CMB maps by high-pass filtering each \Planck foreground-reduced map using apertures $\thetaAP=3$, 4, ..., 20, 22, ..., 48, and 50 arcmin. Note that we do not consider apertures either smaller than 3 arcmin or larger than 50 arcmin because the former are limited by the finite resolution of the \Planck beams, while the latter present strong contamination by primordial CMB anisotropies not subtracted by the filter.


\subsection{Estimating covariance matrices}
\label{sub:resobv_cov}

The extraction of kSZ optical depth measurements from the cross-correlation of ARF and filtered CMB maps via Eq.~\ref{eq:computetau} requires an estimate for the uncertainty in this cross-correlation. In this section, we combine observational and simulated data to estimate the precision of the cross-correlation of maps produced using different samples, selection functions, foreground-reduced maps, and apertures.

The two principal sources of uncertainty in ARF-kSZ tomography are residual CMB contamination not subtracted by aperture photometry and statistical errors arising from the limited number density of sky map tracers. To quantify the impact of these for each cross-correlation, we proceed as follows. We start by producing 1\,000 simulated CMB maps containing just primordial CMB anisotropies. For consistency with observations, we generate these maps by considering random realisations of a Gaussian field characterised by the \Planck temperature power spectrum, we use a pixel resolution analogous to that of \Planck maps ($N_\mathrm{side}=2048$), and we convolve these maps with a Gaussian of FWHM 5 arcmin to mimic the \Planck beams. Then, using each simulated map as input, we generate a specific map for every foreground-reduced \Planck map. To do so, we apply to simulated maps the specific mask and noise pattern of each foreground-reduced map; we estimate the latter from the difference of publicly available foreground-reduced half-mission maps. After that, we produce simulated filtered CMB maps by applying aperture photometry to simulated foreground-reduced maps using each sample, selection function, and aperture considered in \S\ref{sub:resobv_maps}. Lastly, we use the cross-correlation of ARF maps from observations and simulated filtered CMB maps, $S_{\ell, m}^{\mathrm{ARF}-\mathrm{kSZ}}$, to estimate a covariance matrix for each shell, aperture, and foreground-reduced map

\begin{multline}
    \label{eq:compute_cov}
    \mathcal{C}_{\ell\ell'}(\thetaAP) = \frac{1}{M-1} \sum_{m=1}^M \left[S_{\ell, m}^{\mathrm{ARF}-\mathrm{kSZ}}(\thetaAP) - \bar{S}_{\ell}^{\mathrm{ARF}-\mathrm{kSZ}}(\thetaAP)\right]\\ 
    \left[S_{\ell', m}^{\mathrm{ARF}-\mathrm{kSZ}}(\thetaAP) - \bar{S}_{\ell'}^{\mathrm{ARF}-\mathrm{kSZ}}(\thetaAP)\right],
\end{multline}

\noindent where bars indicate an average across simulations, while $M=1000$ is the number of simulated maps. This approach is motivated by the fact that simulated maps do not include kSZ signal, so any departure of the cross-correlation from zero can only result from uncertainties.

We find that the diagonal elements of these matrices decrease very steeply with the filter aperture for $\thetaAP\leq22$ arcmin, while the decline becomes considerably shallower for larger apertures. The trade-off of two effects explains this trend: statistical errors decrease with the filter area due to the increasingly larger number of pixels available, while the level of high-frequency primordial CMB contamination increases. Note that we do not simulate foregrounds to estimate these matrices, and thus errors may be underestimated for large apertures; we study the impact of these and other sources of uncertainty in \S\ref{sec:robustness}.

Extracting kSZ optical depths also requires to compute the inverse of these covariance matrices. To do so, we first use an algorithm based on LU factorisation, and then we account for the limited number of mocks by multiplying the result by the factor $(M-N_\ell-2)/(M-1)$ \citep[e.g.,][]{hartlap07}, where $N_\ell$ is the number of $\ell$-bins.


\begin{figure*}
\begin{center}
\includegraphics[width=0.43\textwidth]{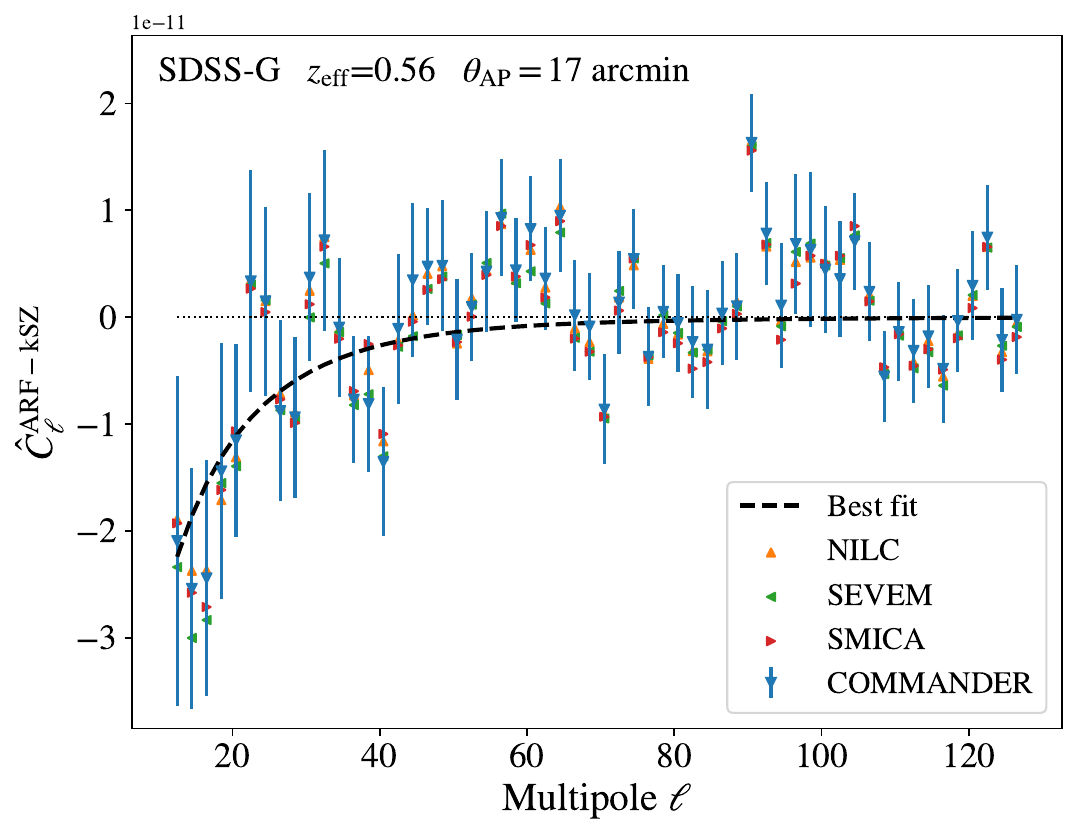} 
\includegraphics[width=0.43\textwidth]{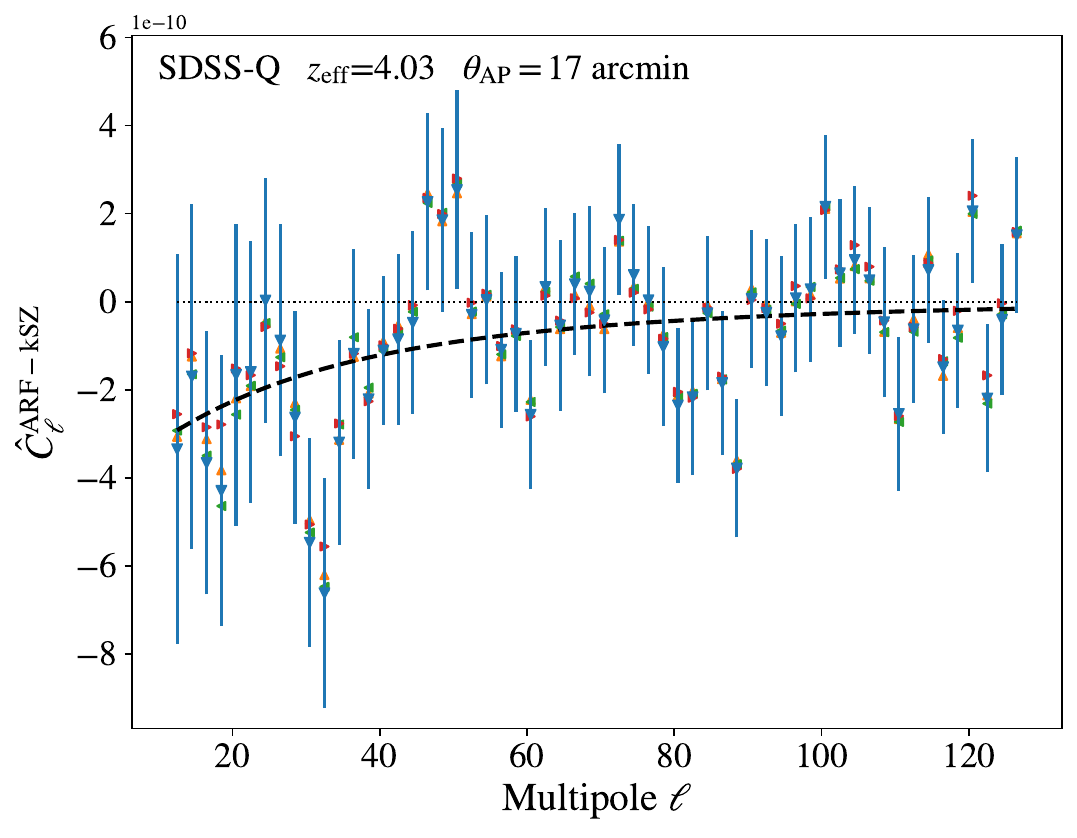} 
\end{center}
\caption{
\label{fig:cls_results}
Cross-correlation of ARF maps and high-pass filtered foreground-reduced \Planck maps produced using BOSS galaxies (left panel) and SDSS quasars (right panel) at $z_{\rm eff}=0.56$ and 4.03, respectively, and an aperture of $\thetaAP=17$ arcmin. Symbols indicate measurements from different foreground-reduced maps, error bars denote $1\sigma$ uncertainties for \Commander, and dashed lines depict the best-fitting model to \Commander data. We find that best-fitting models capture the angular dependence of the signal precisely, which strongly suggests that {\it these cross-correlations departure from zero due to actual measurements of the kSZ effect.}
}
\end{figure*}

\begin{figure}
\begin{center}
\includegraphics[width=0.43\textwidth]{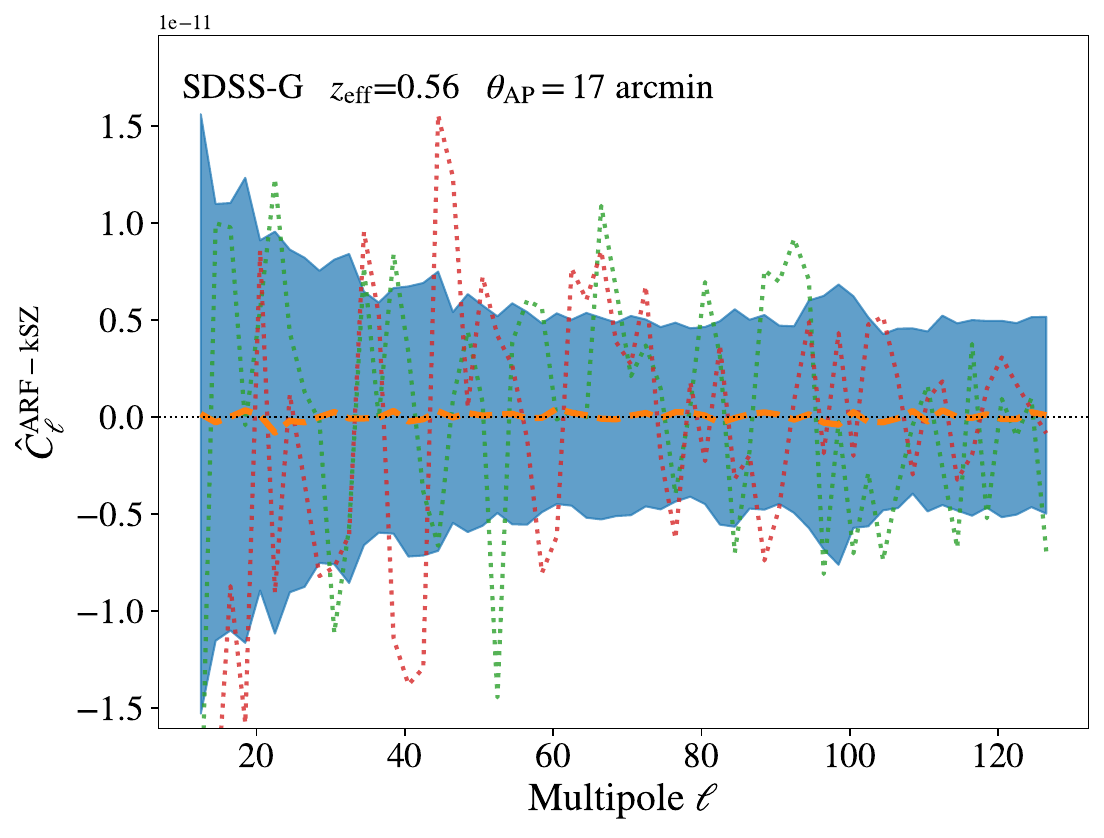} 
\includegraphics[width=0.43\textwidth]{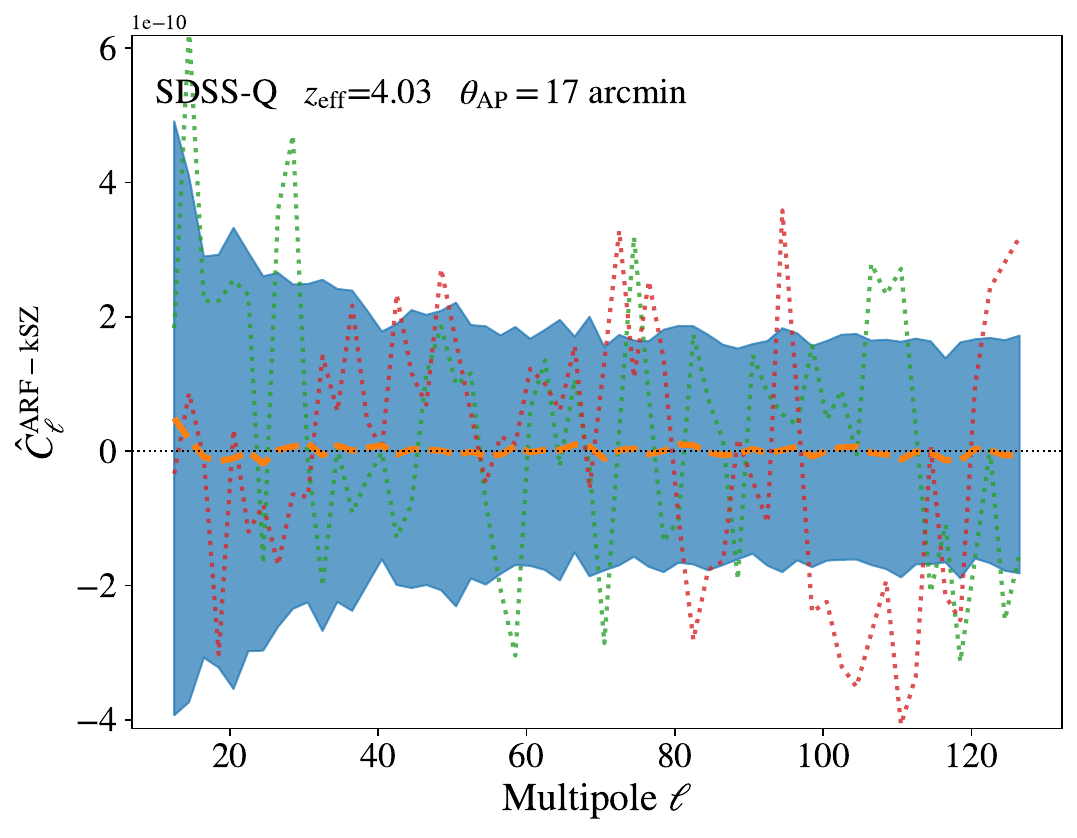} 
\end{center}
\caption{
\label{fig:null_tests}
Cross-correlation of ARF maps from observations and simulated high-pass filtered \Commander maps not including kSZ signal for the same samples, selection function, and aperture as in Fig.~\ref{fig:cls_results}. Orange dashed lines indicate an average across 1\,000 mocks, blue shaded areas denote $1\sigma$ mock-to-mock uncertainties, and red and green dotted lines show two randomly selected cross-correlations. We find no systematic deviation of cross-correlations from zero, which provides further support to the astrophysical origin of the results shown in Fig.~\ref{fig:cls_results}.
}
\end{figure}

\subsection{Cross-correlation of ARF and filtered CMB maps}
\label{sub:resobv_cross}

We can follow two different strategies to extract kSZ measurements from the cross-correlation of ARF and filtered CMB maps: measure kSZ optical depths while holding fixed cosmological parameters and large-scale biases, or set joint constraints on all these quantities. We will follow the first approach prompted by the challenging extraction of the kSZ effect; to do so, we adopt \Planck 2015 cosmological parameters and the large-scale biases quoted in \S\ref{sub:resobv_maps}.

To compute the cross-correlation of ARF and filtered CMB maps, we follow the procedure outlined in \S\ref{sub:foundations_cross} for each sample, selection function, aperture, and foreground-reduced map considered in \S\ref{sub:resobv_maps}; in this manner, we end up with $16\times33\times8=4\,224$ different cross-correlations. Then, to reduce artificial correlations induced by the survey mask, we apply a binning of $\Delta\ell=2$ and we restrict our analysis to multipoles below $\ell=3$ and 13 for the 6dF-G and the SDSS-G and SDSS-Q samples, respectively; this is motivated by the sky area covered by the 6dF and SDSS surveys. We proceed to show a few illustrative examples of these cross-correlations.

In Fig.~\ref{fig:cls_results}, we display the cross-correlation of ARF and filtered CMB maps produced using BOSS galaxies (left panel) and SDSS quasars (right panel) at $z_{\rm eff}=0.56$ and 4.03, respectively, and an aperture of $\thetaAP=17$ arcmin. Symbols indicate measurements from different foreground-reduced maps, error bars denote the diagonal elements of \Commander covariance matrices, and dashed lines depict the best-fitting model to \Commander data. As we can readily see, cross-correlations computed using different foreground-reduced maps show excellent consistency; given that these maps are produced using different foreground-cleaning procedures, this result manifests the weak impact of residual foreground contamination on ARF-kSZ tomography (see also \S\ref{sub:robustness_foreground}).

We can also see that best-fitting models capture the angular dependence of the signal precisely; we produce these models by multiplying the cosmological dependence of the cross-correlation, which is given by Eq.~\ref{eq:cr}, by kSZ optical depths extracted using Eq.~\ref{eq:computetau}, which are $\tauAP=(1.8 \pm 0.6) \times 10^{-3}$ and $0.8 \pm 0.3$ for BOSS galaxies and SDSS quasars, respectively. Note that cross-correlations depart from zero at higher multipoles as redshift increases because the range of scales across which ARF and filtered CMB maps present substantial correlation also grows with redshift.

Even though our theoretical model provides an excellent fit to data, it is still conceivable that cross-correlations could depart from zero due to statistical errors, residual CMB contamination, the impact of the mask, or other uncertainties. Aiming to discard this possibility, we design a null test based on the cross-correlation of ARF maps from observations and the simulated filtered CMB maps produced in \S\ref{sub:resobv_cov}; given that simulated maps do not include kSZ signal, cross-correlations should be compatible with zero. In Fig.~\ref{fig:null_tests}, we display the cross-correlation of ARF and simulated filtered \Commander maps for the same samples, selection function, and apertures as in Fig.~\ref{fig:cls_results}. Orange dashed lines indicate an average across 1\,000 mocks, blue shaded areas denote $1\sigma$ mock-to-mock uncertainties, and red and green dotted lines show two randomly selected cross-correlations. As we can see, there is no systematic deviation of the cross-correlations from zero, which provides further support to the astrophysical origin of the cross-correlations shown in Fig.~\ref{fig:cls_results}. We find similar results for other shells, apertures, and foreground-reduced maps, which reinforces that ARF-kSZ tomography produces unbiased measurements of the kSZ effect.


\begin{figure}
\hspace*{-0.7cm}\includegraphics[width=\columnwidth]{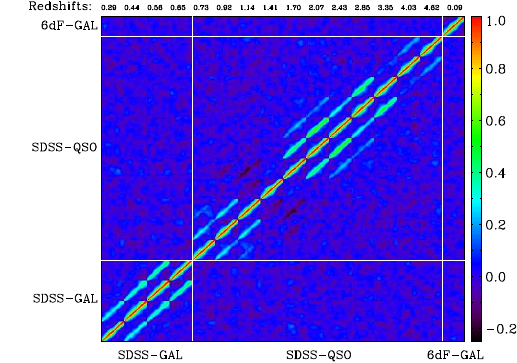}
\caption{
\label{fig:covapz}
Correlation between kSZ optical depths extracted from the \Commander map using different redshifts and apertures. On- and off-diagonal squares indicate the correlation of measurements at the same and different redshift, respectively. As expected, kSZ optical depths extracted using similar apertures at the same or nearby redshifts are correlated.
}
\end{figure}

\begin{figure}
\begin{center}
\includegraphics[width=\columnwidth]{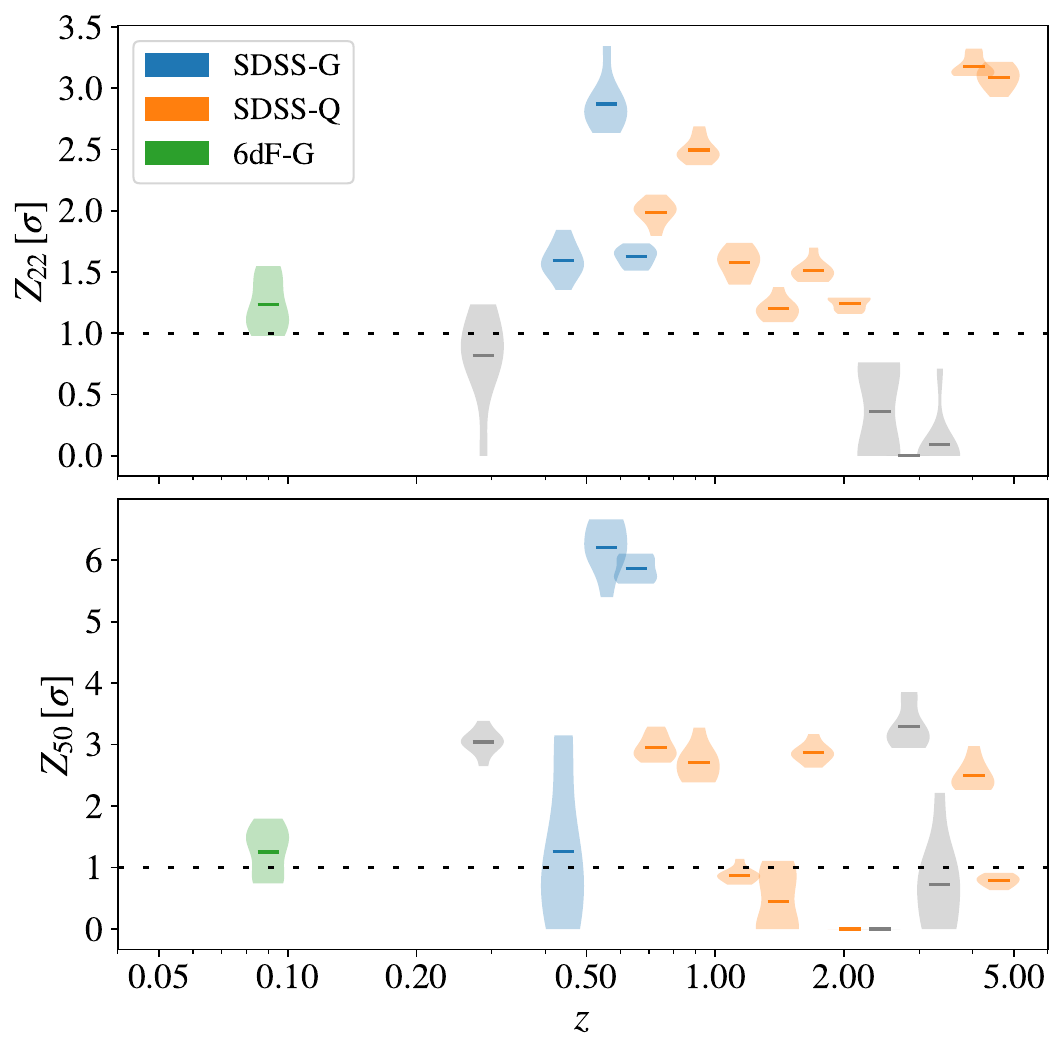}
\end{center}
\caption{
\label{fig:significance}
Joint significance of kSZ measurements from apertures smaller or equal to 22 arcmin (top panel) and 50 arcmin (bottom panel) at the same redshift. The horizontal bar and size of the shaded regions indicate the average and scatter of standard scores computed using different foreground-reduced maps, respectively, green, blue, and orange colours indicate the results for shells containing 6dF galaxies, BOSS galaxies, SDSS quasars, and grey colours denote shells with $Z_{22}<1$. As we can see, we detect the kSZ effect with more significance than 1, 2, and $3\sigma$ on apertures smaller or equal to 22 (50) arcmin for 12, 5, and 2 (10, 8, and 4) redshift shells.
}
\end{figure}

\subsection{Significance of kSZ measurements}
\label{sub:resobv_significance}

In \S\ref{sub:resobv_cross}, we extract kSZ optical depth measurements from the cross-correlation of ARF and filtered CMB maps. In this section, we first study the correlation kSZ measurements from different shells and apertures, and then we compute the joint significance of these.

It is natural to expect correlation between measurements extracted using similar apertures at fixed redshift as these apertures are sensitive to approximately the same gas distribution. We also anticipate correlation between different redshifts due to nearby shells providing similar weights to tracers located between their central redshifts. To estimate these correlations, we begin by extracting simulated kSZ optical depth measurements from the cross-correlation of each of the ARF and simulated filtered CMB maps presented in \S\ref{sub:resobv_cov}; in this manner, we end up with 1\,000 mock kSZ measurements for each shell, aperture, and foreground-reduced map. Then, we use these mock measurements to estimate correlations using the following expression

\begin{equation}
    \label{eq:compute_cov_tau}
    \mathcal{C}_\tau(i, j) = \frac{1}{M-1} \sum_{m=1}^M \left[\simtauAP^m(i) - \bsimtauAP(i)\right] \left[\simtauAP^m(j) - \bsimtauAP(j)\right],
\end{equation}

\noindent where $\simtauAP$ is a vector containing mock measurements from different redshifts and apertures, $i$ and $j$ run over the elements of this vector, and $m$ iterates over mocks. In Fig.~\ref{fig:covapz}, we display the correlation matrix $r_\tau(i,j)=\mathcal{C}_\tau(i,j)/\sqrt{\mathcal{C}_\tau(i,i) \mathcal{C}_\tau(j,j)}$ that we estimate for the \Commander map. Small on- and off-diagonal squares indicate the correlation of measurements at the same and different redshift, respectively. As expected by the physical picture depicted above, we find strong and weak correlation between measurements extracted using apertures of similar sizes from the same and nearby shells, respectively. Note that we find comparable results for other foreground-reduced maps.

We define the joint significance of kSZ measurements from different redshifts and apertures as the signal-to-noise ratio of these weighted by the covariance matrices computed above, $\mathrm{SNR}_\mathrm{kSZ}=\sigma_{\bar{T}}^{}({\bf W}^{\rm T} \mathcal{C}_\tau^{-1}{\bf T})$, where $\mathbf{T}$ is a vector containing the kSZ measurements under consideration, ${\bf W}$ denotes a vector of ones with same length as ${\bf T}$, and $\sigma_{\bar{T}}^2=({\bf W}^{\rm T} \mathcal{C}_\tau^{-1}{\bf W})^{-1}$ indicates the variance of the weighted average. The motivation of this election is twofold: it is model independent and penalises negative kSZ measurements without possible cosmic origin (see \S\ref{sub:theory_ksz}). We use the following expression to express the signal-to-noise in terms of standard deviations:

\begin{equation}
\label{eq:zsig}
Z = \sqrt{2}\,\mathrm{erf}^{-1} \left[ \frac{1}{2} + \frac{1}{2} \mathrm{erf}\left(\frac{\mathrm{SNR}_\mathrm{kSZ}}{\sqrt{2}}\right) \right],
\end{equation}

\noindent where $\mathrm{erf}$ and $\mathrm{erf}^{-1}$ refer to the error function and its inverse, respectively. Note that we set the value of $Z$ to zero for measurements with $\mathrm{SNR}_\mathrm{kSZ}\leq 0$. We check that a standard score of one, two, and three is equivalent to one, two, and three standard deviations using simulated kSZ measurements.

Using the previous expression, we find that the joint significance of kSZ measurements from all redshifts and apertures considered in this work is $Z=11.4\pm{}1.4$, where the first and second number indicates the average and standard deviation of the results for different foreground-reduced maps, respectively. This is highest significance detection of the kSZ effect up to date, which highlights the great potential of ARF-kSZ tomography. When considering the significance of measurements from each sample of tracers separately, we find $Z=1.3\pm{}0.4$, $7.6\pm{}0.4$, and $4.3\pm{}0.6$ for 6dF galaxies, BOSS galaxies, and SDSS quasars, respectively. Note that the scatter across foreground-reduced maps is small in all cases, which provides further evidence of the robustness of ARF-kSZ tomography against residual foreground contamination.

The physical model presented in \S\ref{sec:foundations} predicts that kSZ optical depths increase with aperture, reach a maximum, and decrease thereafter, while the uncertainty in these decreases rapidly and slowly with aperture before and after approximately 22 arcmin, respectively (see \S\ref{sub:resobv_cov}). Taken together, we expect statistically significant measurements on apertures smaller than 22 arcmin at all redshifts. Motivated by this, we compute the significance of kSZ measurements from all apertures smaller or equal to $\thetaAP=22$ and 50 arcmin at each redshift; throughout the remainder of this work, we refer to these as $Z_{22}$ and $Z_{50}$, respectively. 

In Fig.~\ref{fig:significance}, we display the value of $Z_{22}$ and $Z_{50}$ for each redshift shell. The horizontal bar and size of the shaded regions indicate the average and scatter of standard scores computed using different foreground-reduced maps, respectively, green, blue, and orange colours show the results for shells containing 6dF galaxies, BOSS galaxies, SDSS quasars, and grey colours denote shells with $Z_{22}<1$. As we can readily see, we detect the kSZ effect with more significance than 1, 2, and $3\sigma$ on apertures smaller or equal to 22 (50) arcmin for 12, 5, and 2 (10, 8, and 4) redshift shells. Note that standard scores depend weakly on the maximum aperture considered because measurements from different apertures present significant correlation; in particular, we find that the same number of shells present more significance than 1, 2, and $3\sigma$ on apertures smaller or equal to 20, 22, and 24 arcmin.

Interestingly, we find that $Z_{22}$ is greater than $Z_{50}$ for 6 of the 9 shells centred at $z>1$; this is explained by the combination of three effects: residual CMB contamination and other systematics affect more strongly large apertures, the amplitude of the kSZ effect decreases for large apertures as the fraction of new gas entering these is negligible, and the physical scale corresponding to a particular aperture increases with redshift. To alleviate the impact of systematics affecting large apertures, in the next section we analyse kSZ optical depth measurements on apertures smaller or equal to 22 (50) arcmin for shells with $Z_{22}>Z_{50}$ ($Z_{50}>Z_{22}$).


\section{Properties of kSZ gas}
\label{sec:gprop}

In \S\ref{sec:resobv}, we extract kSZ measurements from different shells, apertures, and foreground-reduced maps. In this section, we leverage these measurements to set constraints on the abundance and distribution of cosmic gas from the local universe to redshift $z\simeq5$.


\begin{figure*}
\begin{center}

\includegraphics[width=\wfig,height=\hfig]{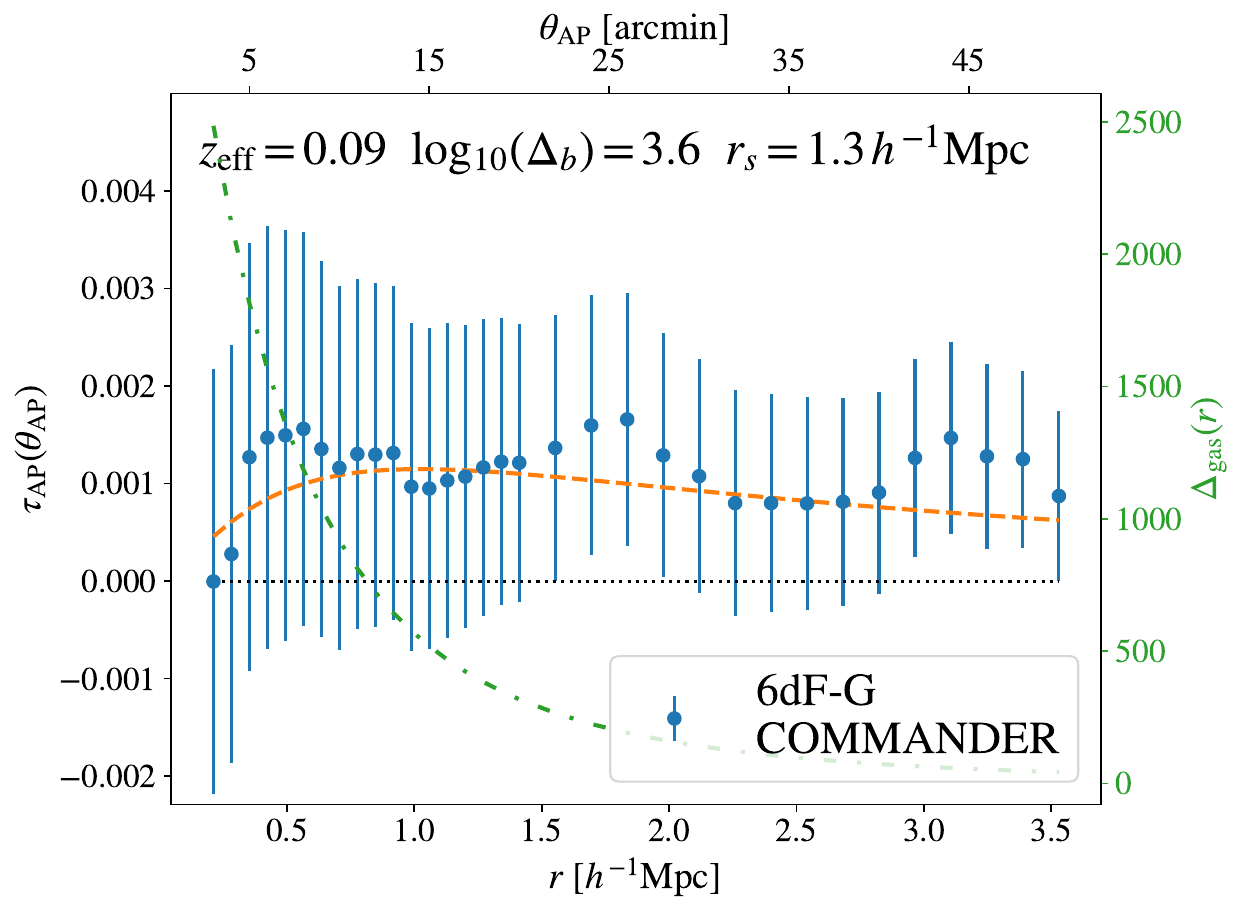} 
\includegraphics[width=\wfig,height=\hfig]{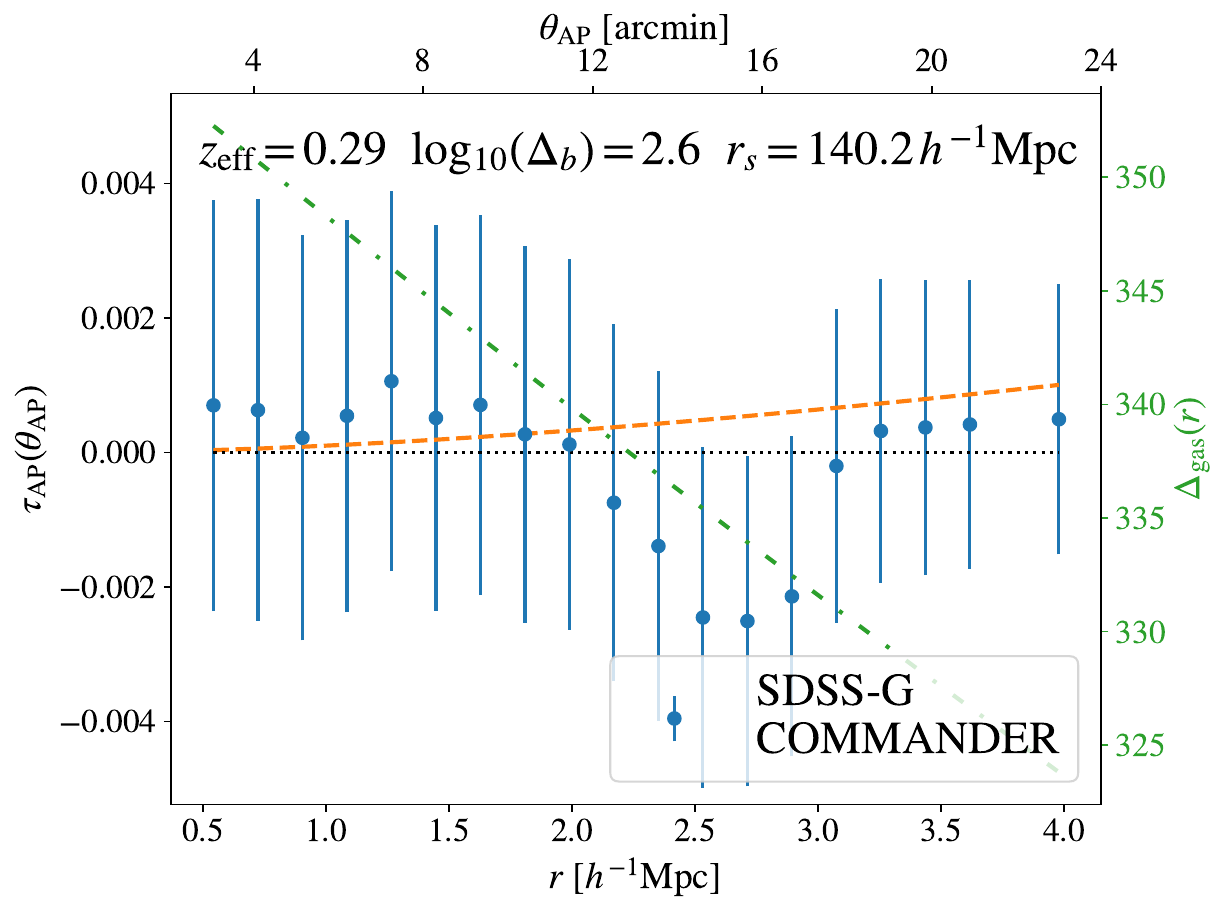} 
\includegraphics[width=\wfig,height=\hfig]{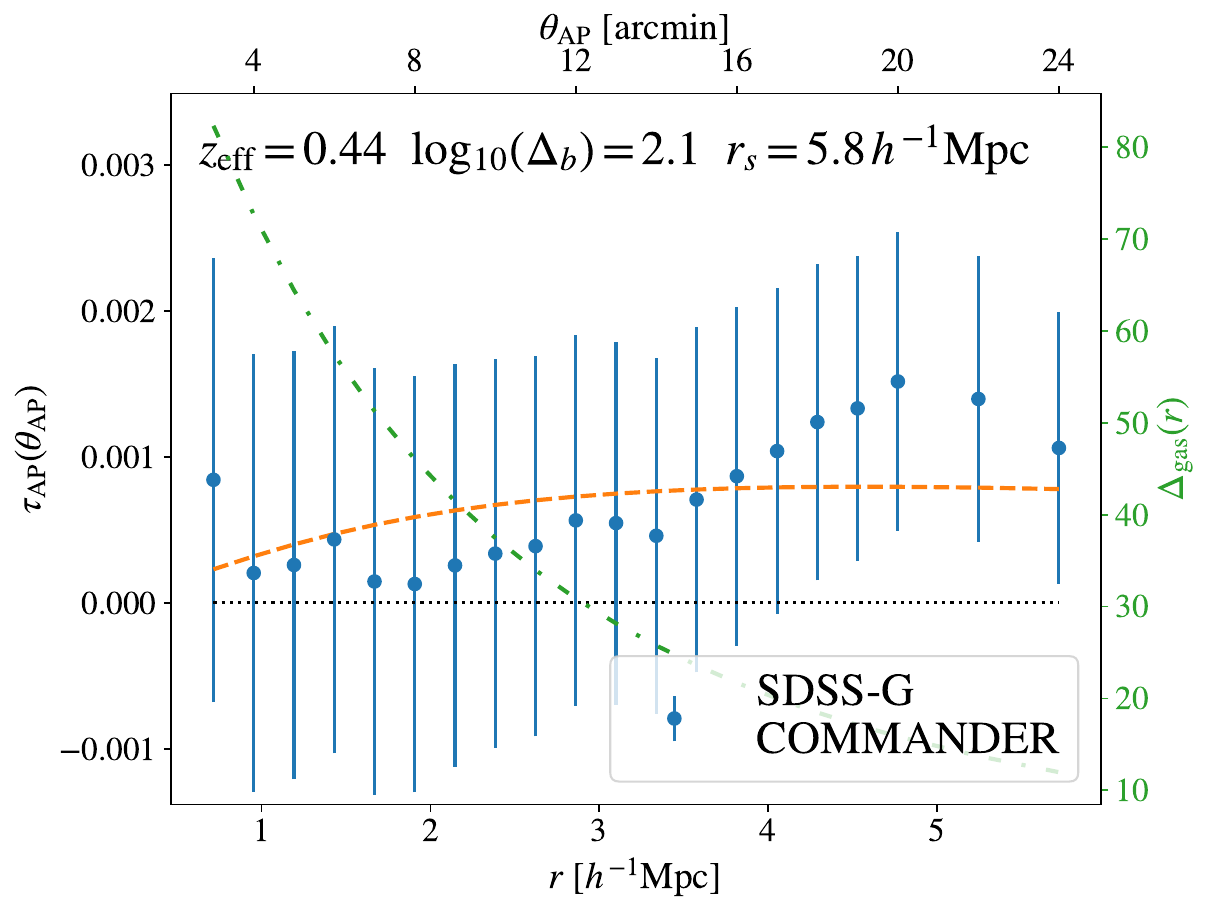} 

\includegraphics[width=\wfig,height=\hfig]{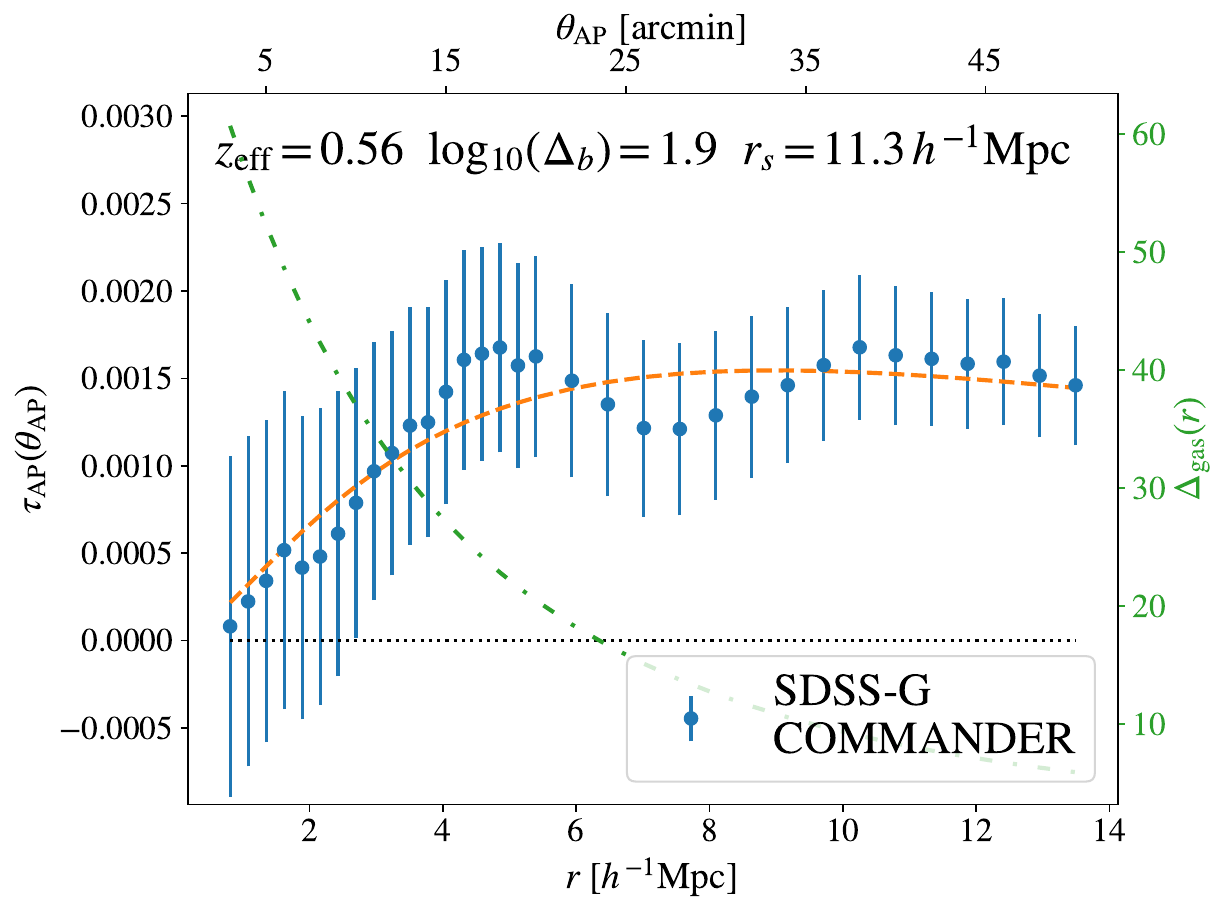} 
\includegraphics[width=\wfig,height=\hfig]{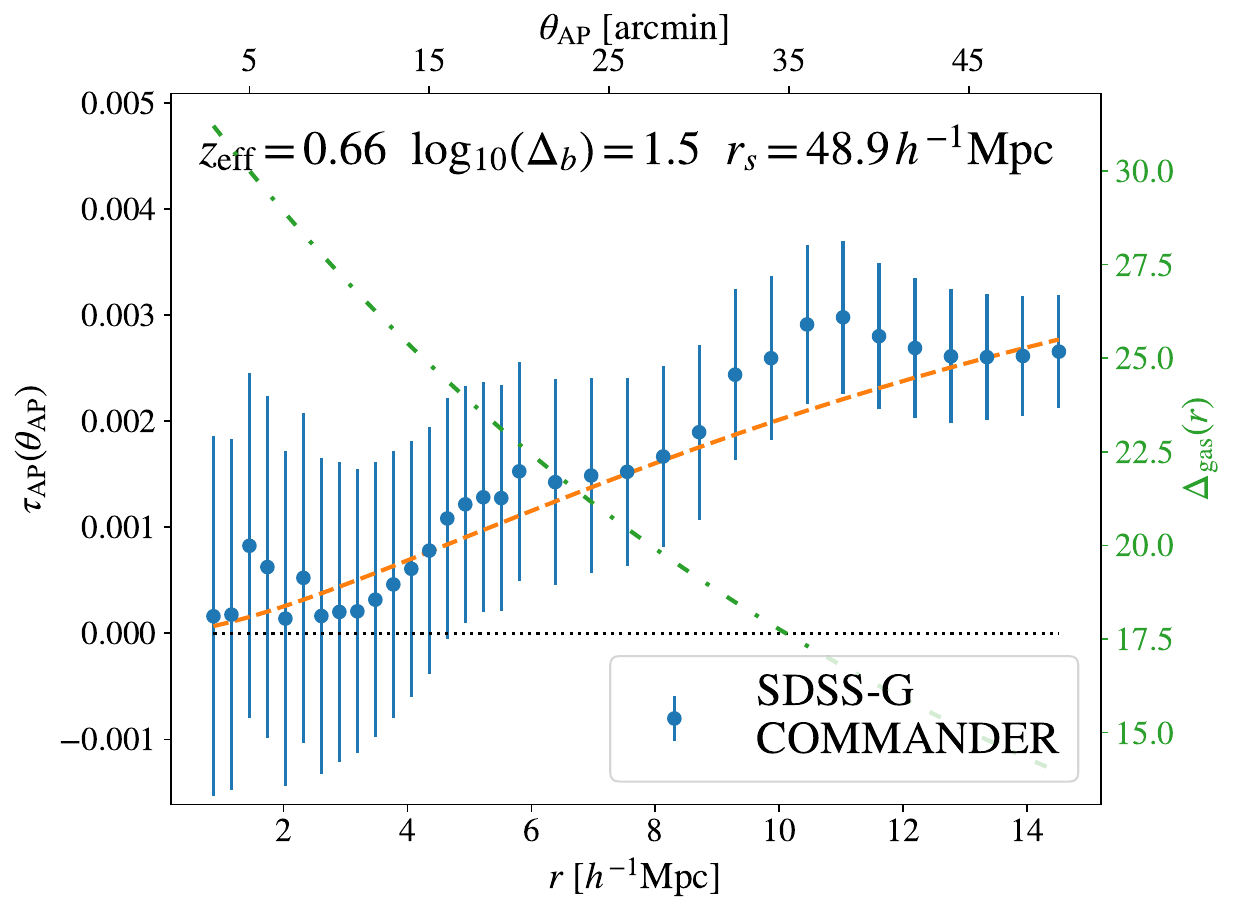} 
\includegraphics[width=\wfig,height=\hfig]{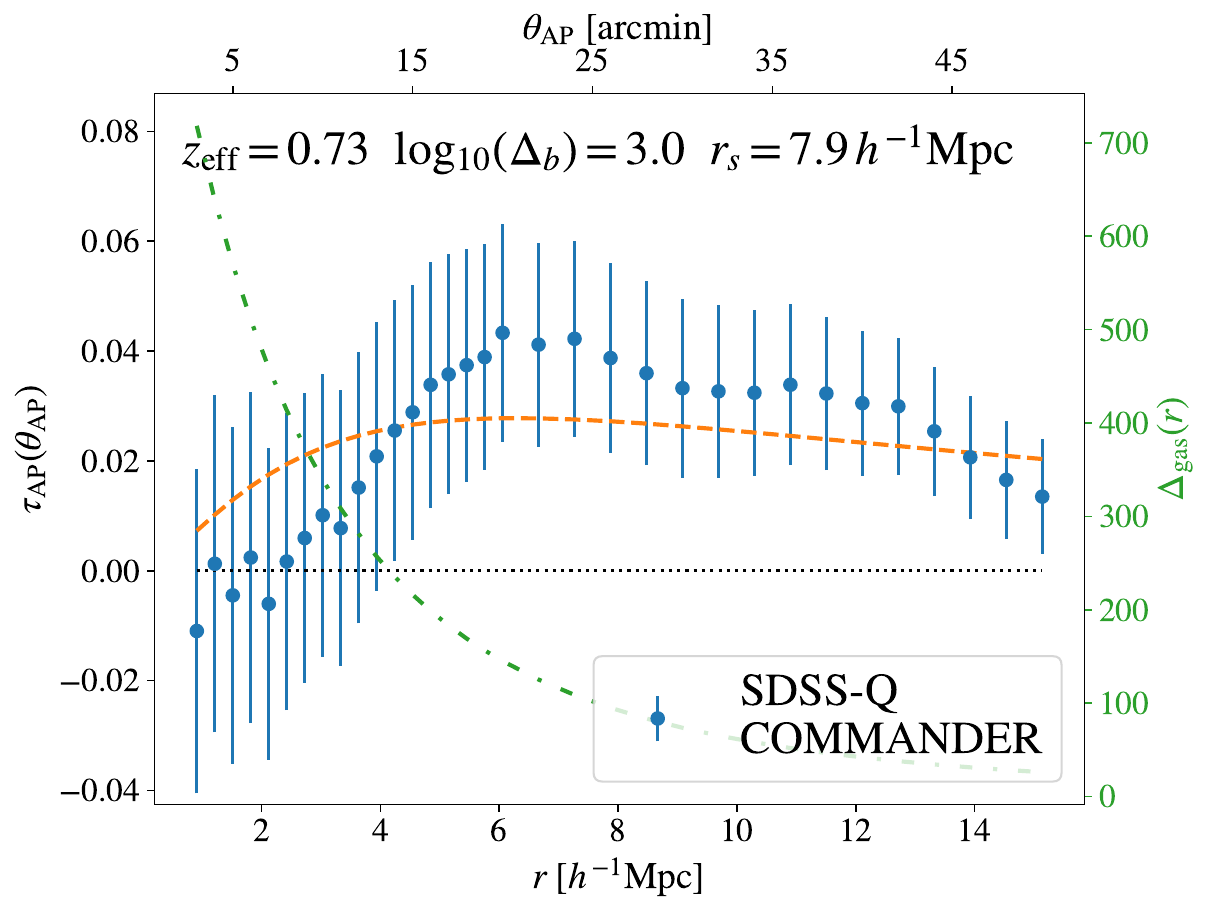} 

\includegraphics[width=\wfig,height=\hfig]{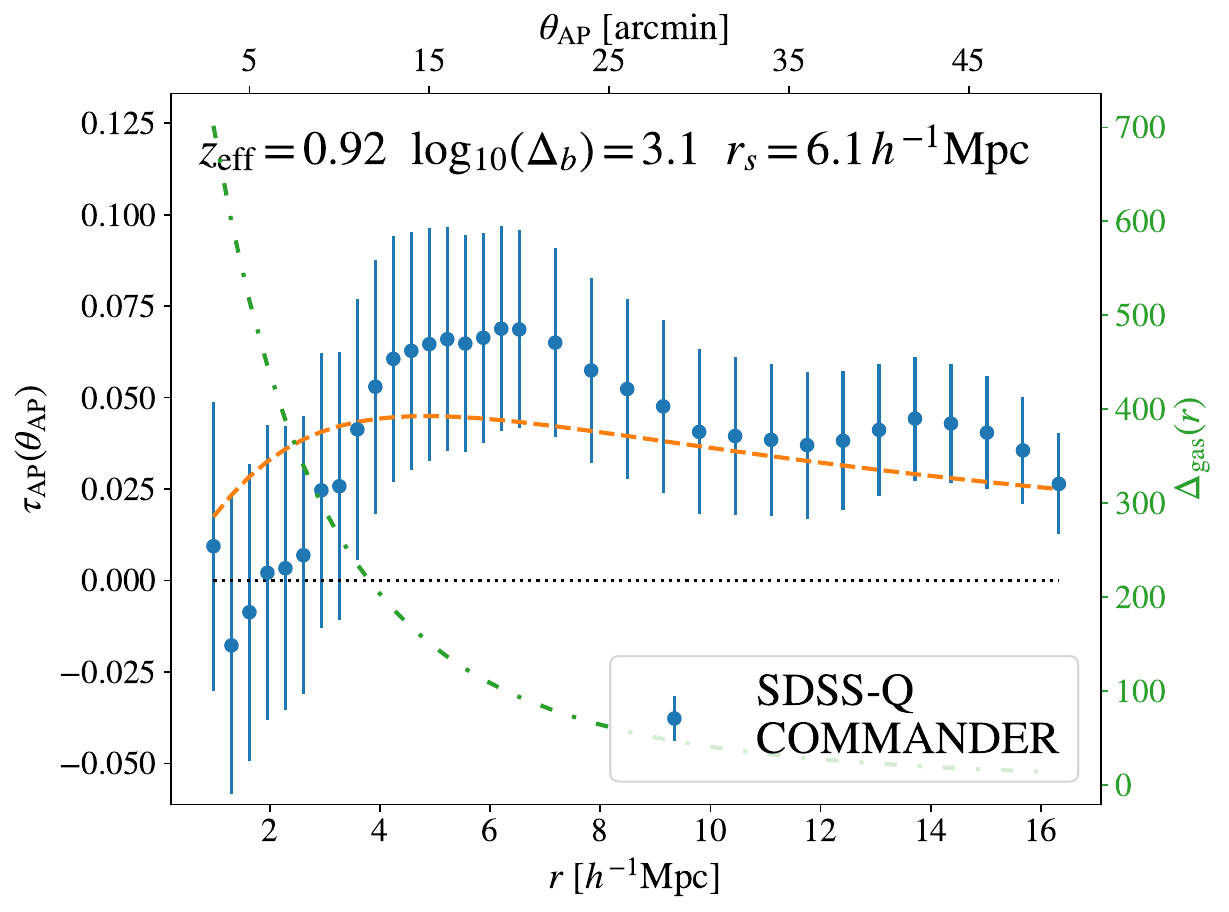} 
\includegraphics[width=\wfig,height=\hfig]{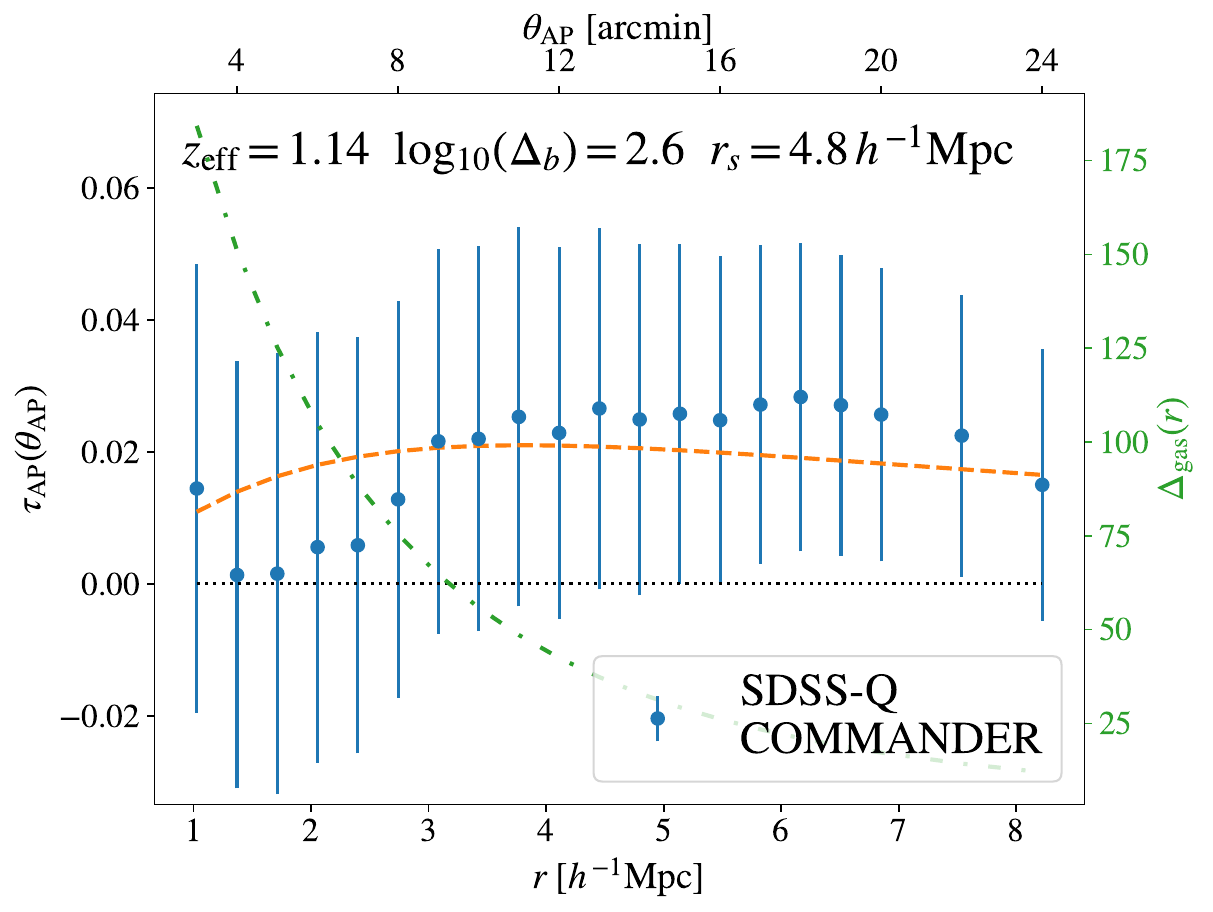} 
\includegraphics[width=\wfig,height=\hfig]{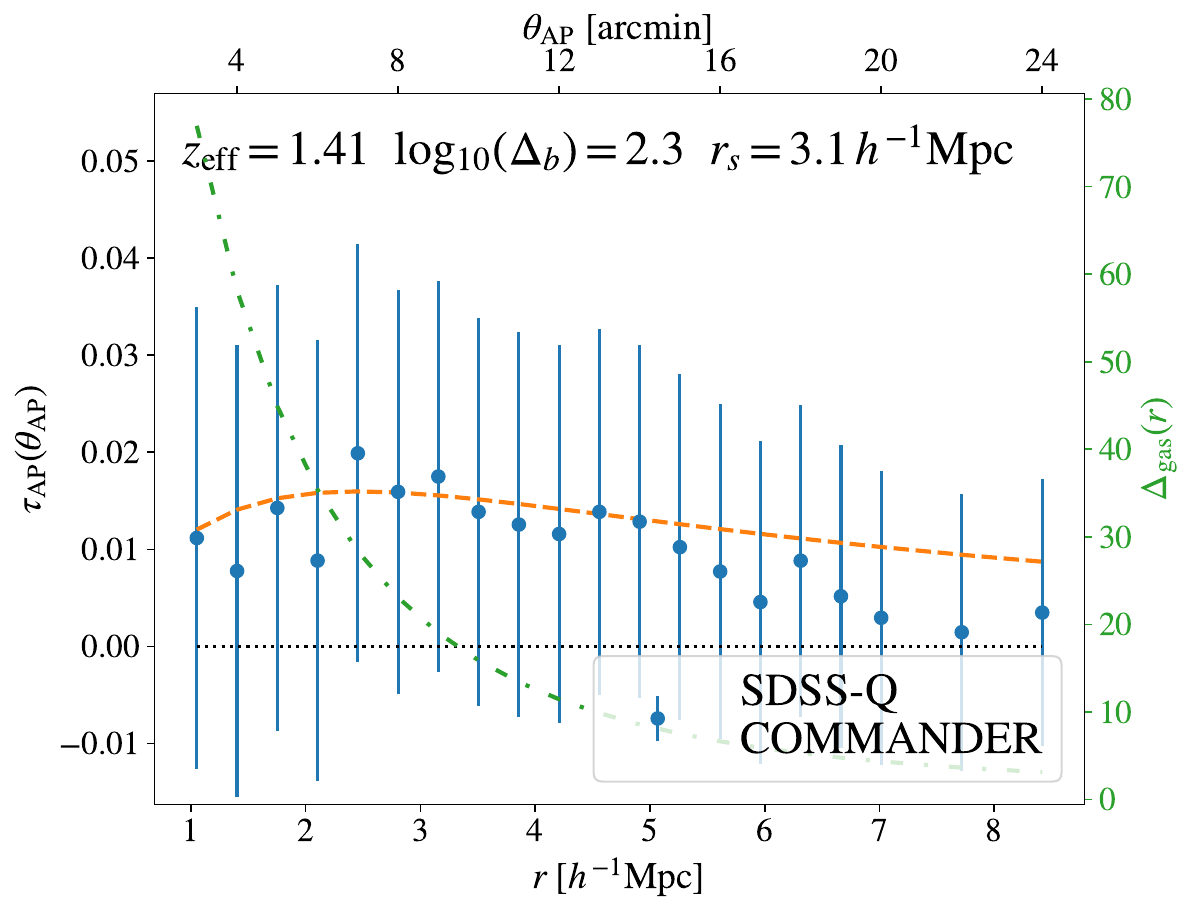} 

\includegraphics[width=\wfigb,height=\hfigb]{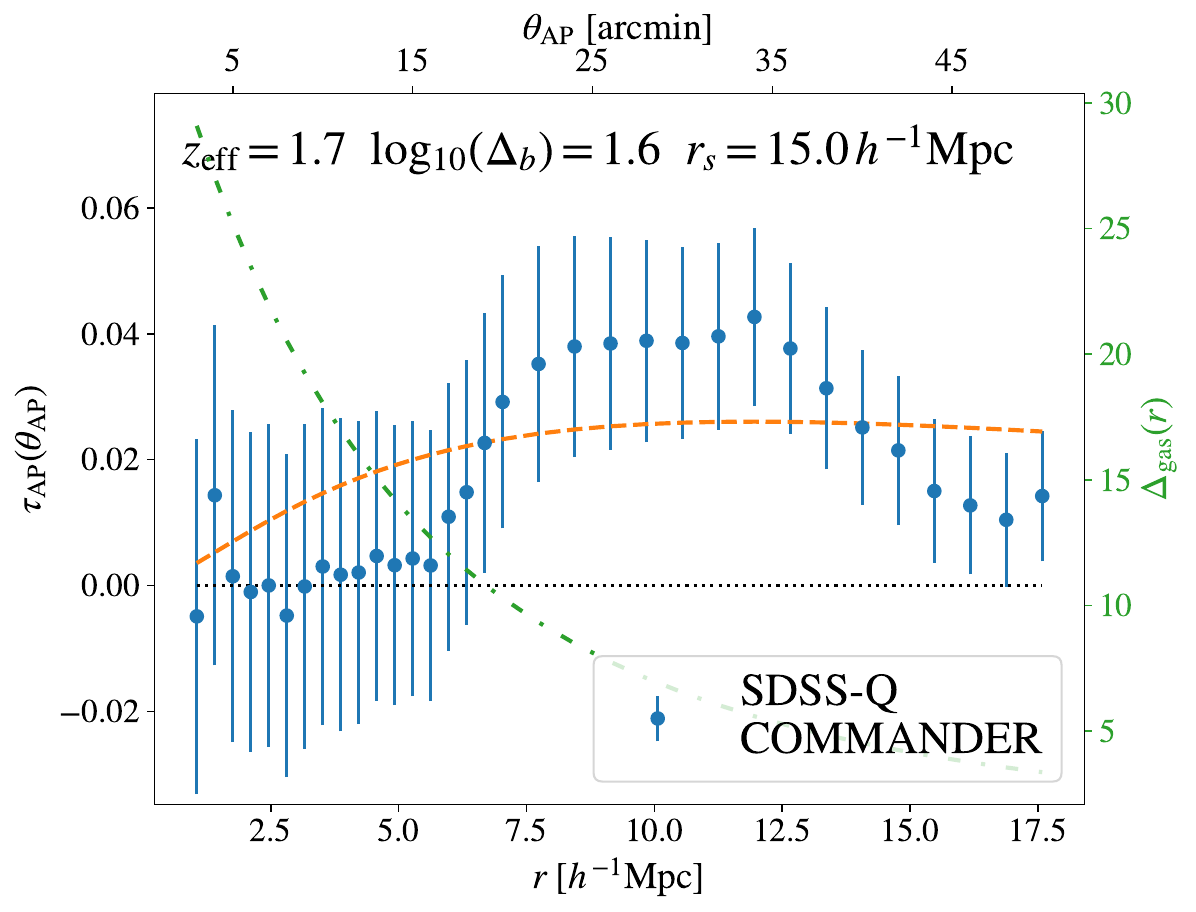} 
\includegraphics[width=\wfigb,height=\hfigb]{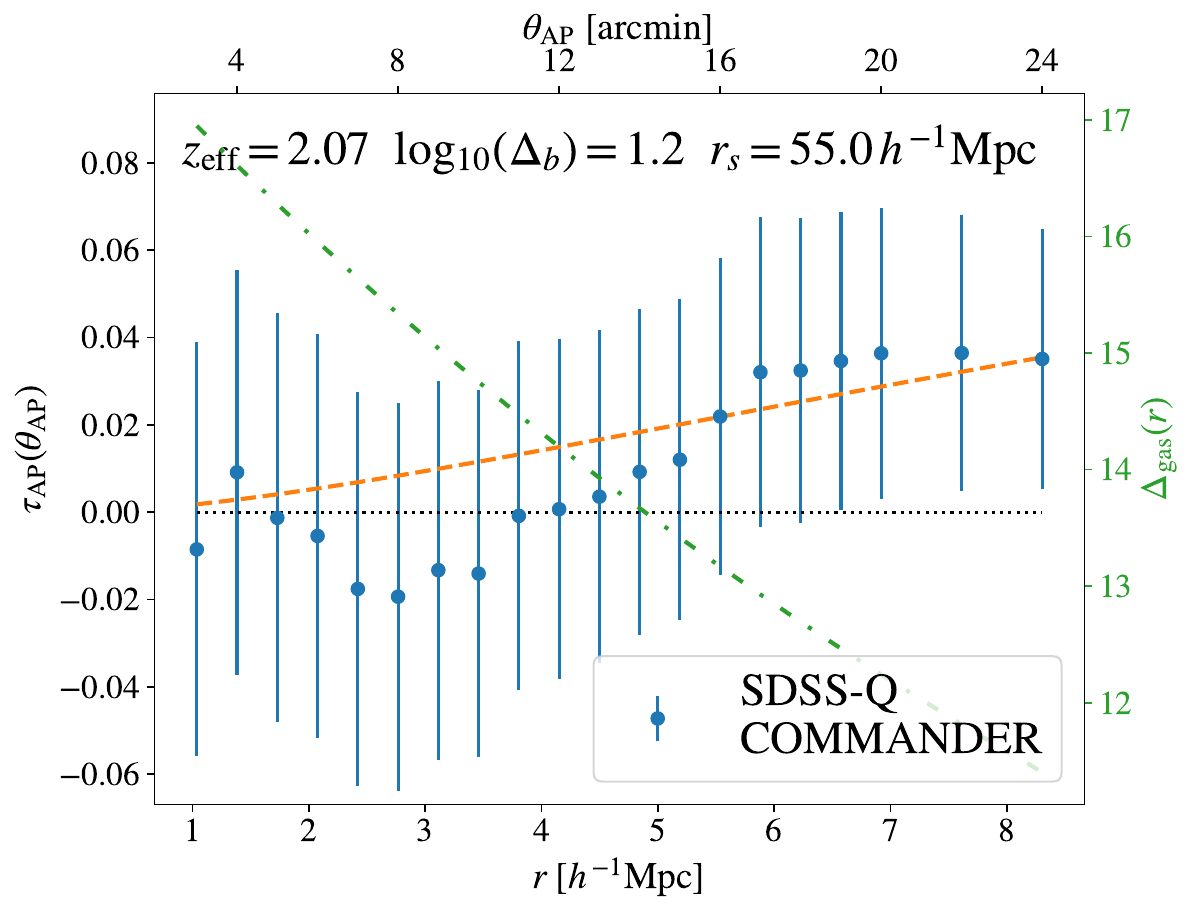} 
\includegraphics[width=\wfigb,height=\hfigb]{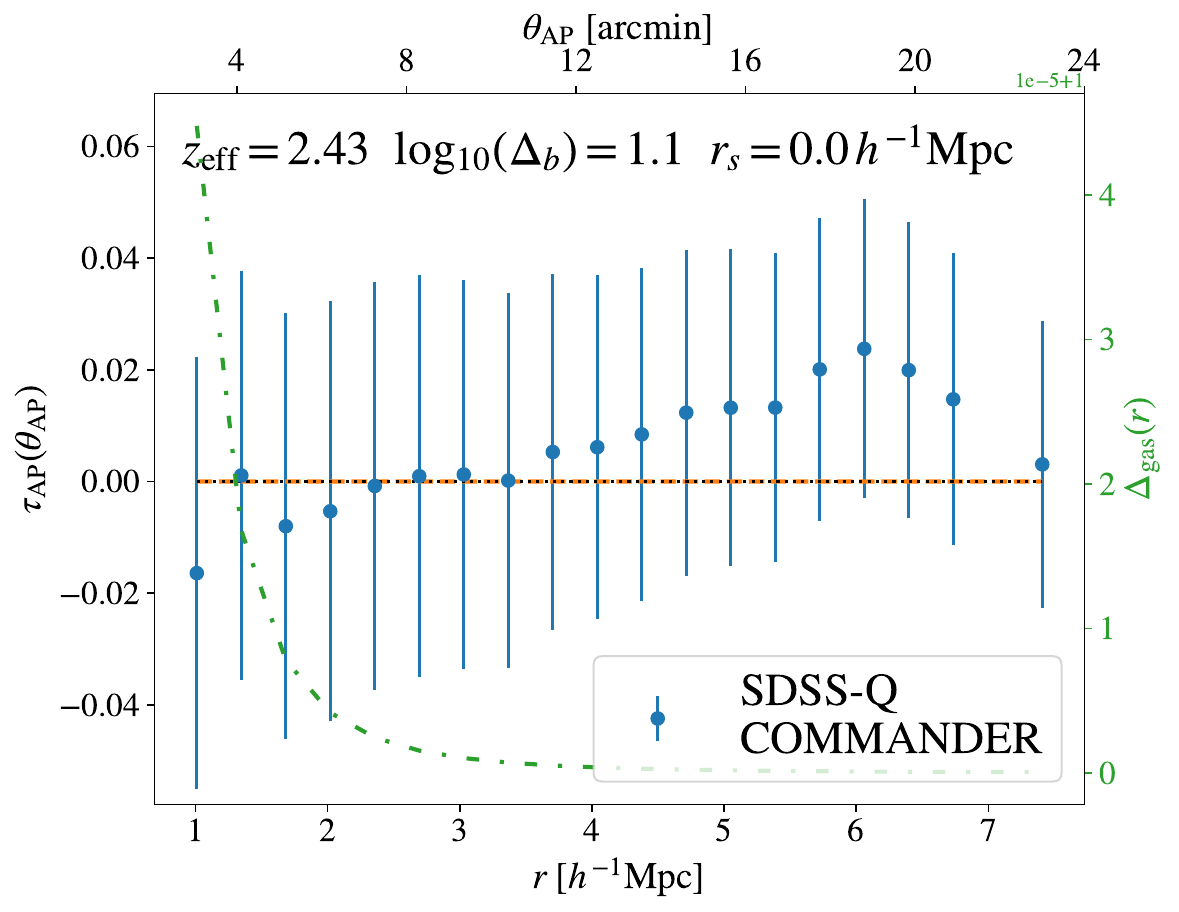}
\includegraphics[width=\wfigb,height=\hfigb]{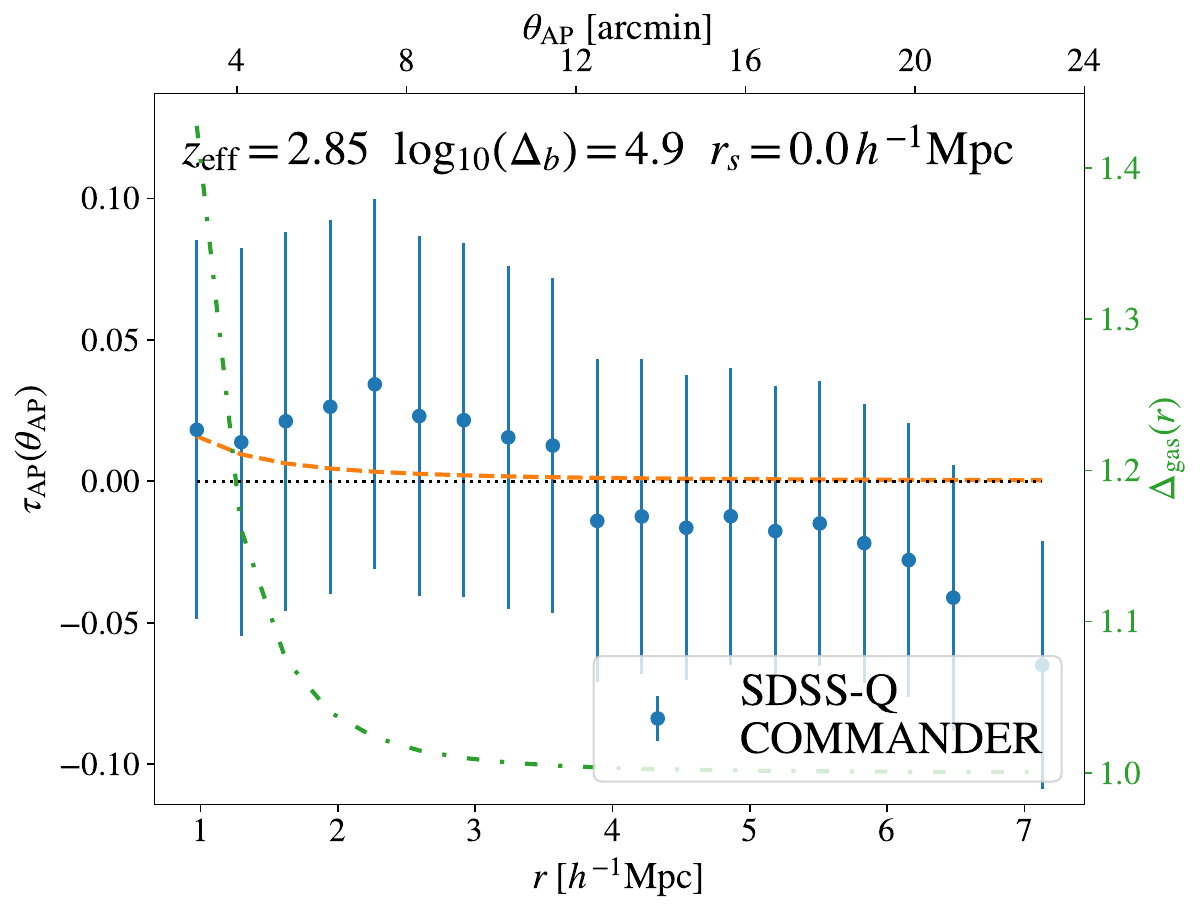} 

\includegraphics[width=\wfig,height=\hfig]{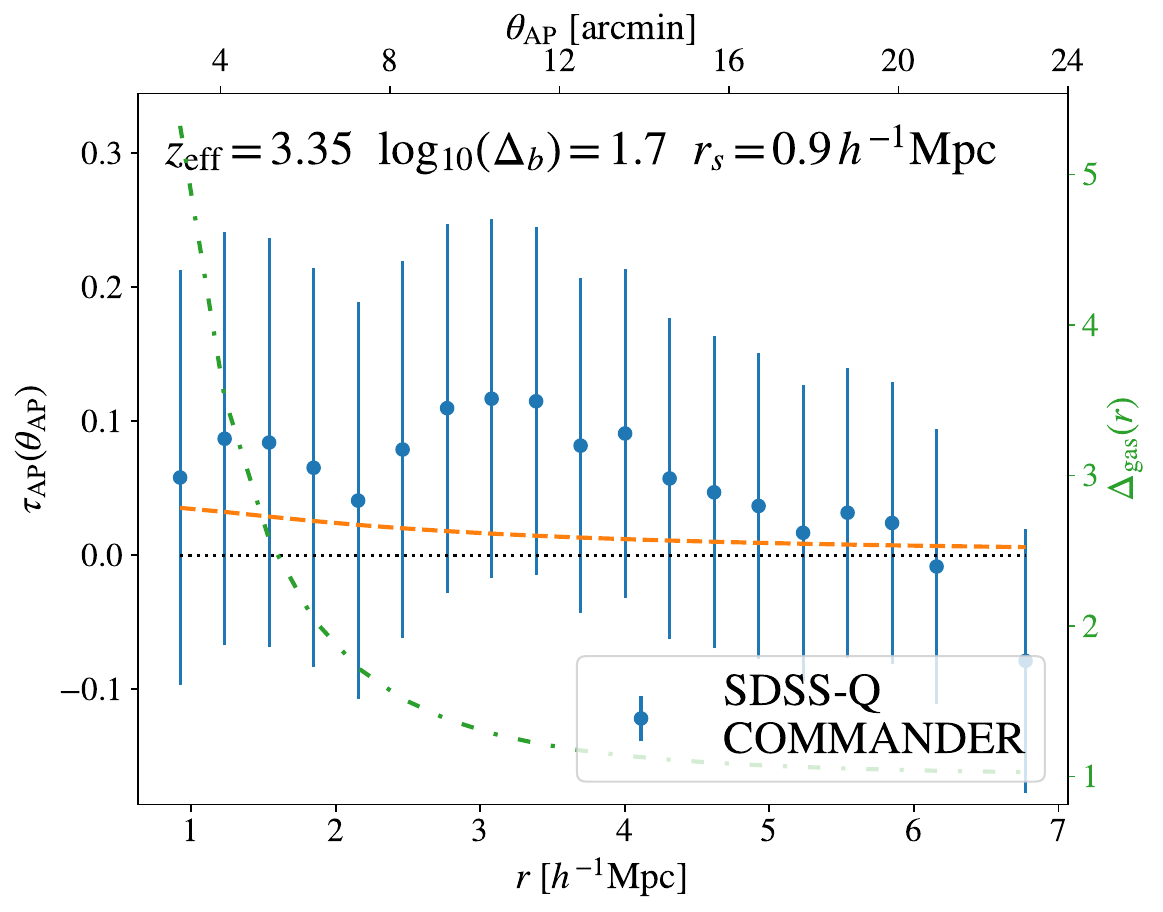} 
\includegraphics[width=\wfig,height=\hfig]{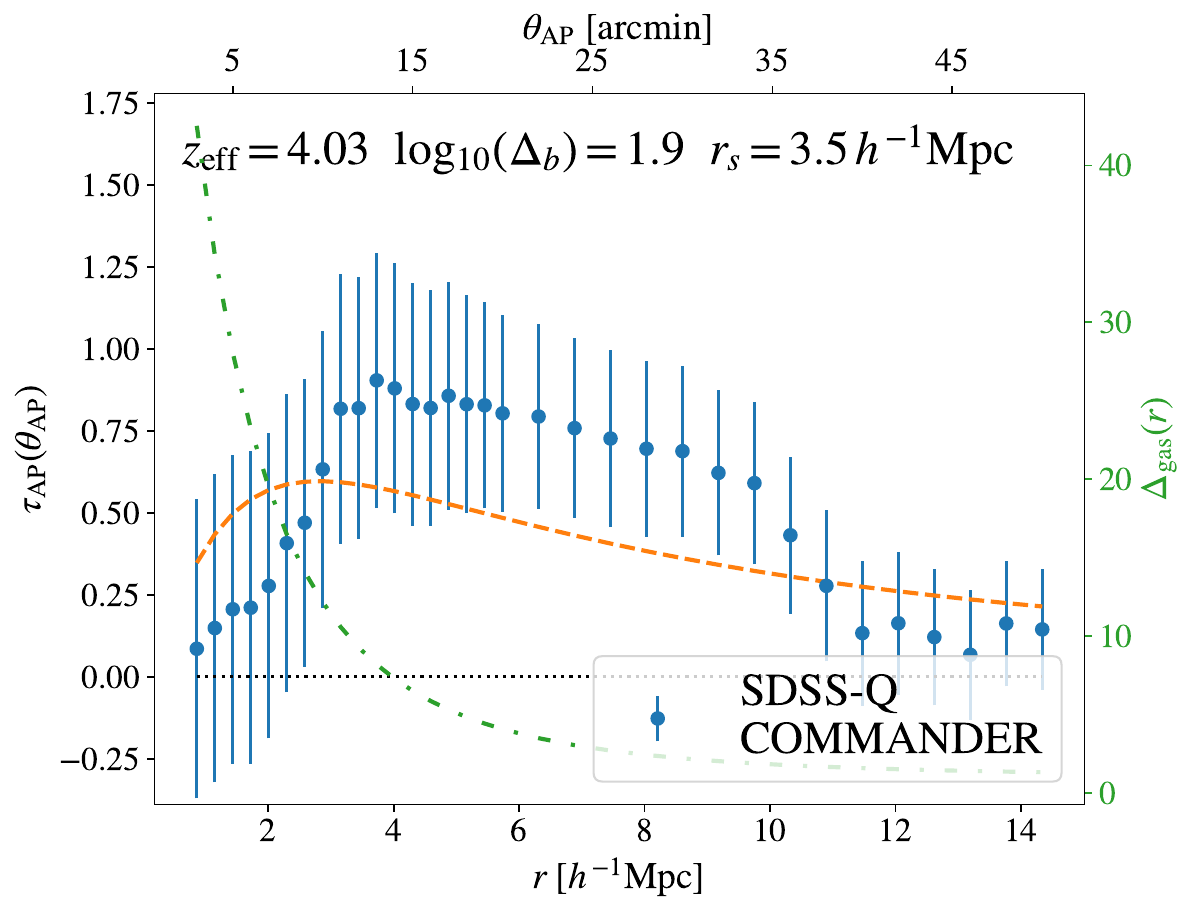}  
\includegraphics[width=\wfig,height=\hfig]{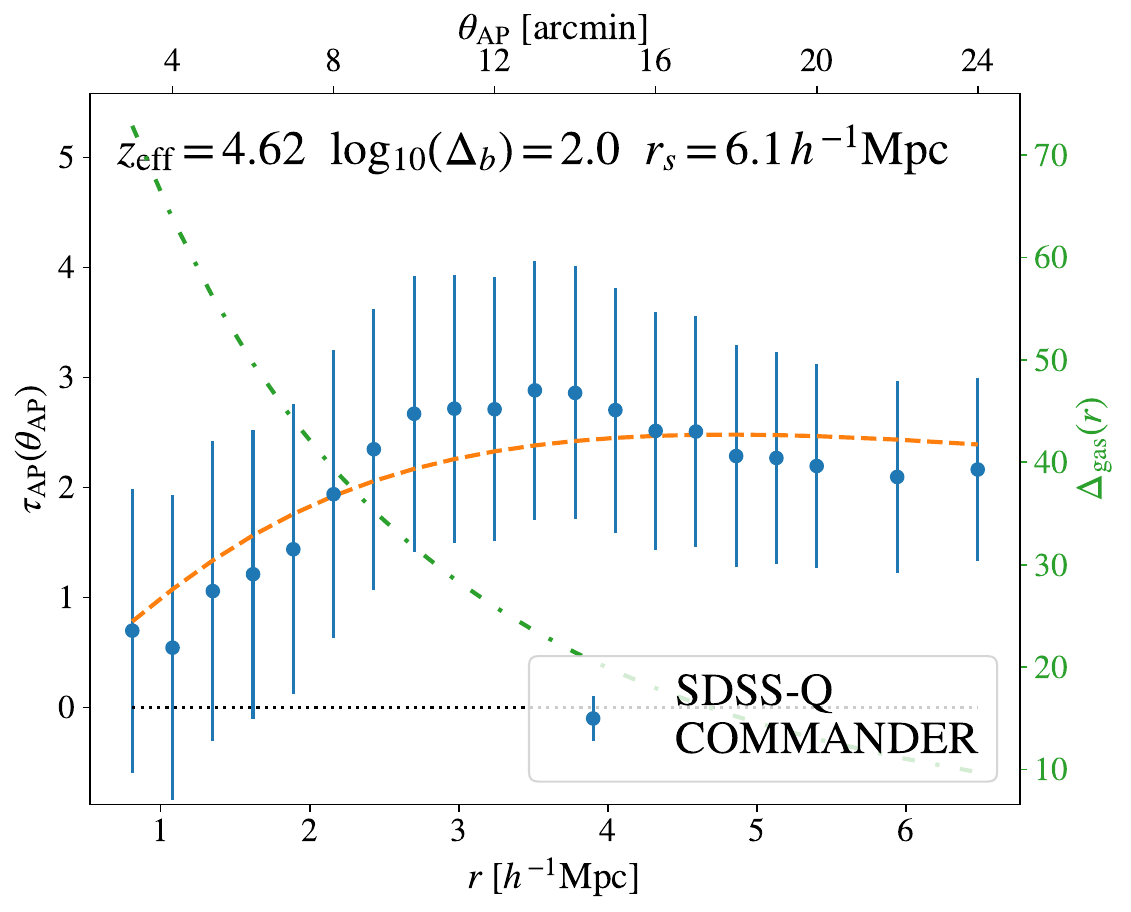}

\end{center}
\caption{
\label{fig:results_tau_ap}
kSZ optical depth measurements extracted from the \Commander map via ARF-kSZ tomography. Each panel displays the results for the shell with effective redshift indicated at its top-left corner, symbols and error bars indicate results and $1\sigma$ uncertainties, respectively, and orange and green lines show best-fitting models and gas profiles. In broad strokes, kSZ measurements grow with aperture, reach a maximum, and decrease thereafter; as we can see, best-fitting models capture this trend precisely for all shells but those at $z_\mathrm{eff}=0.29$, 2.43, 2.85, and 3.35, which present a detection of the kSZ effect less significant than $1\sigma$.
}
\end{figure*}

\begin{figure}
\begin{center}
\includegraphics[width=\columnwidth]{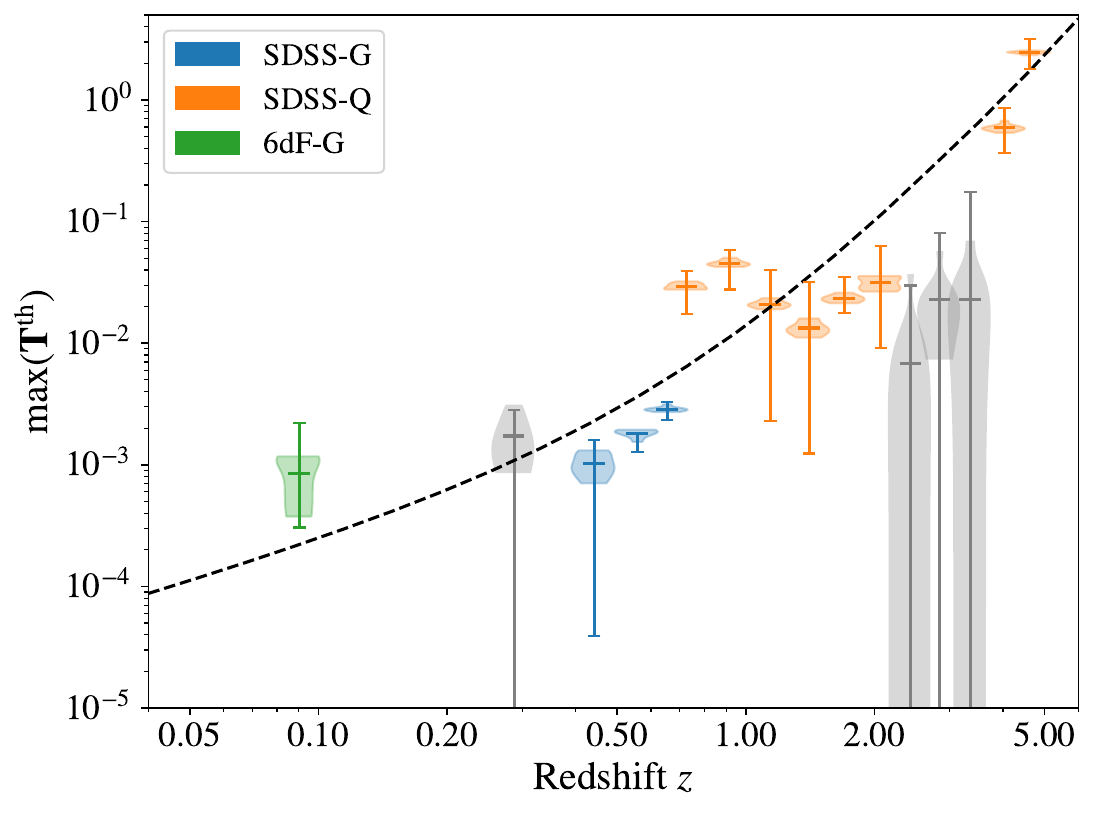}
\end{center}
\caption{
\label{fig:max_tau}
Maximum kSZ optical depth for each redshift shell. We use the same coding as in Fig.~\ref{fig:results_tau_ap} for symbols, error bars denote $1\sigma$ uncertainties for \Commander data, and the dashed line indicates the results predicted by a model assuming the full ionisation of cosmic gas. As we can see, this model captures the redshift evolution of the data precisely.
}
\end{figure}

\subsection{Inferring the properties of kSZ gas}
\label{sub:gprop_inference}

We set constraints on the properties of cosmic gas using kSZ optical depth measurements from different shells separately; this is motivated by the conceivable dependence of such properties on redshift. To do so, we start by creating a vector $\mathbf{T}$ containing all measurements on apertures smaller or equal to 22 and 50 arcmin for shells with $Z_{22}>Z_{50}$ and $Z_{22}<Z_{50}$, respectively (see \S\ref{sub:resobv_significance}). Then, we generate 10,000 logarithmically spaced samples of the two parameters controlling the $\beta$-profile introduced in \S\ref{sub:thsim_gprofile} using Latin Hypercube Sampling \citep[LHS;][]{McKay1979} to within priors $\pi\in\{[1,\,10^6]; [10^{-2},\,10^3]\}$; we check that the results remain unchanged when considering wider priors. After that, we produce theoretical predictions $\mathbf{T}^{\rm th}(\pi)$ for the angular dependence of the kSZ signal at the target redshift by introducing each parameter combination in Eq.~\ref{eq:tauap}. Lastly, we determine the best-fitting model to data by computing the parameters that minimise the absolute value of the log-likelihood $\log \mathcal{L}(\mathbf{T}|\pi) \propto [\mathbf{T}-\mathbf{T}^{\rm th}(\pi)]^{\rm T} C_\tau^{-1} [\mathbf{T}-\mathbf{T}^{\rm th}(\pi)]$.

In Fig.~\ref{fig:results_tau_ap}, we show kSZ optical depth measurements extracted from the \Commander map for each redshift shell considered in this work. Each panel displays the results for the shell with effective redshift indicated at the top-left corner of the panel, symbols and error bars denote results and $1\sigma$ uncertainties, respectively, and orange and green lines depict best-fitting models and gas profiles. The top and bottom x-axes provide conversion between apertures and scales, while the left and right y-axes indicate kSZ optical depths and gas overdensities, respectively. In broad strokes, we find that kSZ optical depths grow with aperture, reach a maximum, and decrease thereafter; as we can see, best-fitting models capture this trend precisely for all shells but those at $z_\mathrm{eff}=0.29$, 2.43, 2.85, and 3.35, which present a detection of the kSZ effect less significant than $1\sigma$.

We find that kSZ optical depths grow with redshift, which is naturally expected by the increase in cosmic density with redshift. For better visualisation of this dependence, in Fig.~\ref{fig:max_tau} we display the maximum of the best-fitting model to each redshift. We use the same coding as in Fig.~\ref{fig:significance} for symbols, error bars show $1\,\sigma$ uncertainties estimated for the \Commander map, and the dashed line indicates theoretical predictions from a model assuming the full ionisation of cosmic gas. As we can see, this model captures the redshift dependence of the data precisely. Note that to estimate error bars, we first select all models to within $1\sigma$ from the best-fitting model using $\log\mathcal{L}(\pi^{\rm best})-\log\mathcal{L}(\pi)<0.5$ \citep{Barlow1993}, where $\pi^{\rm best}$ indicates best-fitting parameters. Then, we compute the upper and lower ends of the error bar by selecting the models with the highest and lowest maximum value, respectively. Note that we obtain approximately the same uncertainties via Markov Chain Monte Carlo sampling; throughout the remainder of this section, we consider this approach for efficiency.

\begin{figure}
\begin{center}
\includegraphics[width=\columnwidth]{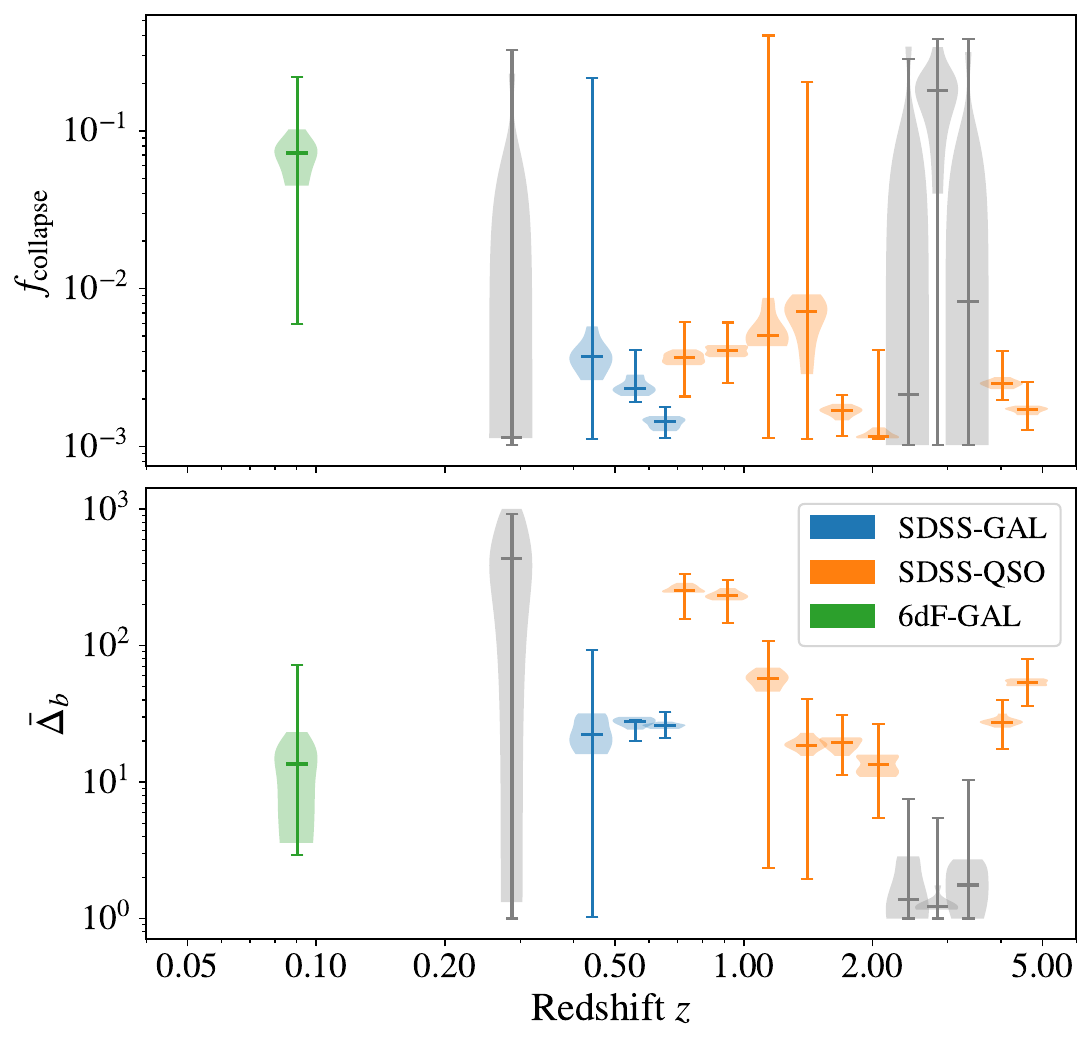}
\end{center}
\caption{
\label{fig:overdensity}
Fraction of kSZ gas to within dark matter haloes (top panel) and average density of kSZ gas outside these structures (bottom panel). We use the same coding as in Fig.~\ref{fig:max_tau}. We find that more than 99\% of kSZ gas resides outside haloes and that the density of this gas ranges from 10 to 250 times the cosmic baryon density, which is in agreement with the density of the interstellar medium according to hydrodynamical simulations.
}
\end{figure}

\subsection{Location and density of kSZ gas}
\label{sub:gprop_overdensity}

It is standard to classify cosmic gas into three main phases according to its location: interstellar medium, which involves gas filling the space between stars, circumgalactic medium, which includes gas outside galaxies but within the virial radius of dark matter haloes, and intergalactic medium, which comprises all gas outside haloes. We compute the fraction of kSZ gas in haloes using $f_{\rm collapse} = \left[\int_0^{r_{\rm in}}r^2\pgas(r, \pi^\mathrm{best})\,\mathrm{d}r\right] / \left[\int_0^{r_{\rm out}}r^2\pgas(r, \pi^\mathrm{best})\,\mathrm{d}r\right]$, where $\pgas(r, \pi^\mathrm{best})$ is the best-fitting gas profile at each redshift, $r_{\rm in}=1\,\Mpch$ indicates an upper limit for the average extension of the halos hosting our tracers, and $r_{\rm out}=10\,\Mpch$ refers to the maximum aperture considered for most redshift shells. Despite the virial radius of haloes hosting 6dF galaxies, BOSS galaxies, and SDSS quasars is on average smaller than $1\,\Mpch$, we select this value to avoid overestimating the fraction of kSZ gas outside haloes.

In the top panel of Fig.~\ref{fig:overdensity}, we display the fraction of kSZ gas in haloes using the same coding as in Fig.~\ref{fig:max_tau}. As we can see, more than 99\% of kSZ gas resides outside haloes for redshift shells with $Z_{22}>1$ at $z>0.1$; therefore, kSZ measurements are practically insensitive to intergalactic and circumgalactic gas. We check that this result presents weak dependence upon $r_{\rm in}$: the average fraction of gas outside haloes is 95.9, 99.1, and 99.8\% for $r_{\rm in}=2$, 1, and $0.5\,\Mpch$, respectively. We also find $f_{\rm collapse}$ is practically independent of the value of $r_{\rm out}$; this is because more than 75\% of kSZ gas resides to within $10\,\Mpch$ from the centre of best-fitting profiles.

In the bottom panel of Fig.~\ref{fig:overdensity}, we display the density of kSZ gas outside haloes relative to the critical density of baryons $\bar{\Delta}_b = 3/(r_\mathrm{out}^3 - r_\mathrm{in}^3) \int_{r_\mathrm{in}}^{r_\mathrm{out}} r^2 \pgas(r, \pi^\mathrm{best})\,\mathrm{d}r$, where we set $r_\mathrm{in}$ and $r_\mathrm{out}$ to the same values as above. As we can see, the density of this gas ranges from 10 to 250 times the cosmic baryon density for shells with $Z_{22}>1$; interestingly, this range of densities agrees with that predicted for intergalactic gas in filaments and sheets by hydrodynamical simulations \citep[e.g.,][]{Martizzi2019}. Taken together with the fraction of kSZ gas outside haloes, these results suggest that ARF-kSZ tomography is mostly sensitive to intergalactic gas in filaments and sheets. Note that we find analogous results when considering other values of $r_\mathrm{in}$ and $r_\mathrm{out}$.


\begin{figure*}
\includegraphics[width=\columnwidth]{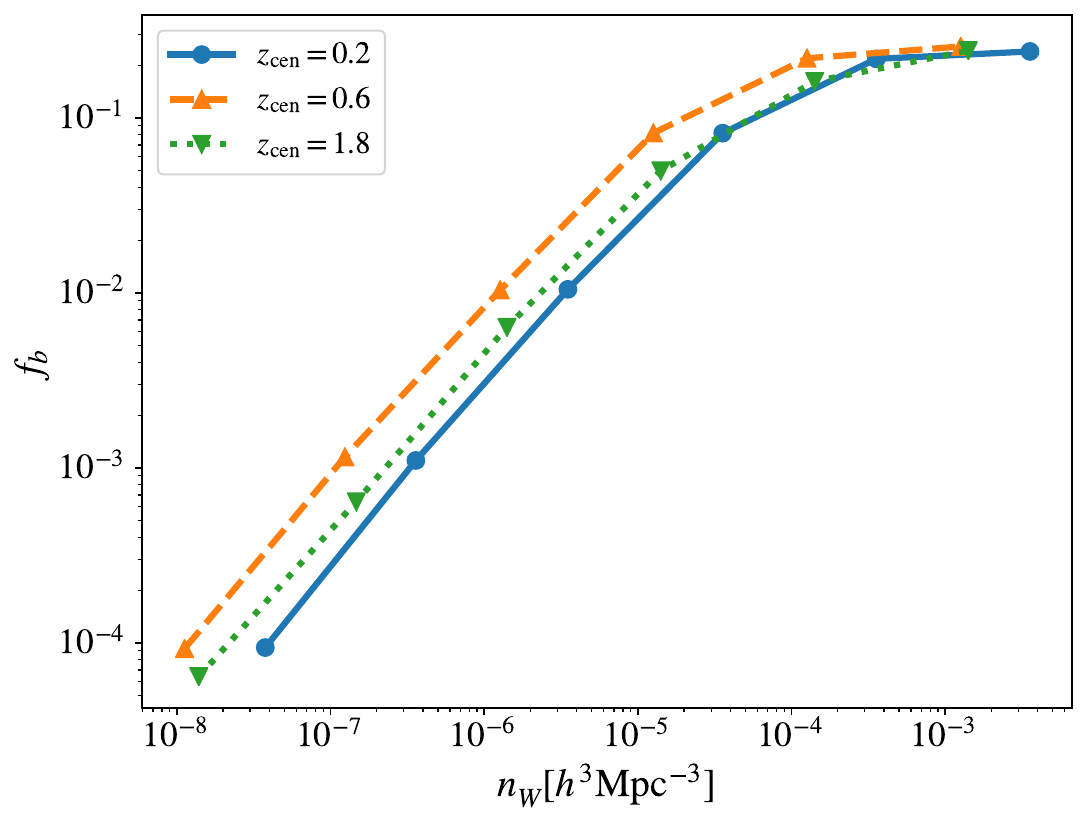}
\includegraphics[width=\columnwidth]{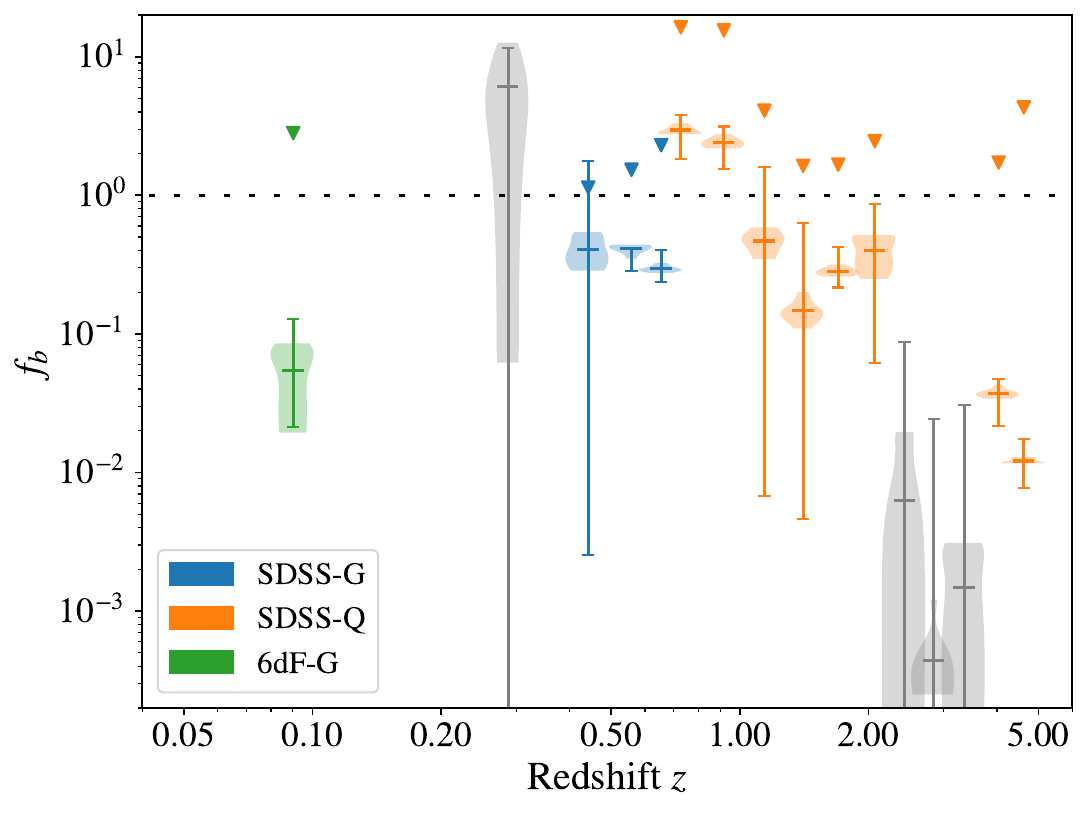}
\caption{
\label{fig:baryons}
{\bf Left panel.} Abundance of detected kSZ gas as a function of number density of tracers. The blue, orange, and green lines indicate predictions from simulations for redshift shells at $z_\mathrm{eff}=0.2$, 0.6, and 1.8, respectively. We find that the quantity of detected kSZ gas increases with the number of tracers and asymptotically approaches a value that depends on the gas profile. {\bf Right panel.} Fraction of the total abundance of cosmic baryons detected at different redshifts. Triangles show predictions for the highest baryon fraction attainable at fixed gas profile. Remarkably, ARF-kSZ tomography is sensitive to approximately 50\% of cosmic baryons, which highlights the efficiency of this technique detecting intergalactic gas.
}
\end{figure*}

\subsection{Abundance of kSZ gas}
\label{sub:gprop_abundance}

Until recently, the percentage of baryons detected at low redshift was well below early universe expectations \citep[e.g.,][]{Shull12, Nicastro18}. To shed new light into this so-called missing baryon problem, in this section we carry out a census of baryons from the local universe to $z\simeq5$ using kSZ optical depth measurements.

To determine the fraction of cosmic baryons that we detect at each redshift, we start by projecting the distribution of gas surrounding each tracer onto a sky map of resolution $N_{\rm side}=1024$, which corresponds to 13 arcmin$^2$. To do so, we assign to each pixel of the map to within 22 arcmin from a tracer the integral of the best-fitting profile to kSZ gas at this redshift over its area

\begin{equation}
  \label{eq:map_gas}
  \mathcal{M}_{\rm gas}(\angr_j) = \int_{A_j} \mathrm{d}A \int \pgas(r-r_i, \pi^\mathrm{best}) W(r_i)\,\mathrm{d}l,
\end{equation}

\noindent where $A_j$ indicates the area of the pixel $j$, $\theta$ and $l$ refer to the axial and vertical coordinates of an imaginary cylinder centred at each tracer, respectively, $\mathrm{d}A \equiv \theta\, \mathrm{d}\theta \,\mathrm{d}\varphi$ denotes the differential area element in cylindrical coordinates, $r = \sqrt{\theta^2 + l^2}$ stands for the radial distance in spherical coordinates, $r_i$ is the distance to the tracer $i$, and the map is in units of gas overdensity times volume. We use 22 arcmin because this is the maximum aperture considered for most shells in \S\ref{sub:gprop_inference}, while we only consider the contribution from the closest tracer because the width of the selection function is of the same order as the range of scales over which ARF and the kSZ effect present substantial correlation. Note that the second criterion also prevents us from accounting for the distribution of gas surrounding tracers hosted by the same halo multiple times. Lastly, we compute the fraction of cosmic baryons that we detect using kSZ measurements from each shell using

\begin{equation}
    \label{eq:fb}
    f_b \equiv \frac{\Omega_b}{\Omega_b^{\rm fid}} = \frac{\sum_j \mathcal{M}_\mathrm{gas}(\angr_j)}{V_\mathrm{shell}},
\end{equation}

\noindent where $\Omega_b^{\rm fid}=0.049$ indicates the cosmic baryon density predicted by early universe studies, $V_\mathrm{shell} = 4 \upi f_{\rm sky} \int \mathrm{d}z' c/H(z') [r(z')]^2 W(z')$ denotes the shell volume, and $f_{\rm sky}$ stands for sky fraction sampled by tracers. We check that baryon fractions do not change significantly when projecting kSZ gas onto all pixels to within 18 or 50 arcmin from tracers or when considering maps with higher resolution.

Given that samples with greater number density access larger cosmic volumes, it is natural to expect correlation between baryon fraction and number of tracers. To estimate this dependence, we generate samples of randomly distributed tracers across redshifts and angles, and then we project these onto sky maps using Eq.~\ref{eq:map_gas}. In the left panel of Fig.~\ref{fig:baryons}, we show the resulting baryon fractions as a function of the number density of tracers weighted by the selection function, $n_W^{} = V_{\rm shell}^{-1} \sum_i W_i$. The blue, orange, and green lines indicate the results for tracers surrounded by gas following a $\beta$-profile with parameters $\Delta_b=100$ and $r_s=10\,\Mpch$ and selected under shells of comoving width $\sigma_r=180\,\Mpch$ centred at $z_\mathrm{eff}=0.2$, 0.6, and 1.8, respectively. As expected, we find that the baryon fraction increases with the angular number density of tracers and that it approaches an asymptotic value for number densities larger than $n_W=10^{-3}h^{-3}\Mpc^{-3}$; throughout the remainder of this section, we refer to this value as the highest detectable baryon fraction. As we can see, the highest detectable baryon fraction is redshift independent; however, we find that it depends upon the properties of the best-fitting gas profile.

In Fig.~\ref{fig:baryons}, we display the cosmic baryon fraction that we detect at each redshift. We use the same coding as in Fig.~\ref{fig:max_tau} for symbols, while triangles show predictions for the highest detectable baryon fraction at fixed gas properties. We can readily see that we detect approximately 50\% of cosmic baryons; given that ARF-kSZ tomography is only sensitive to the $\simeq80\%$ of baryons that reside in the intergalactic medium \citep{Nicastro18}, our measurements are compatible with detecting the majority of intergalactic baryons. This result is in line with kSZ studies that detect practically all baryons surrounding low redshift galaxies \citep{chm15, HillprjkSZ16}. Note that the highest detectable baryon fractions are above unity for all redshifts because we hold fixed the amplitude of gas profiles when computing these values, while this amplitude should decrease for samples with higher number density of tracers as so it does the average large-scale bias of these.

As we can see, we detect a baryon fraction above unity for quasars at $z_\mathrm{eff}=0.73$ and 0.92; this is likely due to random fluctuations as these measurements are correlated and compatible with unity at $1.3\sigma$. On the other hand, we find that the baryon fraction for the shells at $z_\mathrm{eff}=0.09$, 4.03, and 4.62 is significantly lower than others with $Z_{22}<1$; this is likely because the number density of tracers is very low at these redshifts, which is indicated by the large gap between actual measurements and predictions from simulations for an infinite number of tracers.


\section{Robustness of the results}
\label{sec:robustness}

In this section, we study the robustness of ARF-kSZ tomography against foregrounds, tSZ emission, systematic uncertainties affecting the number density and large-scale bias of tracers, and the functional form assumed for the kSZ gas profile.


\begin{figure*}
\begin{center}

\includegraphics[width=0.32\textwidth,height=0.245\textwidth]{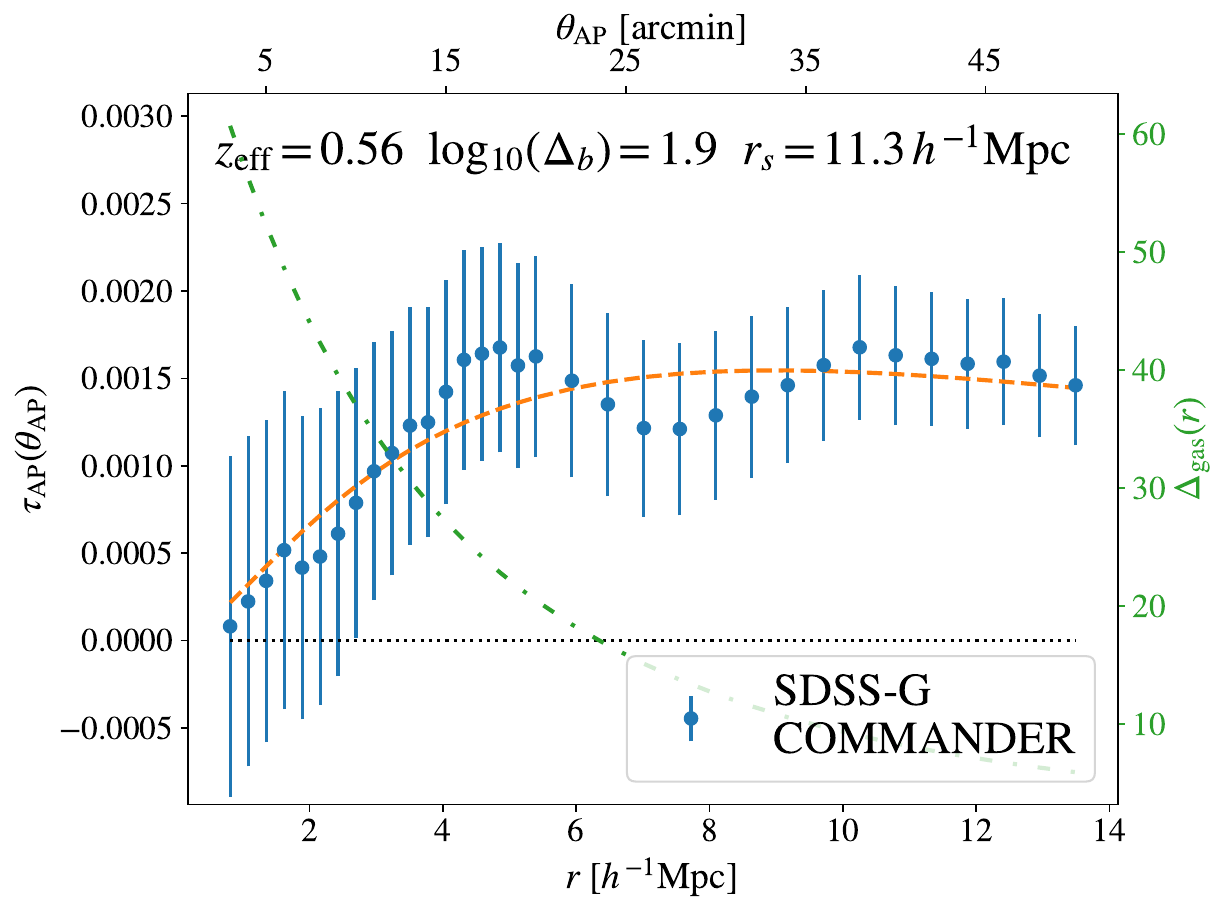} 
\includegraphics[width=0.32\textwidth,height=0.245\textwidth]{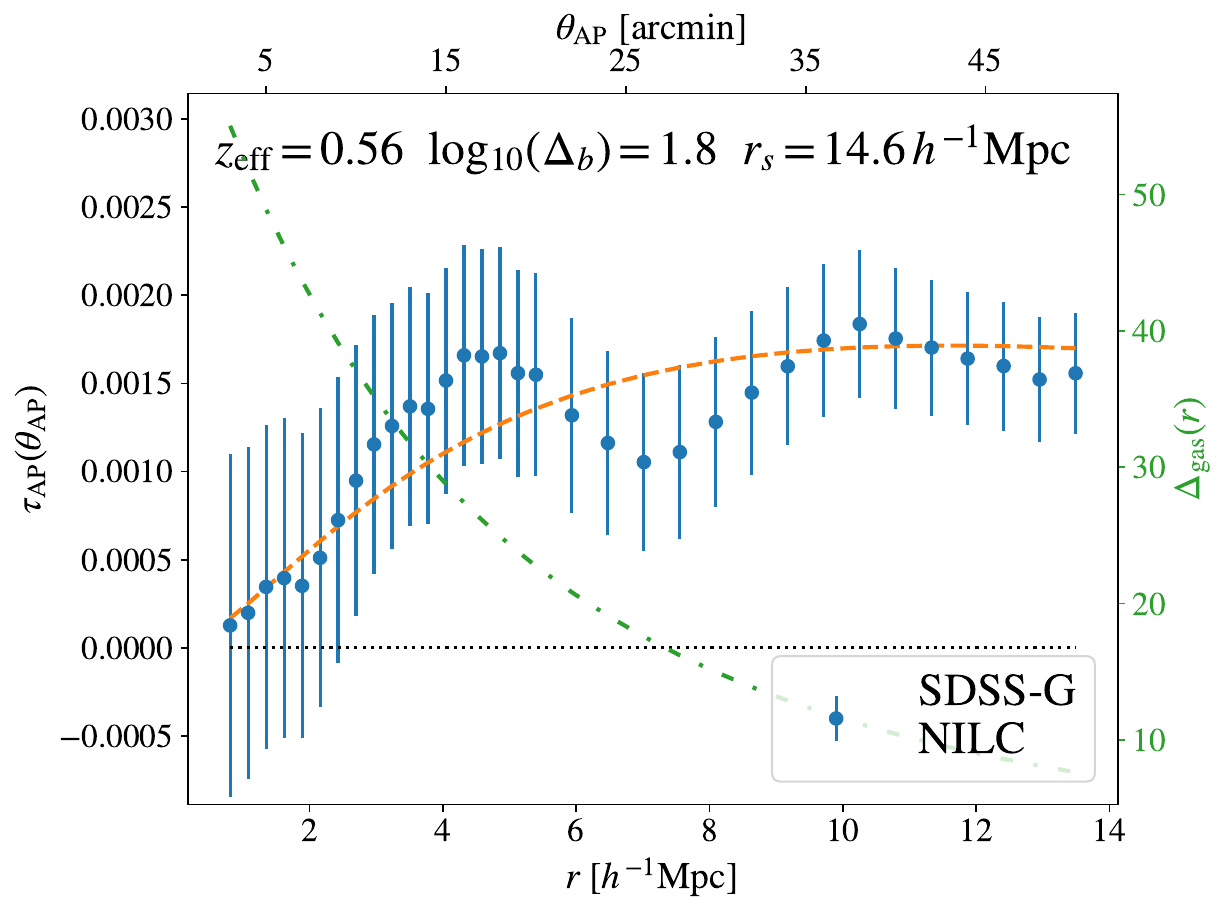} 
\includegraphics[width=0.32\textwidth,height=0.245\textwidth]{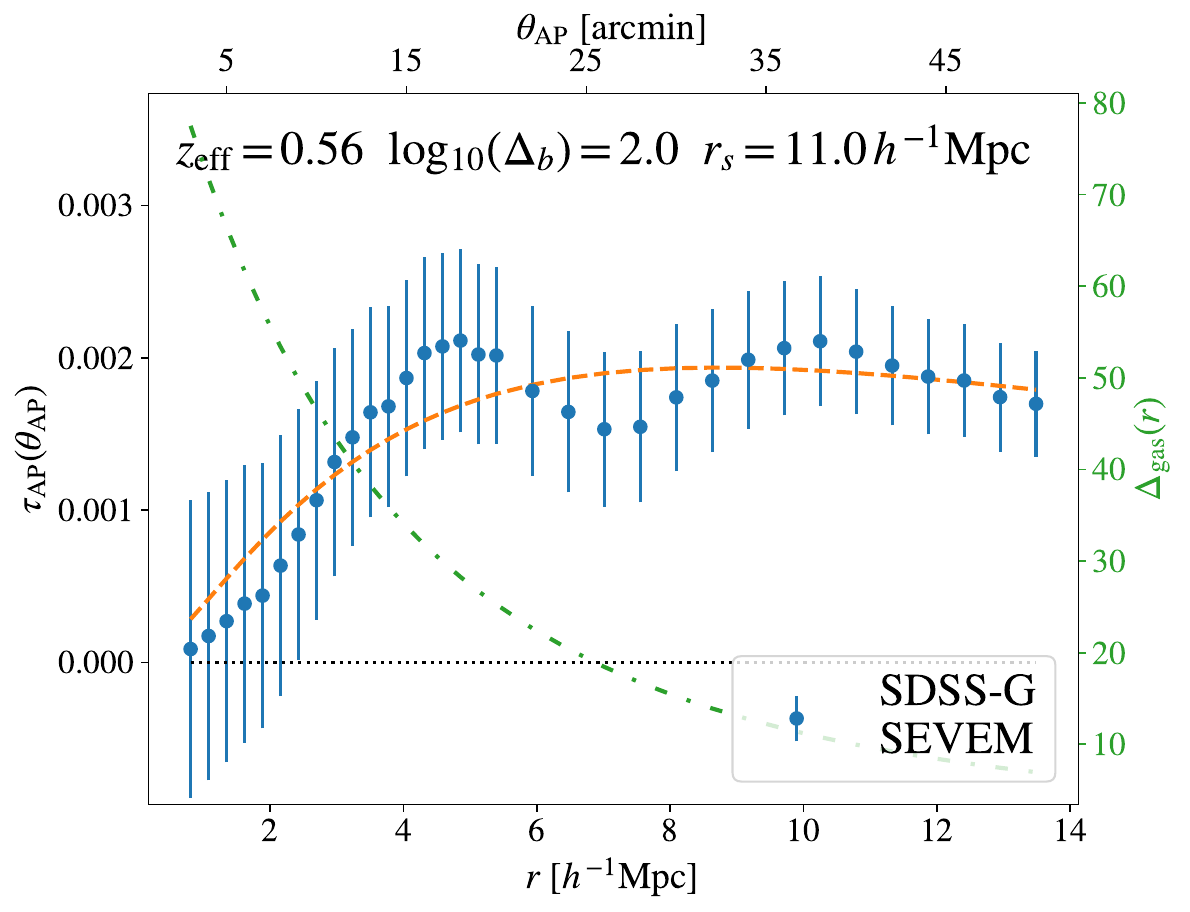} 

\includegraphics[width=0.32\textwidth,height=0.245\textwidth]{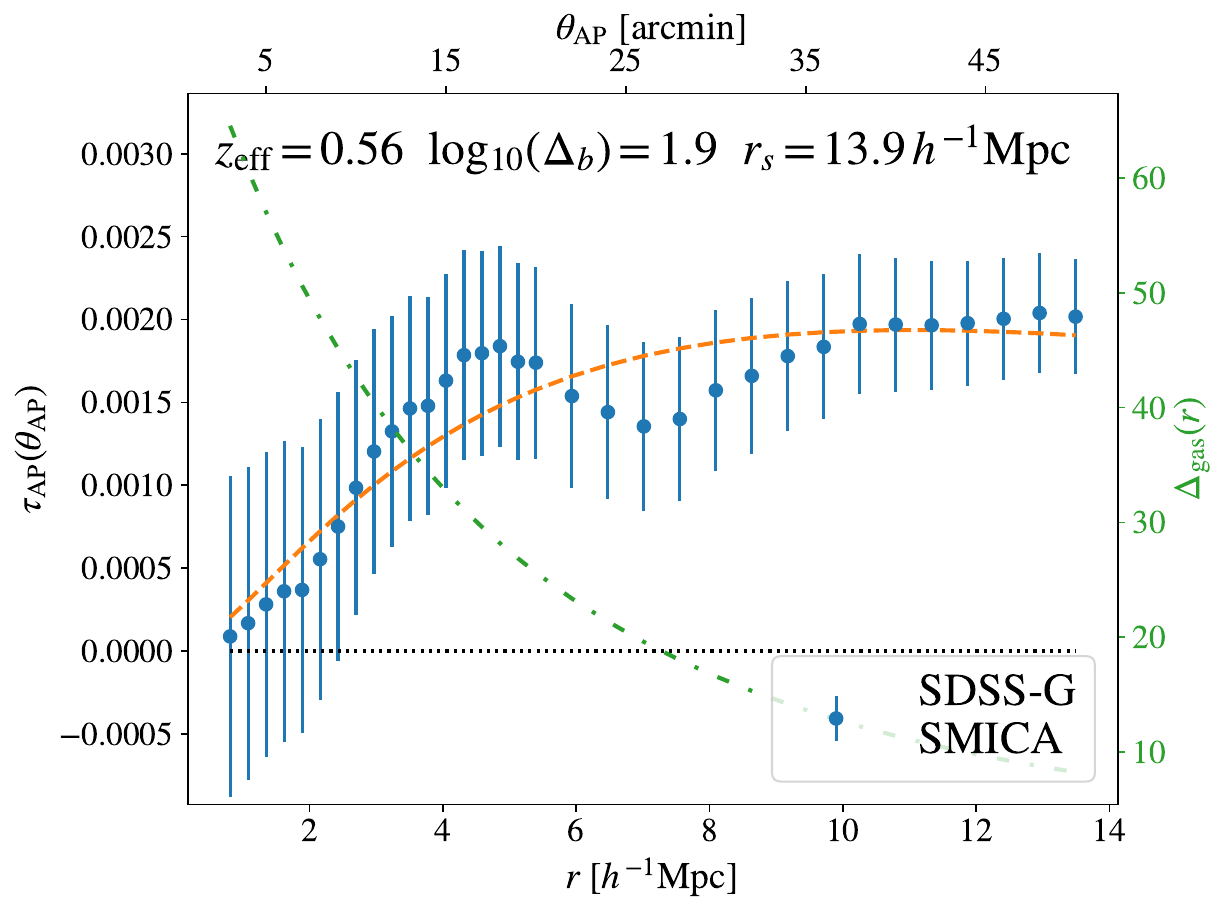}
\includegraphics[width=0.32\textwidth,height=0.245\textwidth]{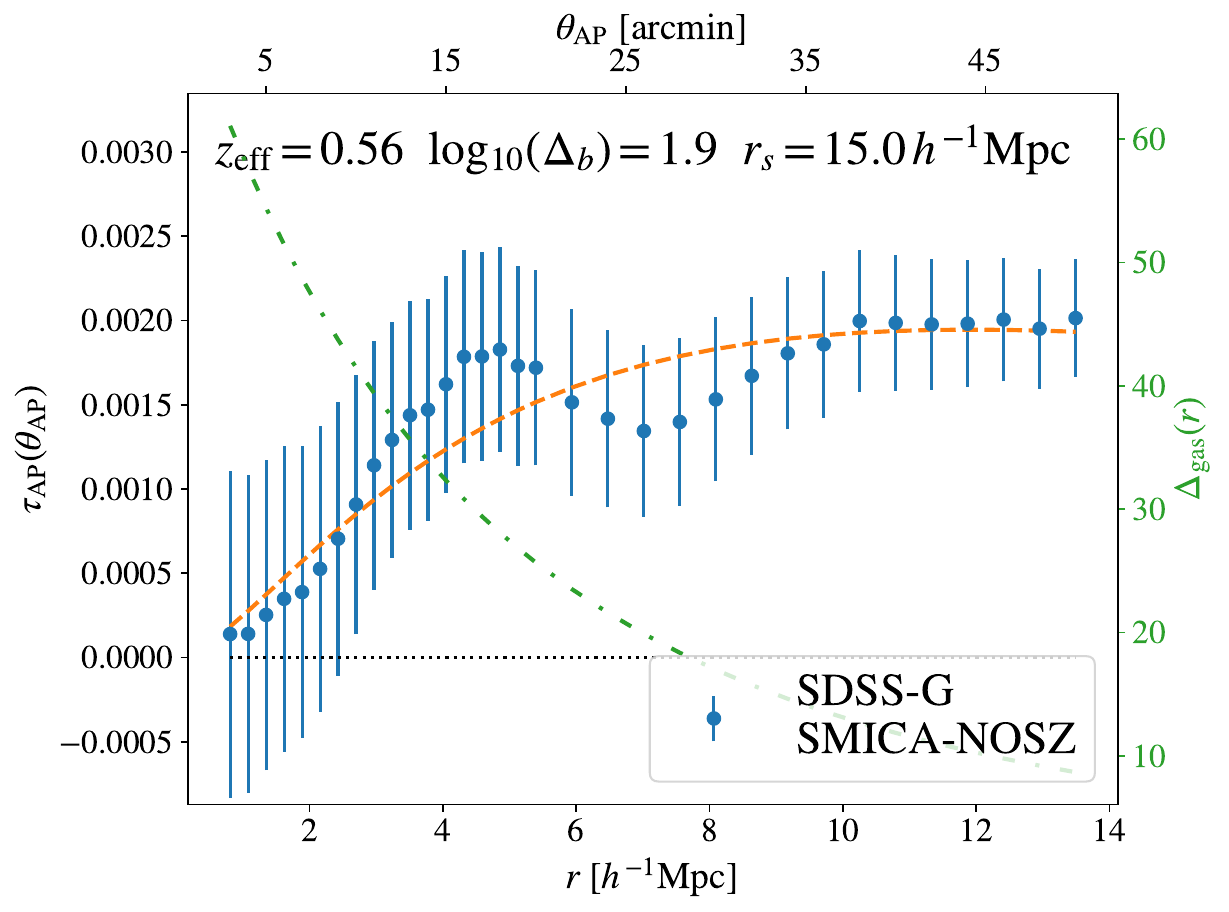}
\includegraphics[width=0.32\textwidth,height=0.245\textwidth]{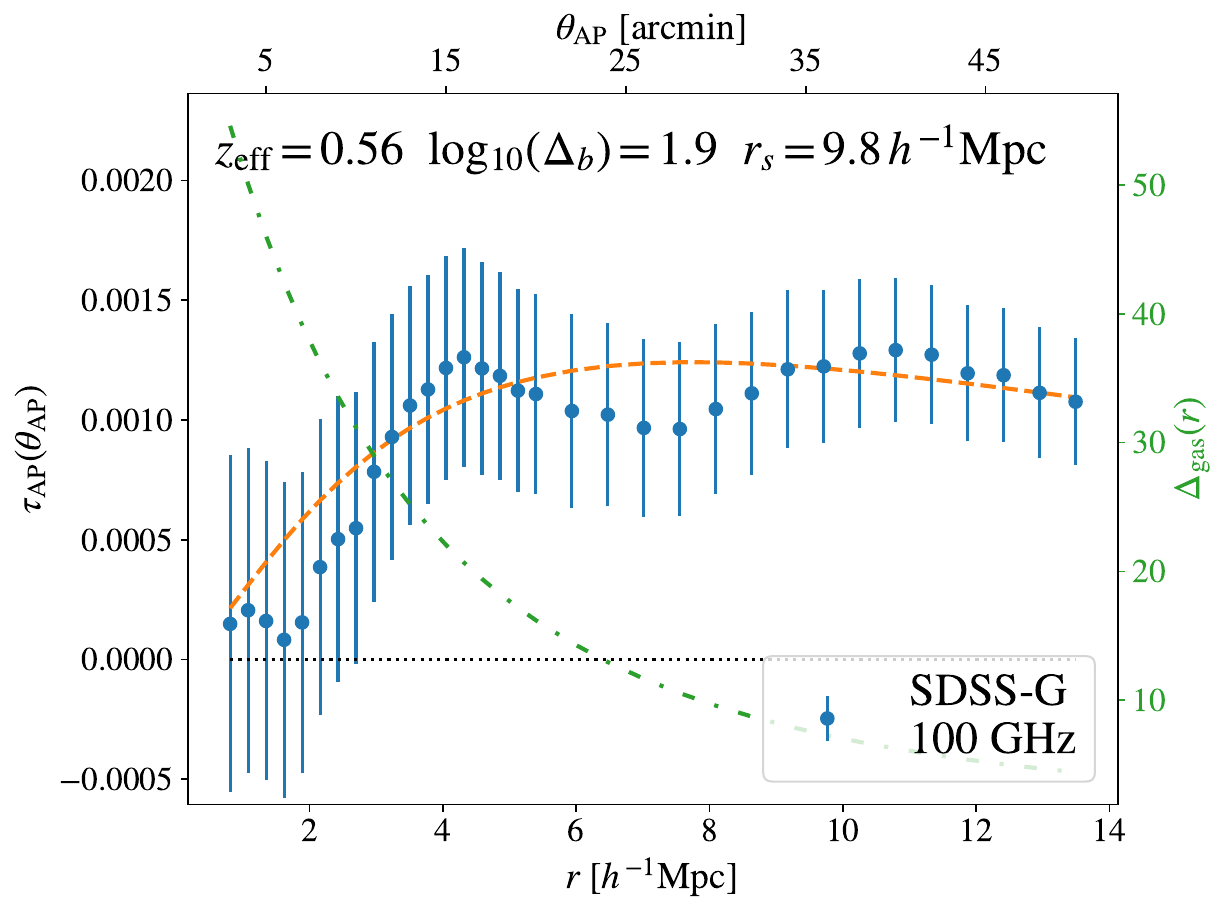}

\includegraphics[width=0.32\textwidth,height=0.245\textwidth]{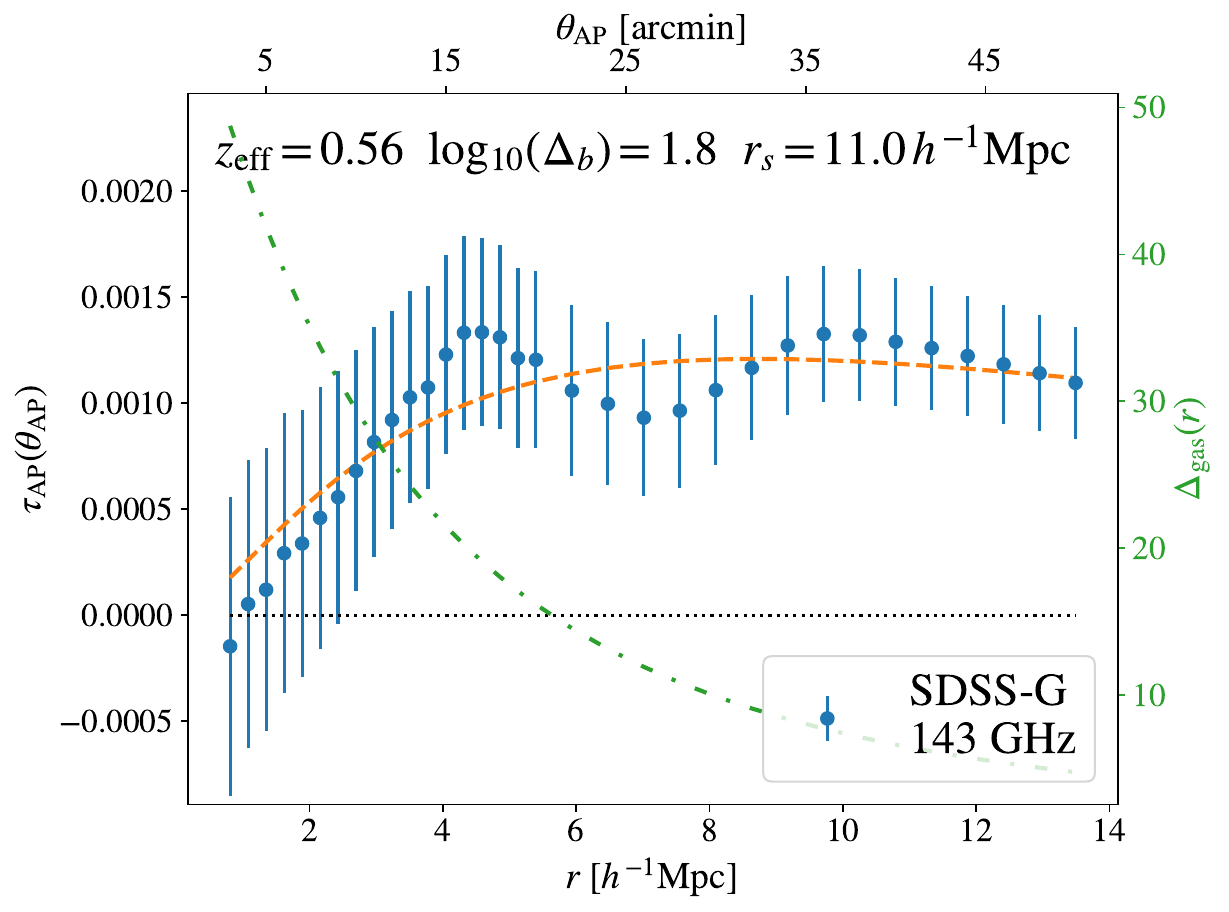}
\includegraphics[width=0.32\textwidth,height=0.245\textwidth]{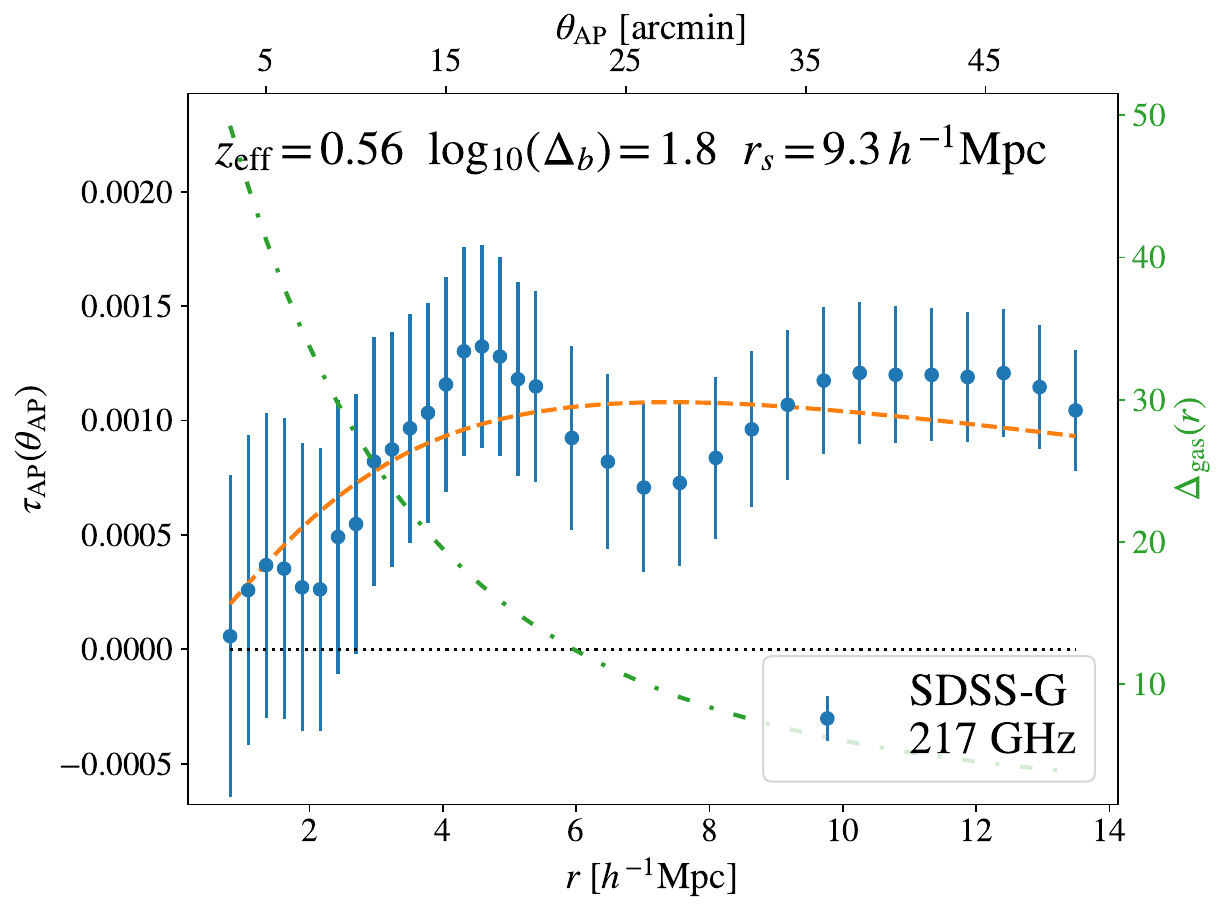}
\includegraphics[width=0.32\textwidth,height=0.245\textwidth]{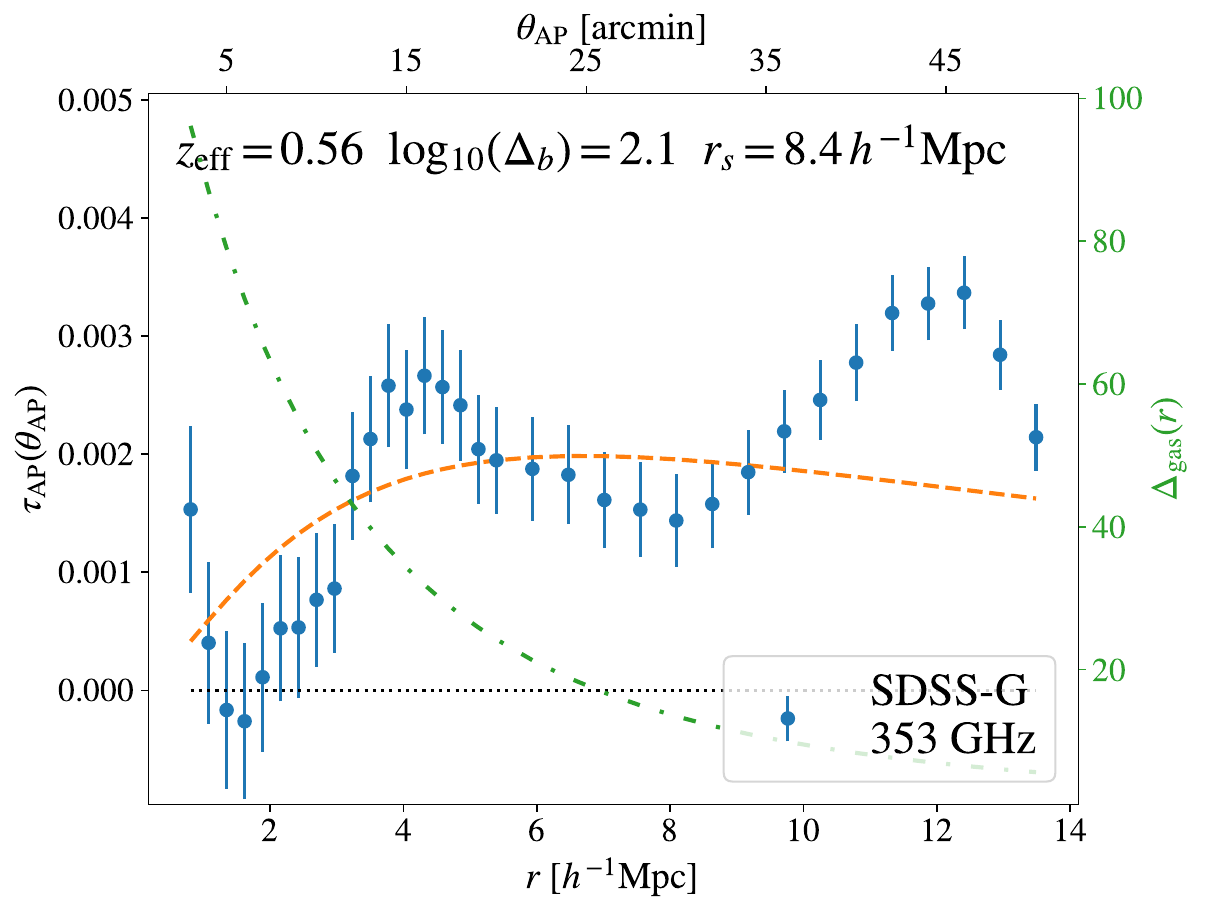}

\end{center}
\caption{
\label{fig:app_tau}
Impact of CMB contaminats on kSZ optical depth measurements extracted from SDSS galaxies at $z_{\rm cen}=0.56$. The first five and last four panels (counting from top left to bottom right) show measurements extracted from different foreground-reduced and raw \Planck maps, respectively. As we can see, kSZ measurements extracted from all but the 353 GHz map present similar amplitude and angular dependence, which proves the robustness of ARF-kSZ tomography against CMB contaminants.
}
\end{figure*}

\subsection{CMB foregrounds and tSZ effect}
\label{sub:robustness_foreground}

The stronger sources of foreground contamination to the CMB at low, intermediate, and high frequencies are synchrotron radiation, free-free emission, and Galactic dust, respectively \citep[e.g.,][]{Tegmark2000}, while the most important source of secondary CMB anisotropies is the tSZ effect \citep{sunyaev70, Sunyaev1972}, which refers to the inverse Compton scattering of CMB photons off hot electrons in haloes and filaments. In this section, we address the impact of these on ARF-kSZ tomography.

Thanks to the different spectral shape of foregrounds, the tSZ effect, and primordial CMB anisotropies, it is standard to leverage radio/millimeter observations from distinct frequency bands to alleviate the impact of these contaminants on CMB studies \citep[e.g.,][]{planck16IX}. In particular, the \Planck collaboration uses the algorithms \Commander, \Nilc, \Sevem, and \Smica to separate foregrounds from primordial CMB signal and an improved version of \Smica to reduce tSZ emission as well \citep{Planck2018IV}. Each of these algorithms follows a different strategy to reduce contaminants, and thus it is conceivable to expect different levels of residual contamination on the resulting maps.

In the first five and last four panels of Fig.~\ref{fig:app_tau} (counting from top left to bottom right), we display kSZ optical depth measurements extracted from foreground-reduced and raw \Planck maps using SDSS galaxies at $z_{\rm cen}=0.56$. We compute error bars for raw maps following the same strategy as for foreground-reduced maps in \S\ref{sub:resobv_cov}; therefore, the size of these is likely to be underestimated because we do not simulate the impact of foregrounds. We find that kSZ measurements extracted from all but the 353 GHz map present similar amplitude and angular dependence, which proves the robustness of ARF-kSZ tomography against CMB contaminants. This robustness is also supported by the fact that symbol sizes in Figs.~\ref{fig:max_tau}, \ref{fig:overdensity}, and \ref{fig:baryons}, which indicate the scatter among results from different foreground-reduced maps, are smaller than error bars, which capture the impact of other sources of uncertainty. Note that measurements extracted from the 353 GHz map are different from others most likely due to Galactic dust contamination, although this difference is moderate. Given that the angular resolution of raw maps varies from FWHM$\simeq 10$~arcmin for the 100~GHz channel to FWHM$\simeq 5$~arcmin for 353~GHz, it is also important to emphasise the little dependence of the results on the angular resolution of \Planck maps.
 
We can also see that measurements extracted from the \Smica maps with and without reduced tSZ emission present great consistency; this is not surprising because we do not expect correlation between ARF, kSZ, and tSZ fluctuations as at fixed frequency the tSZ effect induces same sign anisotropies, while ARF and the kSZ effect produce positive and negative variations.


\begin{figure}
\begin{center}
\hspace*{-1cm}\includegraphics[width=\columnwidth]{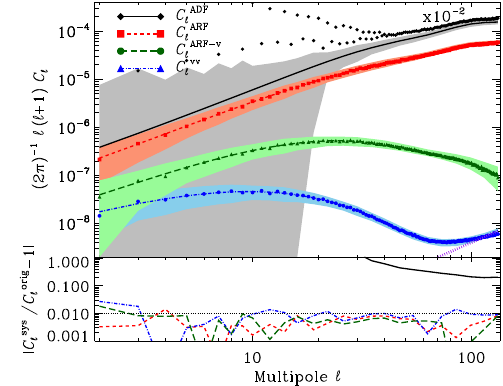}
\end{center}
\caption{
\label{fig:cls_systematics}
Impact of systematics modulating the angular number density of sources on the cross-correlation of ARF and the kSZ effect. We use the same coding as in Fig.~\ref{fig:cls_results}. Symbols indicate measurements from simulations, while lines denote theoretical predictions not accounting for systematics. In the bottom panel, we display the relative difference between the power spectra of maps with and without systematics. As we can see, the impact of this type of systematics on ARF-kSZ tomography is below 1\% across the whole range of multipoles shown.
}
\end{figure}

\subsection{Systematics affecting the angular number density of sources}
\label{sub:robustness_numang}

In \S\ref{sec:foundations}, we argue that ARF and filtered CMB maps are robust against systematic uncertainties modulating the angular number density of sources, which are induced by multiple effects such as seeing, sky background, airmass, galactic extinction, and stellar density \citep[e.g.,][]{ross17}. To study the impact of angular systematics, we produce sky maps following the same procedure as in \S\ref{sub:thsim_precision} while modulating the angular number density of dark matter particles according to a \Planck map of Galactic extinction\footnote{\url{http://pla.esac.esa.int/pla/aio/product-action?MAP.MAP_ID=COM_CompMap_Dust-DL07-AvMaps_2048_R2.00.fits}} \citep{planck14IX}. 

In Fig.~\ref{fig:cls_systematics}, we display the power spectra of ADF, ARF, and radial peculiar velocity maps created following this procedure as well as the cross-correlation of the last two. We remind the reader that such correlation presents the same dependence on cosmology as the cross-correlation of ARF and the kSZ effect (see \S\ref{sub:thsim_precision}). Symbols indicate measurements from simulations, while lines denote theoretical predictions not accounting for systematics. In the bottom panel, we display the relative difference between the power spectra of maps with and without systematics. As we can see, the impact of angular systematics on the power spectra of ARF and radial peculiar velocities and the cross-correlation of these is weaker than 1\%, letting us conclude that ARF-kSZ tomography is very robust against this type of systematics. On the other hand, we find that the impact of angular systematics on the power spectrum of ADF maps is substantial across the whole range of multipoles shown.


\subsection{Large-scale bias of tracers}
\label{sub:robustness_linbias}

In \S\ref{sec:resobv}, we extract kSZ measurements while holding fixed cosmological parameters and the large-scale bias of tracers. Propagating uncertainties in these throughout our methodology is straightforward; nonetheless, large-scale bias uncertainties are not usually provided in the literature. To study the impact of uncertainties in the large-scale bias of tracers, we extract kSZ optical depths from all samples considered in \S\ref{sec:resobv} while assuming slightly different bias values. Overall, we find that an increment in large-scale bias translates into an analogous decrement in kSZ optical depth and that this proportionality is maintained for perturbations modifying the bias up to a factor of two. This is because density terms dominate the cross-correlation of ARF and filtered CMB maps at all redshifts, and these terms are directly proportional to the large-scale bias of tracers (see \S\ref{sec:foundations}). The main consequence of this dependence is that underestimating (overestimating) the large-scale bias of tracers results in overestimating (underestimating) the cosmic baryon fraction.


\subsection{Gas functional form}
\label{sub:robustness_gprofile}

As we discuss in \S\ref{sec:gprop}, setting constraints on the properties of kSZ gas requires adopting a functional form for the distribution of this gas. To study the dependence of constraints on the location, density, and abundance of baryons on the profile functional form, we repeat the same analysis as in \S\ref{sub:gprop_inference} but using a double exponential and a Gaussian profile instead of a $\beta$-profile. We find that the $\beta$- and double exponential profiles result in similar constraints; for instance, the relative difference between the value of $f_\mathrm{collapse}$ that we find for each of these is smaller than 10\%. On the other hand, the Gaussian profile yields noticeably different results. Taken together with the fact that only the first two are flexible enough to capture the distribution of gas surrounding haloes in hydrodynamical simulations (see \S\ref{sub:thsim_gprofile}), we conclude that physically-motivated functional forms lead to similar constraints on the properties of kSZ gas.


\section{Summary and conclusions}
\label{sec:conclusion}

A complete census of baryons in the late universe is challenging due to the intermediate temperature and rarefied character of the majority of the cosmic gas. In this work, set constraints on the location, density, and abundance of this gas using ARF-kSZ tomography, a new technique that leverages the cross-correlation of angular redshift fluctuations and cosmic microwave background observations to extract measurements of the kinematic Sunyaev-Zel'dovich effect. We proceed to summarise our main findings.

\begin{itemize}

    \item In \S\ref{sec:theory} and \ref{sec:foundations}, we derive the dependence of the cross-correlation of ARF and the kSZ effect on cosmology to first order in perturbation theory. Then, we use $N$-body simulations to quantify the precision of our derivations in \S\ref{sec:thsim}, finding that theory and simulations agree to within 5\% across the scales of interest.
    
    \item In \S\ref{sec:resobv}, we produce angular redshift fluctuations and high-pass filtered \Planck maps at 16 different redshifts between $z=0$ and 5 using 6dF galaxies, BOSS galaxies, and SDSS quasars as tracers, and then we cross-correlate these maps to extract kSZ measurements. Remarkably, we detect significant for a wide range of redshifts and filter apertures, yielding a $11\sigma$ detection of the kSZ effect.
    
    \item In \S\ref{sec:gprop}, we leverage kSZ measurements to set constraints on the location, density, and abundance of gas inducing the kSZ effect. In Fig.~\ref{fig:overdensity}, we show that more than 99\% of kSZ gas resides outside haloes and that the density of this gas ranges from 10 to 250 times the cosmic average, which is in agreement the density of the gas in filaments and sheets according to hydrodynamical simulations. Then, in Fig.~\ref{fig:baryons} we show that our kSZ measurements are compatible with detecting 50\% of cosmic baryons. Taken together, these findings indicate that ARF-kSZ tomography provides a nearly complete census of intergalactic gas from $z=0$ to 5.
\end{itemize}

Throughout this work, we extract kSZ measurements while holding fixed cosmological parameters due to the modest signal-to-noise of the cross-correlation of ARF and filtered CMB maps. Nonetheless, upcoming CMB experiments like CMB-S4 and galaxy surveys such as DESI, Euclid, J-PAS, SPHEREx, and WFIRST will deliver much preciser observations, which opens the door to set constraints on cosmology using ARF-kSZ tomography. It is important to note that these constraints will be very reliable due to the robustness of ARF-kSZ tomography against systematics (see \S\ref{sec:robustness}) and because this observable only extracts information from linear, well-understood scales.


\section*{Acknowledgements}

We acknowledge useful discussions with Giovanni Aric\`{o}, Lindsey Bleem, Salman Habib, Andrew Hearin, Guillaume Hurier, Eve Kovacs, and Patricia Larsen. Argonne National Laboratory’s work was supported by the U.S. Department of Energy, Office of Science, Office of Nuclear Physics, under contract DE-AC02-06CH11357. The authors acknowledge support from the Spanish Ministry of Science, Innovation, and Universities through the projects AYA2015-66211-C2-2 and PGC2018-097585-B-C21, and the Marie Curie CIG PCIG9-GA-2011-294183. R.E.A. acknowledges support from the European Research Council through grant number ERC-StG/716151. We gratefully acknowledge use of the science cluster at Centro de Estudios de F\'isica del Cosmos de Arag\'on (CEFCA) and the Phoenix cluster at Argonne National Laboratory, which is jointly maintained by the Cosmological Physics and Advanced Computing group and the Computing, Environment, and Life Sciences Directorate. We thank the \Planck, 6dF, and SDSS collaboration for making their data publicly available. J.C. dedicates this work to the memory of his grandfather, Fernando Montero Eugenio, who inspired his passion for astrophysics.

\bibliographystyle{mnras}
\bibliography{biblio}


\appendix


\bsp
\label{lastpage}
\end{document}